\documentclass[%
 preprint,
 amsmath,amssymb,
 aps,
 prd,
 floatfix,
 sort&compress,
 merge,
 longbibliography
]{revtex4-2}

\usepackage{graphicx}
\usepackage{bm}
\usepackage{color}
\usepackage{amsmath}
\usepackage{bbm}
\usepackage{subfigure}
\usepackage[bookmarks=true,colorlinks=true,linkcolor=blue,unicode=true]{hyperref}

\begin{document}

\preprint{APS/123-QED}

\title{Simple Routes to Turing-Like Patterns}%

\author{Zhan Zhao}
\email{zhaozhan061300@163.com}
\author{Xiang-Ning Li}
\email{2788249040@qq.com}
\author{Wen-Wen Li}
\email{lww1514148@163.com}
\author{Ji-Chong Yang}%
\email{yangjichong@lnnu.edu.cn}
\thanks{Corresponding author}
\affiliation{%
 Liaoning Normal University, Dalian 116029, China
}%

\date{\today}

\begin{abstract}
Spots, stripes, and networks recur in systems with very different microscopic dynamics, suggesting that the competing processes behind Turing-like patterns may be easy to realise. We ask whether local amplification coupled to suppression over a wider region can generate these geometries in simple systems beyond chemical reaction--diffusion models. We test this idea in two settings, a binary cellular automaton obtained by adding a distant neighbour count to Conway's Game of Life and a self-gravitating toy universe described by particle motion and two continuous fields. The automaton produces spots, stripes, and coexistence across its control plane, while a rule retaining only the two opposing counts produces islands, stripes, and holes. Directional correlations and intrinsic dimension identify the same region of orientational order. In the gravitational system, convergence increases local density while mass transport depletes the surroundings. The two-field and particle densities have a smoothed spatial correlation of $0.962$ at scale factor $a=1$ under the original force law, and strengthening its long-wavelength component drives both descriptions toward fewer dense condensations. Together, these results show how simple counting rules and redistribution of a conserved quantity can realise competing growth and suppression, providing two concrete routes by which Turing-like geometries can recur across different dynamics.
\end{abstract}

\maketitle

\section{\label{sec:1}Introduction}

Two fields that react with one another and spread at different rates can settle into a stationary spatial pattern instead of into a uniform mixture.
Turing's 1952 argument made the conditions precise. 
A state that is stable against every spatially uniform disturbance may still be unstable against a disturbance of one particular finite wavelength~\cite{Turing1952}.
Three ingredients are involved.
Something has to amplify itself wherever it is already abundant. 
Something has to oppose that amplification. 
And the two have to act on different lengthscales, with the opposition reaching further than the amplification.
The linear growth rates favour a band of wavelengths. The boundary conditions restrict the allowed modes, and nonlinear evolution and the initial condition can select among them.
Such a pattern is self-organised.
Because the list of ingredients is so short, the mechanism has been worked out in a large number of model systems~\cite{Gierer1972,Gray1983} and has become the standard account of a wide class of natural textures, from animal coats to chemical reaction fronts~\cite{Kondo2010,Vittadello2021}.
Its canonical expression, however, is a pair of coupled partial differential equations~\cite{Cross:1993zz,Murray2003}, and it is worth asking how much of that expression belongs to the mechanism itself and how much belongs only to the way it is usually written down.

The visible content of a Turing pattern is small enough to list, and the list is worth stating because it keeps recurring.
Spots are isolated maxima of the patterned quantity, in the simplest cases arranged on a triangular array. 
Stripes, or rolls, are parallel bands of high and low value. 
A labyrinth is the maze-like variant in which the bands wander instead of running straight. 
Holes are spots with the two roles exchanged. 
Coexistence denotes two of these forms occurring side by side in the same sample.
Which of them occurs is decided by a small number of parameters, so that one and the same system can be walked from spots through a labyrinth to stripes and back.

These forms appear in systems that have neither chemistry nor diffusion.
A magnetic skyrmion is a whirl-like configuration of the local spin direction, protected against unwinding by its topology. 
Where many of them condense they form a periodic lattice, and stripe-like domains occur alongside it in the same materials~\cite{Nagaosa:2013ftn,Tokura2014}.
In QCD the colour-electric field of a static quark--antiquark pair is not spread through space but gathered into a narrow tube, and several static sources are joined by networks of nearly straight segments~\cite{Isgur:1983wj,Bali:2000gf}.
On the largest scales that can be observed, matter is arranged in a web of filaments and sheets enclosing voids~\cite{Bond:1995yt,Springel:2006vs}.
At the level of appearance, and only at that level, these are the elements of the list above. Round objects on a regular array, parallel bands, a connected network of bands and nodes, and empty regions between them.

It is this resemblance, and nothing more, that the present work is about, so we should be exact about what a resemblance can carry.
Shape is cheap. 
Any system with a preferred length and a tendency to fill space produces bands and blobs, and the same picture can arise from unrelated causes.
What would be remarkable is not the shared vocabulary but a shared set of requirements (short-range self-amplification, a longer-range opposition, and a separation between the lengthscales on which the two act).
The question is therefore not whether the systems above resemble Turing patterns, which they do, but whether the Turing ingredients are the ones actually operating in them.

To answer that question directly one would have to identify the two fields and exhibit the instability system by system, which is beyond the scope of this paper and, in some of the cases above, beyond the present understanding of the system itself.
We therefore invert it. 
Instead of asking whether the Turing mechanism is at work in a given system, we ask how much that mechanism demands.
The distinction matters because the two questions carry opposite logical weight.
If the mechanism needs the whole apparatus of coupled reaction--diffusion equations (continuous amplitudes, a diffusion operator, two diffusion constants with the inhibiting one the larger, and a uniform steady state that is unstable only to a finite wavelength), then every system in which it operates is special, and its appearance in a confining gauge theory or in the large-scale distribution of matter has to be argued for case by case.
If instead it survives the removal of all of that apparatus, then finding it anywhere ceases to be surprising, because so little is required of the host system for it to be there.
The inversion is limited in a way that should be stated plainly. 
Demonstrating that a mechanism is easy to realise cannot prove that any particular system realises it.
What it does is change how surprising such a claim would be, and tell us what kind of system is worth looking at.

We look for the mechanism in the least promising kind of system we know how to build, a cellular automaton. 
A cellular automaton assigns a state from a finite set to each lattice site and updates all sites simultaneously from their neighbourhoods~\cite{Berlekamp1982}. 
Conway's Game of Life uses two states, alive and dead, and the same eight surrounding sites for both birth and survival~\cite{Gardner1970,Berlekamp1982}. It therefore provides a simple starting point for testing what changes when the rule acquires a second interaction range.
Modifying the rule is not itself new. The Larger than Life family generalises nearest-neighbour outer-totalistic rules to large neighbourhoods~\cite{Evans2001,Evans2003}, and Turing patterns have been placed on discrete supports, where a network or lattice replaces the continuous domain while the reaction--diffusion dynamics is kept~\cite{Muolo2024}. 
The network reaction--diffusion models retain continuous state variables. 
Here we add a more distant neighbour count to a binary update rule, with two integer thresholds controlling its contributions to birth and survival. 
We also test a simpler rule in which the nearby count favours activity and the distant count suppresses it. 
The resulting spots, stripes, and labyrinths are characterised by six observables and by the intrinsic dimension, an estimate of the number of independent parameters needed to describe the set of long-time configurations. 
We map their dependence on the thresholds, including a region where stripes align with either lattice axis. 
We use phase-transition language to describe abrupt morphological changes, coexistence, and slow dynamics. Establishing thermodynamic transitions would additionally require a systematic study of lattice size.

Because the inverted question yields only indirect evidence, we also take one of the systems listed above and put the original question to it.
The cosmic web is one case in which the question can be posed self-containedly, and the one we take up here is a box of particles that attract one another and do nothing else, with no pressure, no chemistry and no imposed template.
Gravity is purely attractive, so an activator--inhibitor reading of it is the last thing one would expect, yet mass conservation supplies both halves by itself. 
A growing concentration takes its matter from its surroundings, and two concentrations drawing on the same surroundings suppress one another.
In Sec.~\ref{sec:5} we compare particle trajectories with a two-field description of the same mass transport, test the roles of convergence and transport, and change the force to strengthen its long-wavelength component. Together with the automaton, this provides a second test of whether simple dynamics can generate the competition and morphological changes motivating the paper.

The remainder of this paper is organised as follows.
Sec.~\ref{sec:2} states the Turing mechanism and the Game of Life in the terms used below.
Sec.~\ref{sec:3} defines the modified rule and the patterns it produces.
Sec.~\ref{sec:4} scans the two thresholds, defines the six quantities used to classify the resulting configurations, and presents the morphology map of the threshold plane.
Sec.~\ref{sec:5} examines whether gravitational mass transport in a minimal self-gravitating toy universe admits an effective activator-depleted description, tested in matched particle and continuum calculations.
Sec.~\ref{sec:6} summarises the results.

\section{\label{sec:2}Turing patterns and Conway's Game of Life}

Turing patterns and the Game of Life are usually read as unrelated. 
One is a pair of coupled differential equations for continuous concentrations, the other a binary rule on a lattice. 
They agree in that whatever order appears on a large scale is produced by rules that act only between neighbours, and they differ in how many spatial scales those rules speak about. 
The Turing mechanism needs a short range of amplification and a longer range of opposition, whereas the Game of Life has one scale only. 
This section sets out each ingredient and fixes the terms used below.

\subsection{The Turing mechanism}\label{sec:turing}

Turing's 1952 insight~\cite{Turing1952} was that a uniform mixture of two chemicals can develop spatial structure on its own. 
The mechanism needs two species that act on different ranges. 
An `activator' promotes its own production, and an `inhibitor' that suppresses the activator and diffuses fast enough to reach further than the activator spreads. 
A Turing instability occurs when diffusion destabilises a spatially uniform steady state that is stable without diffusion~\cite{Turing1952}. Consider two reacting and diffusing concentrations,
\begin{align}
\frac{\partial u}{\partial t} &= D_u \nabla^2 u + f(u,v), \label{eq:rd1}\\
\frac{\partial v}{\partial t} &= D_v \nabla^2 v + g(u,v), \label{eq:rd2}
\end{align}
where $u(\mathbf r,t)$ and $v(\mathbf r,t)$ depend on position $\mathbf r$ and time $t$, $D_u,D_v>0$ are constant diffusion coefficients, $\nabla^2$ is the spatial Laplacian, and the differentiable functions $f$ and $g$ give the local production and removal rates. 
We use an unbounded domain for the mode calculation below.

Let $(u_0,v_0)$ satisfy $f(u_0,v_0)=g(u_0,v_0)=0$. 
The matrix of reaction-rate derivatives at this state is
\begin{equation}
J = \begin{pmatrix} f_u & f_v \\ g_u & g_v \end{pmatrix}
\label{eq:jac}
\end{equation}
where subscripts denote partial derivatives evaluated at $(u_0,v_0)$. 
Write the perturbations as the real parts of $(u-u_0,v-v_0)=\epsilon(U,V)\mathrm{e}^{\lambda t+\mathrm i\mathbf q\cdot\mathbf r}$, where $\epsilon\ll1$, $U,V$ are fixed amplitudes, $\lambda$ is the growth rate, and $\mathbf q$ is a fixed wave vector with $q=|\mathbf q|$ and wavelength $2\pi/q$ for $q>0$. 
Substituting this form into Eqs.~\eqref{eq:rd1} and \eqref{eq:rd2}, keeping only terms of order $\epsilon$, and using $\nabla^2\mathrm{e}^{\mathrm i\mathbf q\cdot\mathbf r}=-q^2\mathrm{e}^{\mathrm i\mathbf q\cdot\mathbf r}$, we obtain an eigenvalue problem $\lambda\begin{pmatrix}U \\ V\end{pmatrix}=L(q)\begin{pmatrix}U \\ V\end{pmatrix}$, where $L(q)=J-q^2\operatorname{diag}(D_u,D_v)$ and $\operatorname{diag}(D_u,D_v)$ is the diagonal matrix with those two entries. 
Its eigenvalues satisfy $\lambda^2-\operatorname{tr}L\,\lambda+\det L=0$. 
At $q=0$, stability therefore requires $\operatorname{tr}J=f_u+g_v<0$ and $\det J=f_ug_v-f_vg_u>0$. 
Under these conditions, $\operatorname{tr}L(q)=\operatorname{tr}J-q^2(D_u+D_v)<0$, so a growing mode requires $\det L(q)<0$. 
Expanding this determinant gives
\begin{equation}
 {\det L(q)=D_uD_vq^4-(D_ug_v+D_vf_u)q^2+\det J.}
 \label{eq:rd_determinant}
\end{equation}
As a quadratic in $q^2$, this expression has its minimum at $q^2=(D_ug_v+D_vf_u)/(2D_uD_v)$. Requiring that minimum to occur at $q^2>0$ and be negative gives the two conditions
\begin{align}
D_u g_v + D_v f_u &> 0, \label{eq:tdi}\\
\left(D_u g_v + D_v f_u\right)^2 &> 4D_u D_v\det J . \label{eq:turing}
\end{align}
Both inequalities are necessary. For a self-amplifying component with $f_u>0$ and a locally damped second component with $g_v<0$, Eq.~\eqref{eq:tdi} requires $D_v/D_u>|g_v|/f_u>1$, where the last inequality follows from $\operatorname{tr}J<0$. Thus diffusion must spread the opposing effect sufficiently rapidly. On a finite domain, at least one mode allowed by the boundary conditions must also have $\det L(q)<0$.
These linear conditions determine the onset of growth. 
The final amplitude and morphology require the nonlinear dynamics and do not follow from the instability criterion alone.

Well-known models that exhibit Turing patterns include the Gierer--Meinhardt model~\cite{Gierer1972} and the Gray--Scott model~\cite{Gray1983}.
Taking the latter as an example,
\begin{align}
\frac{\partial u}{\partial t} &= D_u\nabla^2 u + u^2v - (F+k)u, \label{eq:gs1}\\
\frac{\partial v}{\partial t} &= D_v\nabla^2 v - u^2v + F(1-v), \label{eq:gs2}
\end{align}
written in dimensionless variables. Here $u$ is the activator and $v$ the inhibitor, $F>0$ is the feed and dilution rate, and $k>0$ is the additional removal rate of $u$. The reaction $u^2v$ produces $u$ while consuming $v$, so growth reduces the supply available to support further growth. The opposing effect is therefore depletion of $v$ rather than a direct suppression of $u$. At $u,v>0$, the derivatives $f_v=u^2>0$ and $g_u=-2uv<0$ express the promotion of $u$ by $v$ and the consumption of $v$ by $u$, respectively. The criterion above applies at a uniform steady state satisfying $\operatorname{tr}J<0$ and $\det J>0$.

\subsection{Conway's Game of Life}\label{sec:gol}

Conway's Game of Life is a cellular automaton on a square lattice~\cite{Gardner1970,Berlekamp1982}. At each site the state $s$ is either alive ($1$) or dead ($0$). 
Let $t$ now count integer update steps and $N$ be the number of live sites among its four axial and four diagonal neighbours. 
All sites are updated simultaneously from the states at step $t$ according to
\begin{equation}
s(t+1)=\begin{cases}
1, & s(t)=0 \text{ and } N=3,\\
1, & s(t)=1 \text{ and } N\in\{2,3\},\\
0, & \text{otherwise}.
\end{cases}
\label{eq:gol}
\end{equation}
This is commonly abbreviated as rule B3/S23. 
A dead site is `born' if it has exactly three live neighbours, and a live site `survives' if it has two or three.

The rule lacks the three features the mechanism asks for. Its states are binary, so no continuous amplitude is left for diffusion to smooth. Every site reads the same eight neighbours, so the rule carries one spatial scale only. The discrete states admit no infinitesimal perturbation, while the only spatially uniform fixed point is the absorbing empty configuration, which no finite wavelength can destabilise. The binary states do bound the amplitude, but a bound is not a balance between production and removal. It is these absences that make the rule a useful place to add one ingredient at a time.

The decisive absence is the second. Birth and survival read the same eight neighbours, so nothing in the rule acts beyond that range. Sec.~\ref{sec:3} adds a count over a more distant ring, so the rule combines a short and a long range, and asks how much of the mechanism survives without the differential equations.

\section{\label{sec:3}A modified Game of Life}

The Game of Life supplies one range of interaction where the mechanism asks for two, so the modification is to add a second count over a farther ring and to add nothing else. 
The rule remains a deterministic cellular automaton (binary states, one update applied to every site at once, no fields and no differential equations). 
The question we put to it is the one Sec.~\ref{sec:2} leaves open. 
Does that second range do the job the Turing mechanism asks of it?

\subsection{The modified rule}\label{sec:rule}

The stage is a square lattice of $L\times L$ sites with periodic boundary conditions.
The site $(i,j)$ is identified with $(i+L,j)$ and with $(i,j+L)$, so the lattice is a torus, every site has the same surroundings, and no edge can supply the order we are looking for. The examples in this section use $L=128$. 
The scan of Sec.~\ref{sec:4} uses $L=80$. 
A configuration assigns to each site a state $s_{i,j}\in\{0,1\}$, one being live and zero dead, and the starting configuration is drawn by setting each site live independently with probability $\rho_0$, which we call the initial density.

\begin{figure}[htbp]
\centering
\includegraphics[width=0.85\linewidth]{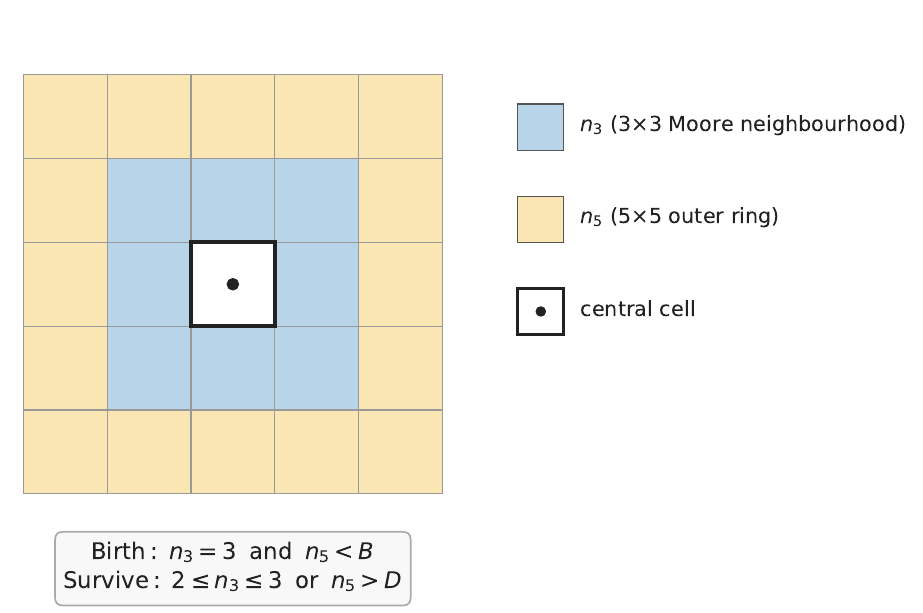}
\caption{Definition of the modified rule.
The blue cells contribute to the short-range count $n_3$ and the yellow ring to the long-range count $n_5$.
Birth requires the standard Game-of-Life condition $n_3=3$ and additionally $n_5<B$. 
Survival requires $2\le n_3\le 3$ or $n_5>D$.}
\label{fig:schematic}
\end{figure}

Following the recipe of the Turing mechanism (Sec.~\ref{sec:turing}), each site keeps the short-range count of the Game of Life and gains a second count over the next ring out (Fig.~\ref{fig:schematic}),
\begin{align}
n_3(i,j) &= \sum_{(k,l)\in\mathcal{N}_3(i,j)} s_{k,l}, \\
n_5(i,j) &= \sum_{(k,l)\in\mathcal{N}_5(i,j)} s_{k,l},
\end{align}
where $\mathcal{N}_3$ is the set of eight nearest neighbours and $\mathcal{N}_5$ the set of sixteen cells forming the outer ring of the $5\times5$ block centred on the site. The two sets are disjoint and together exhaust that block apart from its centre, so the rule reads the twenty-four cells that surround a site and nothing beyond them.
The modified rule is parameterised by two integers $(B,D)$,
\begin{align}
\text{birth:}   \quad & s=0 \to 1 \quad \text{if } n_3=3 \text{ and } n_5<B, \\
\text{survival:}\quad & s=1 \to 1 \quad \text{if } 2\le n_3\le 3 \text{ or } n_5>D.
\end{align}
All other cells die.

The distant count thus enters twice, with opposite signs, and the two must be separated before any pattern is shown. 
At birth it opposes. 
A cell with exactly the right number of near neighbours is still denied if the ring about it is busy, so distant activity suppresses new life, and $B$ is how much of that activity is tolerated. At survival it supports. 
A cell that the Game of Life would kill is kept alive once enough of the ring is live, so distant activity reinforces what is already there, and $D$ is how much reinforcement counts. 
One added range therefore supplies both halves of the vocabulary Sec.~\ref{sec:turing} asks for, and which of the two is at work is a property of the point in the $(B,D)$ plane rather than of the rule. That is a nuisance for interpretation, and Sec.~\ref{sec:pure} removes it by discarding the Game-of-Life conditions and keeping the opposing term alone.

Two limits fix the relation to the original game. The count $n_5$ cannot exceed sixteen, so any $B\ge17$ leaves the birth condition as the Game of Life has it and any $D\ge17$ leaves the survival condition as it is. 
At $(B,D)=(17,17)$ the rule is exactly B3/S23. The scan of Sec.~\ref{sec:4} covers $0\le B,D\le17$ and so runs from the unmodified game to the point where the distant count binds at every opportunity.

\subsection{Stripe formation}\label{sec:stripe}

The first question is whether the added range does anything at all, and the least it can do is hold up a state that the unmodified game cannot. Because the update is deterministic and the number of configurations is finite, a trajectory must eventually return to a configuration it has already visited. From that step onward it repeats, cycling with some period $p$. 
Let $T_{\rm rec}$ be the first step at which a configuration repeats and $p$ the period of the resulting cycle. The cycle begins at $T_{\rm rec}-p$. We call that cycle the equilibrium and save the configuration at $T_{\rm rec}-1$, the last distinct state before the repeat, as its equilibrium configuration.
These are the objects measured throughout the paper. 
An equilibrium is a cycle, including a fixed configuration when $p=1$, but for the short cycles found here a single snapshot already carries the morphology.

\begin{figure}[htbp]
\centering
\includegraphics[width=0.95\linewidth]{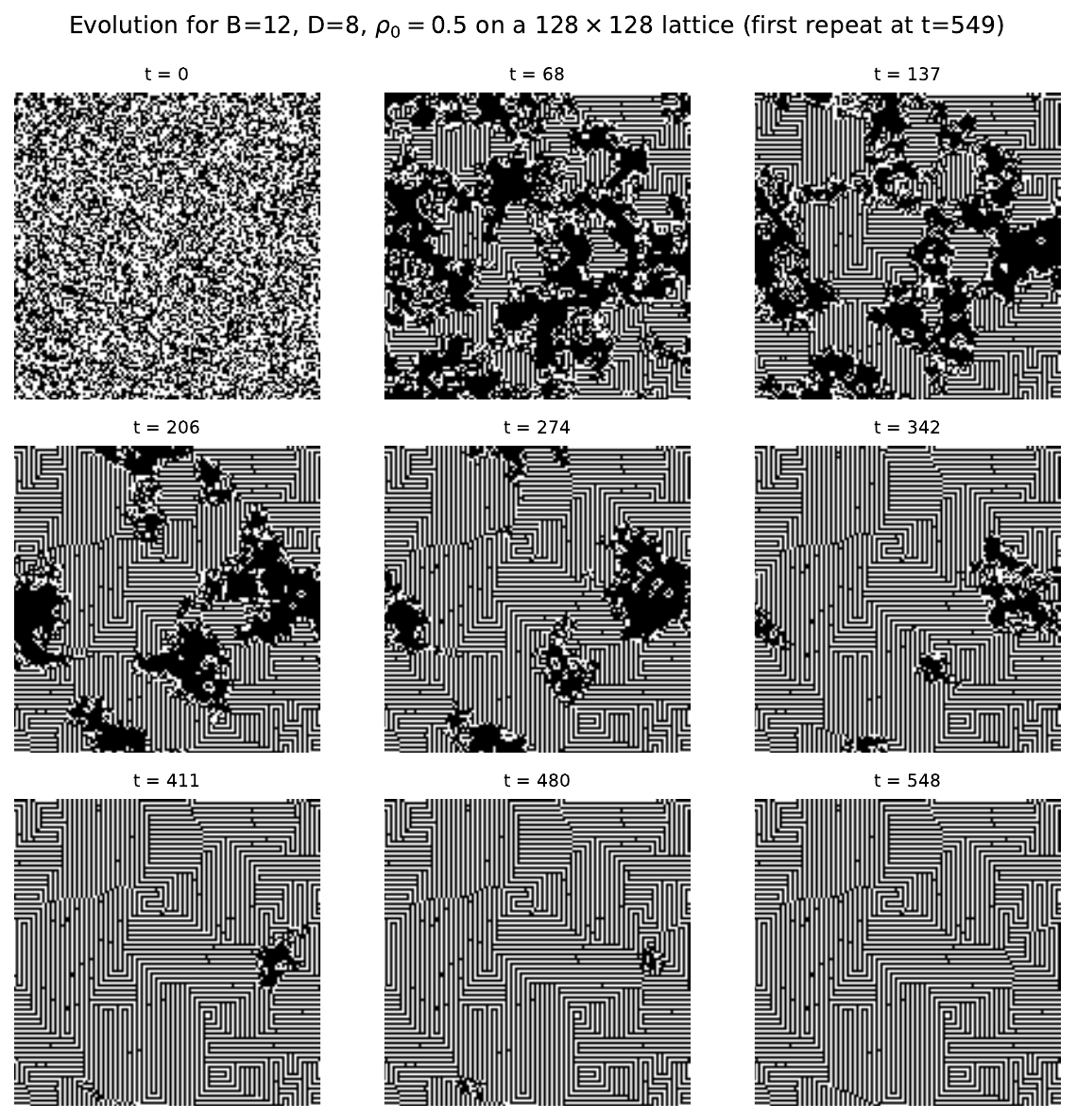}
\caption{Evolution from a random initial state to an equilibrium stripe pattern for $B=12$, $D=8$, $\rho_0=0.5$ on a $128\times128$ lattice with periodic boundary conditions.
Snapshots are taken uniformly from $t=0$ to $548$. The period-six cycle begins at $t=543$ and the first repeated configuration occurs at $t=549$.
Domains of parallel stripes are present within the first hundred steps. 
The remainder of the trajectory is the competition between the two orientations and the annealing of the defects they leave behind.}
\label{fig:evolution}
\end{figure}

Fig.~\ref{fig:evolution} shows one such run, at $(B,D)=(12,8)$ from $\rho_0=0.5$, whose first repeated configuration occurs at $t=549$. 
Within the first hundred the random field has resolved into domains of parallel stripes, but they run in both directions, broad featureless regions remain, and the density is still climbing ($0.34$ at $t=68$ against $0.49$ at the end). Over the four hundred steps that follow the two orientations compete and the defects are expelled one by one. 
The trajectory enters a cycle of period six at $t=543$, and the final displayed snapshot is at $t=548$. Between consecutive cycle states, $33$--$36$ of the $16384$ sites change, or $0.20$--$0.22\%$.
The stripe array is static, and what blinks is the residue of the defects.

What the run selects, and what it does not, is worth separating now because the mechanism cares about only one of the two. The direction is selected in the sense the continuous theory means. 
The rule treats the two lattice axes alike, and this run took one of them. 
The spacing is not. 
The stripes are a single live column separated by a single dead one, two lattice spacings apart, which is the finest a stripe can be on this lattice, so at this point of the plane the scale belongs to the lattice rather than to the rule. Sec.~\ref{sec:pure} returns to the spacing at parameters where the thresholds do set it.

That the pattern is the added range's doing and not the game's can be checked on the same configuration. Started from the equilibrium of Fig.~\ref{fig:evolution}, the unmodified B3/S23 rule falls in 300 steps from $\rho=0.49$ to $\rho=0.07$, about $14.5\%$ of the live cells remaining, and the stripes with them, while the modified rule stays on the cycle it was already on.

\subsection{Controlling the stripe direction}\label{sec:aniso}

A stripe has a direction, and a system that takes one of the two available can be asked what did the taking. The question has a standard answer for reaction--diffusion systems, and the automaton has a counterpart that is read off the same feature of the rule.

\begin{figure}[htbp]
\centering
\includegraphics[width=\linewidth]{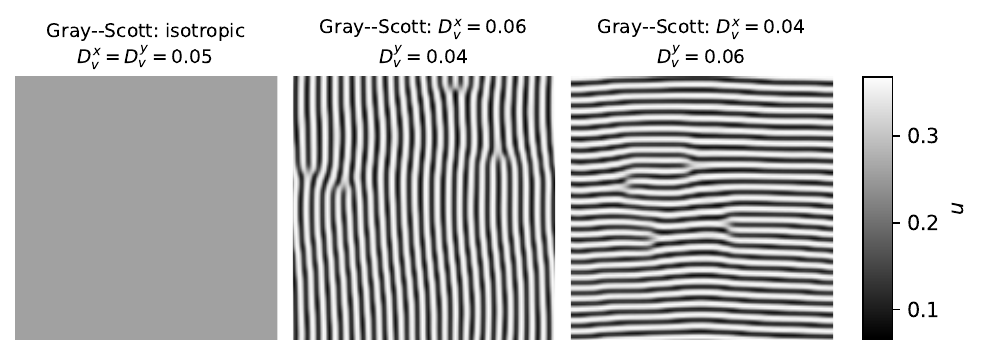}
\caption{The same selection in the continuous model. 
The panels show solutions of the anisotropic Gray--Scott model, Eq.~\eqref{eq:gs_aniso}, shown as the activator $u$ on the scale at the right, with $D_u=0.02$, $F=0.04$, $k=0.057$ and one shared random initial configuration. 
These are parameters at which the nonzero uniform state $(u_0,v_0)=(0.25631,0.37845)$ is stable to every wave number when the diffusion of the inhibitor is isotropic, so that the anisotropy alone can destabilise it. Left $(D_v^x,D_v^y)=(0.05,0.05)$, which stays uniform. 
Middle $(0.06,0.04)$ and right $(0.04,0.06)$ give bands running along $y$ and along $x$.}
\label{fig:aniso_gs}
\end{figure}

In a continuous system the rotational symmetry can be broken by the diffusion itself. Eq.~\eqref{eq:tdi} compares the reach of the inhibition against the strength of the amplification, and the reach of the inhibition along any one direction is set by how far the inhibitor travels in that direction. Make the diffusion direction dependent and the comparison is won first for a modulation along the axis on which the inhibitor travels fastest. The pattern varies most rapidly along that axis, so its bands run parallel to the other. The anisotropic Gray--Scott model supplies the check,
\begin{align}
\frac{\partial u}{\partial t} &= D_u\nabla^2 u + u^2v - (F+k)u, \\
\frac{\partial v}{\partial t} &= D_v^x\frac{\partial^2 v}{\partial x^2} + D_v^y\frac{\partial^2 v}{\partial y^2} - u^2v + F(1-v),
\label{eq:gs_aniso}
\end{align}
where $u$ is the activator and $v$ the inhibitor, as in Eqs.~\eqref{eq:gs1} and \eqref{eq:gs2}. Here $x$ and $y$ are the two spatial coordinates, and $D_v^x$ and $D_v^y$ are the diffusion coefficients of the inhibitor along them. We keep $D_u$ isotropic and vary the two inhibitor coefficients at fixed sum. In Fig.~\ref{fig:aniso_gs}, an equal split leaves the state uniform, while $D_v^x>D_v^y$ makes the bands vertical and exchanging the two coefficients turns them horizontal.

The automaton has no diffusion to make anisotropic, but it has the object that stands in for its range, which is the ring the distant count reads. Restricting that ring to one axis is the discrete counterpart of anisotropic diffusion, and it is a restriction of exactly the kind the criterion cares about, since it changes how much the long range sees in one direction and not in the other. We call the `horizontal' mask the one that counts only the middle three sites of each of the left and right columns of the $5\times5$ ring, and the `vertical' mask the one that counts only the middle three sites of each of the top and bottom rows. Each reads six of the sixteen cells, so the two are matched to one another in every respect but the axis they sense, and either may stand as the other's control.

\begin{figure}[htbp]
\centering
\includegraphics[width=\linewidth]{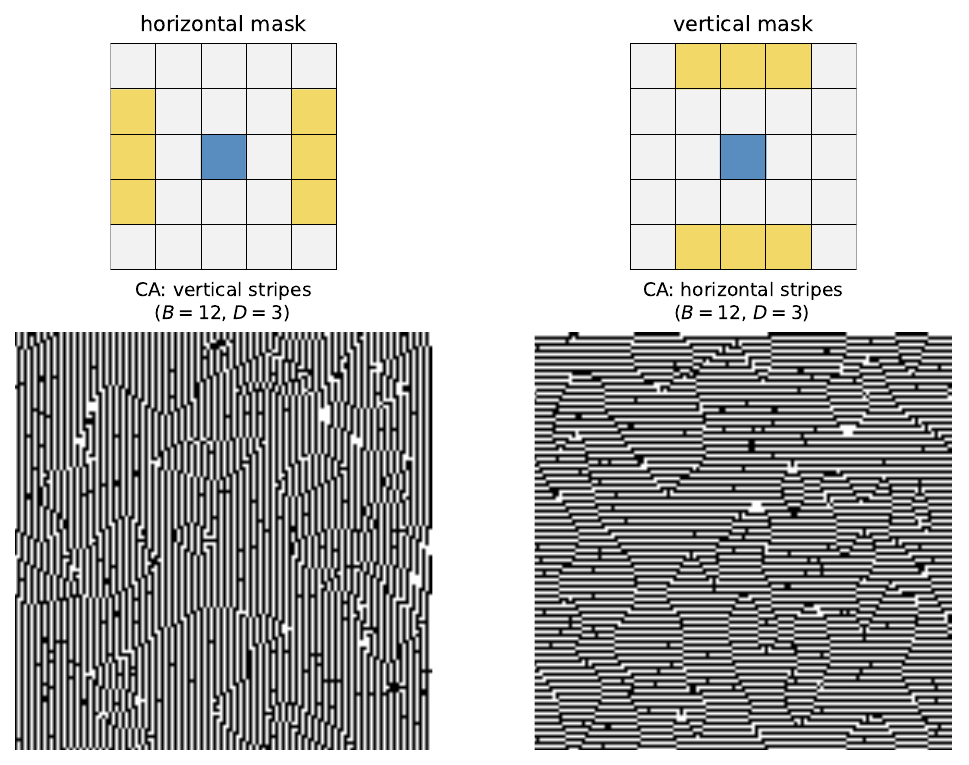}
\caption{The direction of the long-range interaction in the automaton.
Top row shows the two direction-dependent $n_5$ masks, each counting six cells of the $5\times5$ ring, the middle three sites of each of the left and right columns in the horizontal mask, and of each of the top and bottom rows in the vertical one.
Bottom row shows equilibrium configurations of the modified Game of Life at $(B,D)=(12,3)$ and $\rho_0=0.5$ on a $128\times128$ lattice, each run from its own random start. The horizontal mask gives bands running along $y$, the vertical mask bands along $x$. At these parameters the masked count cannot bind the birth gate and acts through the survival clause.}
\label{fig:aniso_ca}
\end{figure}

At the parameters of Fig.~\ref{fig:aniso_ca}, $(B,D)=(12,3)$, the masked count is by construction too small for the birth gate to bind, since six is all the mask can return and $B=12$, so the direction is carried entirely by the survival clause, where the distant count supports what is already alive. 
The orientation rule comes out the same as in the continuous case for all that. 
The horizontal mask, which reads along $x$, gives bands running along $y$, and the vertical mask gives the transpose. Whether the same rule holds when the long range is given nothing but the opposing sign is the question of Sec.~\ref{sec:pure}, and it is answered there on the pure rule rather than on this one.

\subsection{Activation and inhibition alone}\label{sec:pure}

So far the distant count has been bolted onto the Game of Life, whose own short-range arithmetic (birth on exactly three neighbours, survival on two or three) has been kept untouched. To find out how much of that inheritance is needed we delete it and keep only the two counts. A site is live at the next step in
\begin{equation}
n_3 > A \quad \text{and} \quad n_5 < I,
\label{eq:pure}
\end{equation}
where $A$ is the short-range threshold and $I$ the long-range one, and that is the whole rule. There is no separate survival clause. The same two inequalities decide whether a dead site is born and whether a live site persists, so the next state of a site depends on its neighbourhood alone and not on its own state. Of the two counts only $n_3$ can favour growth, and $n_5$ can only oppose it, so the double sign noted in Sec.~\ref{sec:rule} is gone and the long range in this rule is inhibition and nothing else. Eq.~\eqref{eq:pure} is therefore the test the preceding sections have been building towards. If the two ranges make the patterns, Eq.~\eqref{eq:pure} must still make them. If the game's particular short-range arithmetic makes them, it will not.

\begin{figure}[p]
\centering
\includegraphics[width=0.95\linewidth]{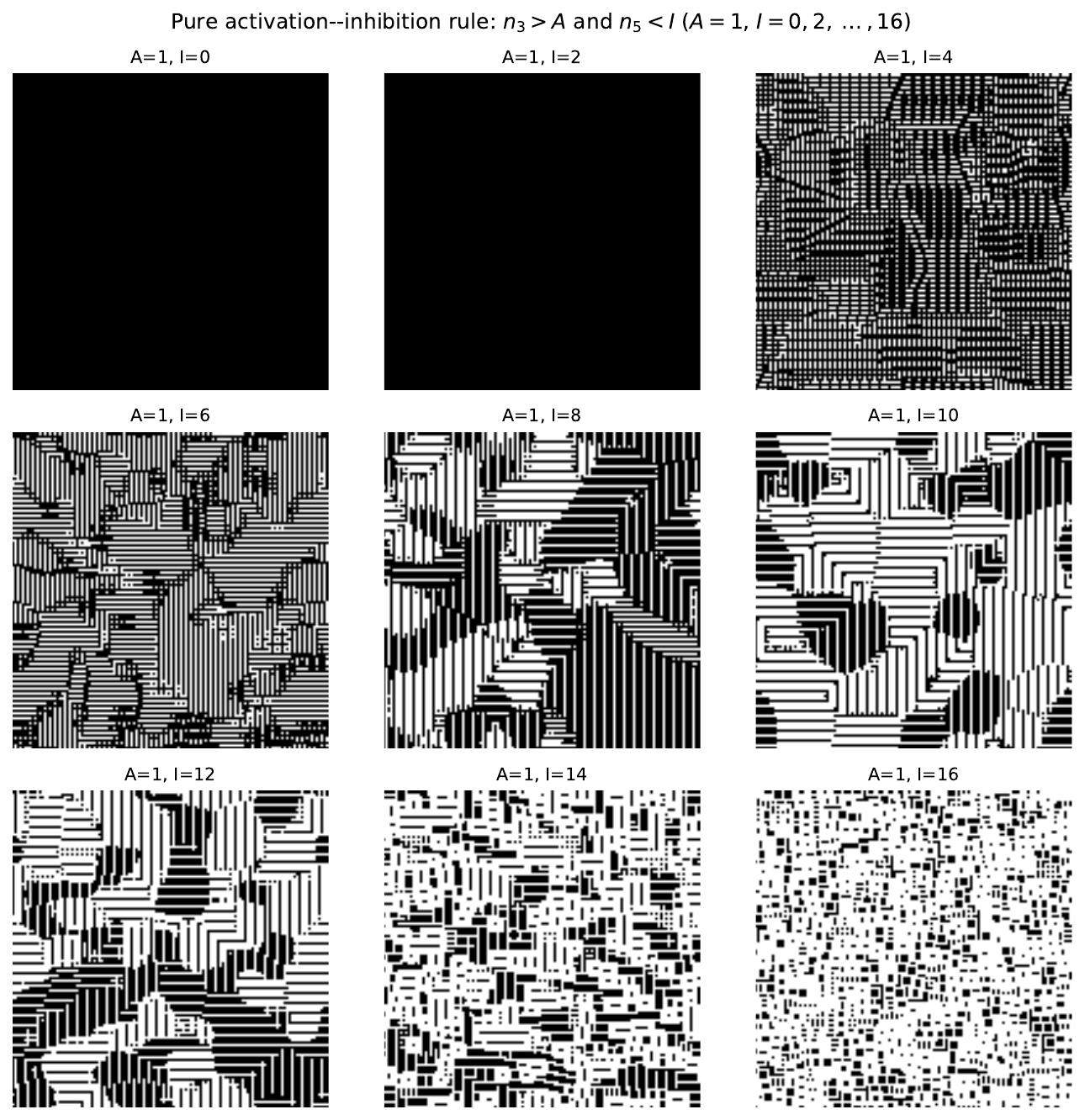}
\caption{Recurrent states of the pure activation--inhibition rule of Eq.~\eqref{eq:pure} at $A=1$ and $I=0,2,\dots,16$ (left to right, top to bottom), on a $128\times128$ lattice with periodic boundary conditions, from random starts at $\rho_0=0.5$, one start for each $I$.
The first two panels are the extinct cases. The panels that follow run from live islands through stripes to holes, and the last is one step short of the point where the long-range condition stops binding altogether.
}
\label{fig:pure}
\end{figure}

Fig.~\ref{fig:pure} shows the recurrent states of Eq.~\eqref{eq:pure} at $A=1$ for $I=0,2,\dots,16$, on the same $128\times128$ lattice with periodic boundary conditions, from independent random starts at $\rho_0=0.5$, one start for each $I$. Two of the nine cases are empty. At $I=0$ the condition $n_5<I$ cannot be met by any site and the lattice is dead after a single step, and at $I=2$ the live population dies out within two steps. At the other end, $n_5$ is at most sixteen, so at $I=17$ the long-range condition is vacuous and the rule reduces to $n_3>A$. From a random start drawn the same way, that rule drives the lattice to $\rho=1$ in three steps and holds it there. Everything between those two limits is patterned. The density grows from $0.29$ at $I=4$ to $0.78$ at $I=16$ as the inhibition is relaxed, and it does so unevenly. It holds at $0.42$ from $I=6$ to $I=8$ and dips to $0.55$ at $I=12$, each panel being a single realisation, so the run-to-run spread that Sec.~\ref{sec:4} measures systematically is not sampled here.

\begin{figure}[htbp]
\centering
\includegraphics[width=0.95\linewidth]{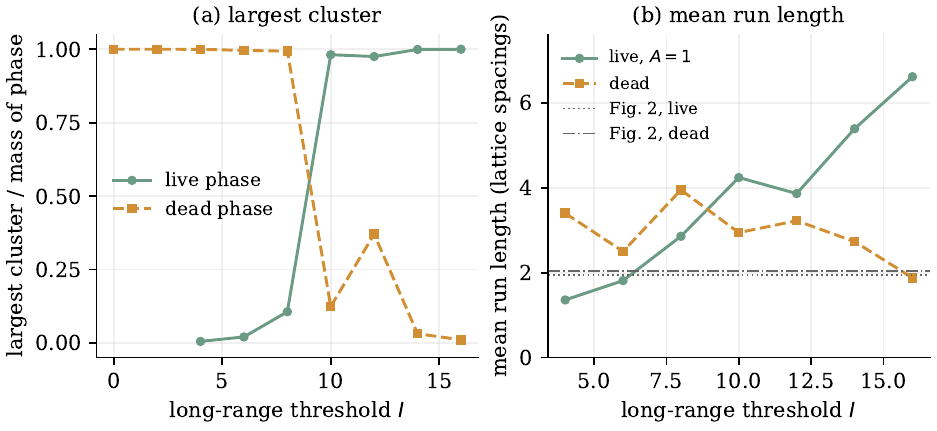}
\caption{Two measurements on the nine configurations of Fig.~\ref{fig:pure}, computed from those same configurations.
Left panel gives the mass of the largest eight-connected cluster of live cells, and of dead cells, divided by the size of the population it is drawn from. A live phase of spots gives a small live fraction with the dead phase in one piece, and a phase of holes gives the reverse. The two curves cross between $I=8$ and $I=10$, which is where the phases exchange roles.
Right panel gives the mean number of consecutive live cells along a row or a column of the periodic lattice, and the same for dead cells, which is the real-space size of the features. The dotted and dash-dotted lines mark these two lengths for the stripe state of Fig.~\ref{fig:evolution}, $(B,D)=(12,8)$, at $1.95$ and $2.03$ lattice spacings. The extinct cases at $I=0$ and $2$ have no live cells and are omitted from this panel.
The power-spectrum maxima occur at different wave numbers across these configurations, as specified in the text.}
\label{fig:pure_morphometry}
\end{figure}

To say which of the forms of Sec.~\ref{sec:1} a configuration shows without appealing to the eye, a criterion is needed. Two sites in the same state belong to the same `cluster' if they touch horizontally, vertically or diagonally, and every configuration then has a largest live cluster and a largest dead cluster. We quote each of the two as a fraction of the population it is drawn from. A live phase broken into spots gives a small live fraction against a dead fraction near one, because the dead cells then form one connected background, and a phase of holes exchanges the two roles. Fig.~\ref{fig:pure_morphometry}(a) plots the two fractions against $I$, and they trace the sequence. At $I=4$ and $6$ the largest live cluster holds $0.5\%$ and $2\%$ of the live cells while the dead cells are almost all in one piece ($99.98\%$ and $99.6\%$), live islands in a connected dead sea. At $I=8$ the live cells have stretched into the long parallel stripes of Fig.~\ref{fig:pure} but have not joined up, the largest holding $10.6\%$ against the dead phase's $99.4\%$. Between $I=8$ and $I=10$ the two populations exchange roles. At $I=10$ and $12$ the live cells are one connected body ($98.1\%$ and $97.5\%$) and the dead cells are the broken phase ($12.4\%$ and $37.1\%$), and at $I=14$ and $16$ the live cells are essentially all in a single cluster ($99.9\%$ and $100\%$) with the dead left as sparse holes ($3.2\%$ and $1.1\%$). The sequence with increasing $I$ is thus islands, then stripes, then holes, and finally a nearly uniform live state. This is the family of forms Sec.~\ref{sec:turing} associates with the mechanism, from two thresholds and nothing else.

The second measurement takes up the point Sec.~\ref{sec:stripe} left open. There the stripes were one live column alternating with one dead one, the finest a lattice can make, so their spacing was the lattice's rather than the rule's. In Eq.~\eqref{eq:pure} the thresholds do set a size, and Fig.~\ref{fig:pure_morphometry}(b) gives it as a real-space length. The mean number of consecutive live cells along a row or a column of the torus, and the same for the dead cells. The live runs grow from $1.4$ lattice spacings at $I=4$ to $6.6$ at $I=16$ while the dead runs shrink from $3.4$ to $1.9$. The state of Fig.~\ref{fig:evolution}, measured the same way, gives $1.95$ and $2.03$. For the seven patterned pure-rule configurations, the radial Fourier-power maximum lies in shell $1$ at $I=4,8,10,12$, shell $63$ at $I=6$, and shell $2$ at $I=14,16$. Here shell $j$ contains wave vectors with $j-1/2<L|\mathbf q|/(2\pi)\le j+1/2$, for $j=1,\ldots,64$.

\begin{figure}[htbp]
\centering
\includegraphics[width=0.8\linewidth]{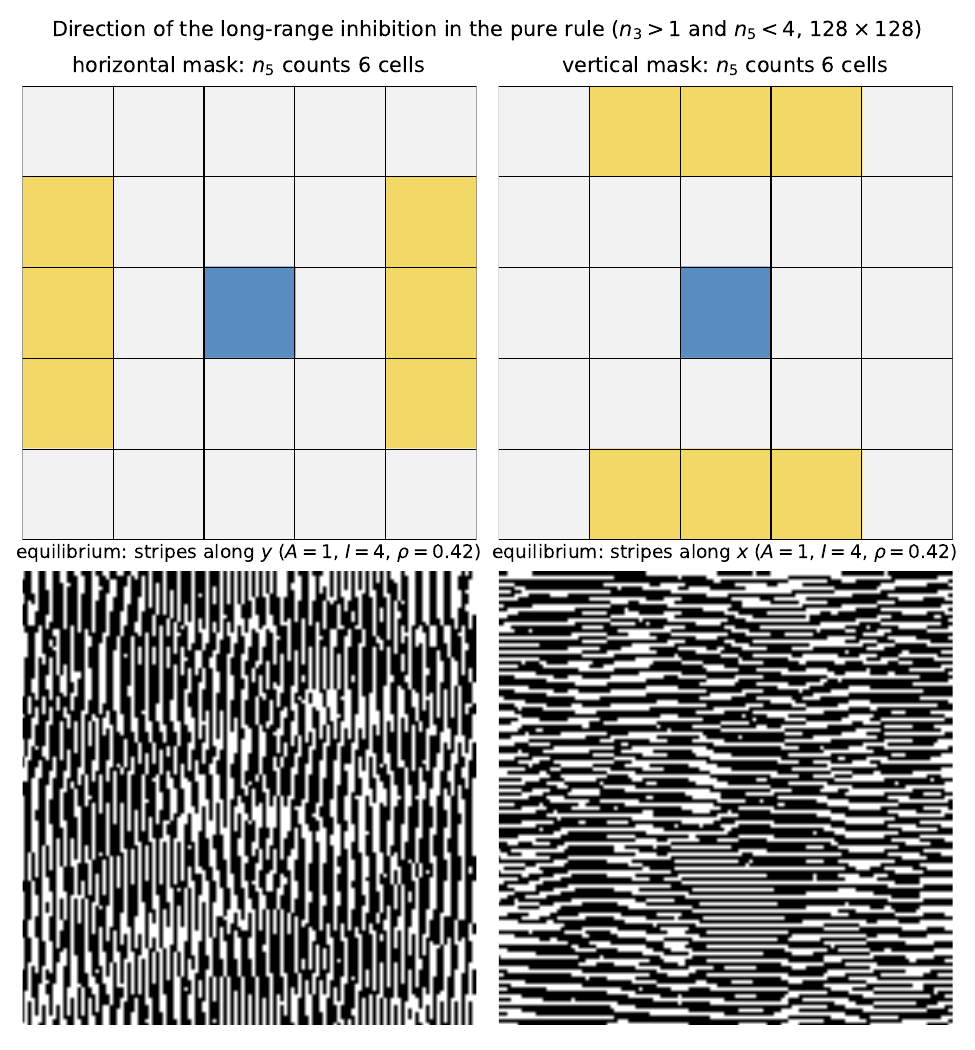}
\caption{The mask experiment of Fig.~\ref{fig:aniso_ca} repeated inside the pure rule of Eq.~\eqref{eq:pure}, at $A=1$ and $I=4$ on a $128\times128$ lattice with $\rho_0=0.5$, both masks run from a common initial configuration.
Top row shows the two direction-dependent $n_5$ masks, each reading six of the sixteen ring cells.
Bottom row shows the recurrent configurations. The horizontal mask gives bands running along $y$ and the vertical mask bands along $x$, at densities that differ by less than $0.2\%$. In this rule the long-range count can only oppose growth, so the orientation is set by inhibition alone.}
\label{fig:pure_aniso}
\end{figure}

Eq.~\eqref{eq:pure} also answers the question Sec.~\ref{sec:aniso} leaves open. There the direction of the bands was carried by the survival clause, because a six-cell mask cannot reach the birth threshold $B=12$. In Eq.~\eqref{eq:pure} the long range has only the opposing sign, which makes this rule the place to ask whether the same orientation rule holds. We repeat the mask experiment inside the pure rule, at $A=1$ and $I=4$, with the same two masks on the same lattice and, unlike Fig.~\ref{fig:aniso_ca}, where each mask was run from its own random start, from one shared initial configuration. Fig.~\ref{fig:pure_aniso} is the result. The horizontal mask, which reads to left and right, gives bands running along $y$, and the vertical mask gives the transpose. The run lengths say it numerically. With the horizontal mask the live cells run $1.4$ spacings along $x$ and $9.4$ along $y$, and $88\%$ of the pattern's Fourier power sits at wave vectors with $|q_x|>|q_y|$, where $q_x,q_y$ are their components. With the vertical mask the two lengths are $10.0$ and $1.4$, and that fraction is $11\%$. The two equilibria are otherwise the same state, their densities of $0.4158$ and $0.4164$ differing by less than one part in five hundred. Only the recurrence time, $49$ steps against $59$, and the period of the cycle reached, $2$ against $4$, differ alongside them. The orientation rule of Sec.~\ref{sec:aniso} thus does not depend on which sign carries the long range. It holds when the distant count supports what is alive, and when it opposes growth outright.

What Eq.~\eqref{eq:pure} establishes, then, is that the two ranges do the work on their own. Neither the birth on exactly three neighbours nor the survival on two or three is needed to hold a live population on a lattice, to arrange that population into islands, stripes, and holes, to let the thresholds set the size of its features, or to let the direction of the inhibition choose the orientation of the pattern. All that is needed is the second count, which is also the only thing the modification of Sec.~\ref{sec:rule} added. Every state in this section, however, was looked at one parameter pair at a time. We took $(12,8)$ in Sec.~\ref{sec:stripe}, $(12,3)$ in Sec.~\ref{sec:aniso}, and a line of $I$ values here. Whether these are isolated curiosities or the surface of a systematic behaviour is a question about the whole plane of thresholds, and Sec.~\ref{sec:4} scans it.

\section{\label{sec:4}Morphological transitions across the $(B,D)$ plane}

Sec.~\ref{sec:3} fixed the two thresholds at one point of the $(B,D)$ plane and found stripes there. A continuous Turing system is not satisfied with a single morphology. As the control parameters are varied the pattern changes, and the family of forms it passes through (spots, stripes, and the textures in between) is the clearest evidence that one mechanism stands behind all of them. The question for our automaton is therefore not whether it makes a pattern but whether that pattern changes, and whether the changes are organised in the same way. We answer it by scanning $B$ and $D$ over the whole range the rule allows and looking at what the dynamics leaves behind.

Two cautions about the status of the language used in this section. 
The automaton is finite, discrete, and deterministic, and its long-time behaviour is a repeating cycle rather than a thermodynamic steady state, so nothing below is a transition in the thermodynamic sense. What makes the analogy worth drawing is that the run from the initial configuration to that cycle plays the role of a thermalization. The path is not an observable, the endpoint is, and the endpoint is selected by the control parameters. The change of morphology with $(B,D)$ is then a change of the equilibrium state, and it can be described with order parameters and drawn as a map of the plane. We call that a `morphology map'. Its entries are the patterns the two thresholds produce, read off the endpoint. We take the vocabulary that far and no further.

The scan gives a sequence. At fixed $B$, increasing $D$ carries the equilibrium state from a dense state whose only structure is at the scale of the lattice(which we call noise) through stripes, to dilute spots. The scan also gives one thing that the isotropic continuous system does not. There is a line of the $(B,D)$ plane on which the stripes lock onto one of the two lattice axes, and on which the choice of axis differs from one realisation of the rule to the next. Sec.~\ref{sec:aniso} asked what fixes an orientation. On that line nothing does, and the orientation is broken spontaneously.

\subsection{The parameter scan}

We scan the control parameters $B,D\in\{0,1,\dots,17\}$.
For each pair $(B,D)$ we run $2000$ independent realisations on an $80\times80$ lattice at $\rho_0=0.5$ (the effect of $\rho_0$ is examined separately in Sec.~\ref{sec:density}). Each run is evolved until the configuration repeats, which it must because the lattice is finite and the update is deterministic, and we keep the last distinct configuration before the first repeat, the `equilibrium configuration' defined in Sec.~\ref{sec:stripe}. The five geometric observables below are computed from the saved equilibrium configurations. The sixth observable, recurrence time, is obtained from the preceding trajectory.

\subsection{A tour of the morphologies}

The $(B,D)$ plane is not uniform. How the morphology changes with $D$ depends on $B$. We go through the different $B$ regions one by one and describe, by eye, what the equilibria look like and where the morphology changes. Sec.~\ref{sec:observables} turns these observations into numbers. Each panel of Figs.~\ref{fig:phase_low}--\ref{fig:phase_high} shows the first of the $2000$ realisations of that cell, so the panels are samples of an endpoint ensemble rather than the ensemble itself. Where a panel is quoted below, the corresponding ensemble mean is quoted with it.

\subsubsection{A coexistence transition at low $B$}

\begin{figure}[htbp]
\centering
\includegraphics[width=0.95\linewidth]{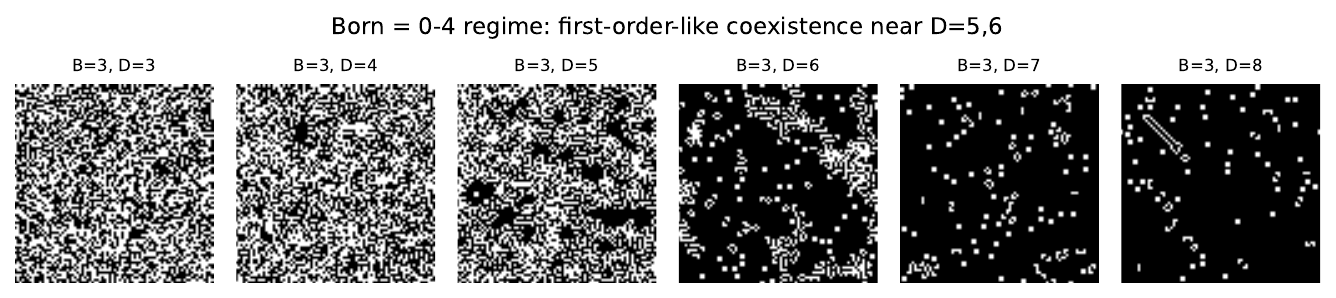}
\caption{Equilibrium configurations for $B=3$, $D=3,\dots,8$ in the low-$B$ regime. Each panel is the first realisation of the $2000$-member ensemble at that $(B,D)$.
The $B=3$ row is representative of the whole $B=0$--$4$ range, whose equilibrium is dense up to $D=4$ and dilute from $D=7$ on.
The dense and dilute regions present together in the $D=5$ and $D=6$ panels indicate a first-order-like transition near $D=5,6$.}
\label{fig:phase_low}
\end{figure}

For $B=3$ and $D$ increasing from 3 to 8 the system passes from a dense state through a coexistence region to a dilute spot morphology, as shown in Fig.~\ref{fig:phase_low}. At $D=3,4$ the equilibrium is dense ($\rho=0.50$ and $0.48$) and its only structure is at the scale of the lattice. At $D=5,6$ one and the same configuration carries both morphologies. Dense regions of the $D=4$ kind float in a sea of empty lattice, and the mean density falls through $0.43$ to $0.17$. By $D=7$ the dense regions are gone and what is left is a dilute gas of spots, $\rho=0.068$ there and $0.046$ at $D=8$.

The passage is a change of state rather than a continuous softening. The density drops by nearly an order of magnitude between $D=4$ and $D=7$, and at each individual $D$ the $2000$ realisations give the same density to within their standard deviation, which is $0.008$ at $D=4$, $0.017$ at $D=5$ and $0.025$ at $D=6$. The two morphologies therefore coexist within a configuration at $D=5,6$ rather than between different realisations of it. 
Two states meeting inside one sample, with the density changing rapidly between them, is what first-order-like behaviour looks like in this sequence.

No stripe morphology appears in this regime. None of the panels of Fig.~\ref{fig:phase_low} contains the extended, parallel, lattice-spaced bands that appear at larger $B$ in Fig.~\ref{fig:phase_710}.

\subsubsection{A sharper version at intermediate $B$}

\begin{figure}[htbp]
\centering
\includegraphics[width=0.95\linewidth]{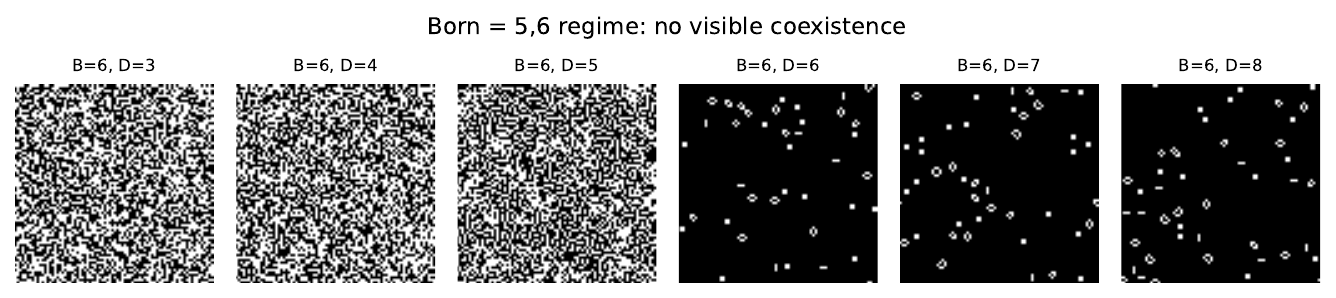}
\caption{Equilibrium configurations for $B=6$, $D=3,\dots,8$ in the intermediate-$B$ regime. Each panel is the first realisation of its ensemble.
This row is representative of $B=5$--$6$. The equilibrium is dense at $D=3,4,5$ and dilute from $D=6$ on.
The change from $D=5$ to $D=6$ is sharp and no coexistence is visible in single samples.}
\label{fig:phase_56}
\end{figure}

For $B=6$ the same walk from a dense equilibrium to a dilute one occurs, but the coexistence region is no longer visible in single realisations (Fig.~\ref{fig:phase_56}). The mean density is $0.48$ at $D=5$ and $0.031$ at $D=6$, a factor of fifteen in one step of $D$, with no intermediate value in between.
Two readings of this are available at a finite lattice and the scan does not separate them. The transition may remain first-order-like with the two morphologies sharing a region too narrow in $D$ to fall on a grid point, or $D=5$ and $D=6$ may simply lie on opposite sides of a boundary that passes between them.
The $D=5$ panel is the one place in this regime where the dense state is not salt-and-pepper noise but a coarse, labyrinth-like texture, and Sec.~\ref{sec:observables} returns to the difficulty of telling the two apart by number.

\subsubsection{Noise, stripes, and spots at moderate $B$}

\begin{figure}[htbp]
\centering
\includegraphics[width=0.95\linewidth]{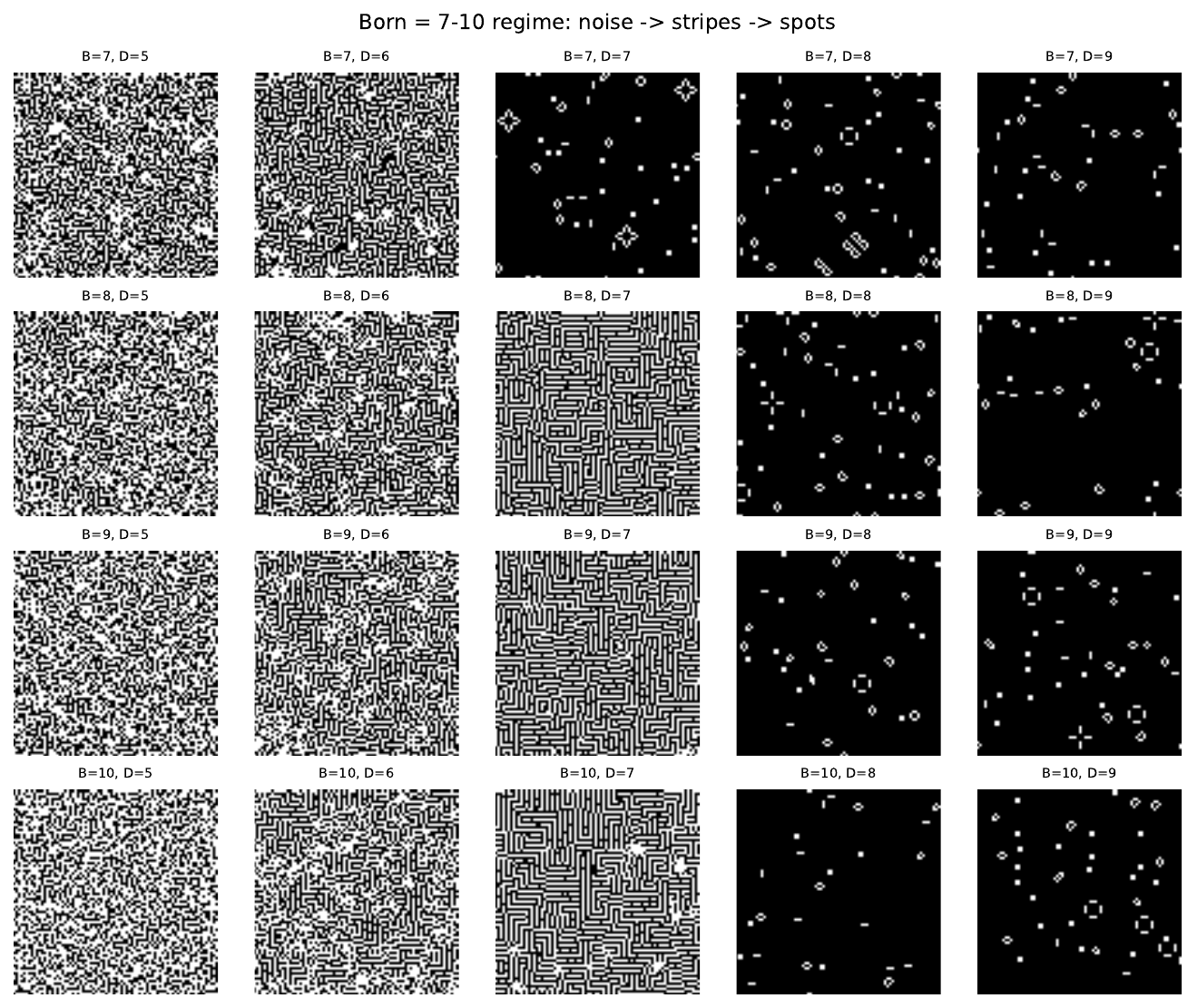}
\caption{Equilibrium configurations for $B=7,\dots,10$ and $D=5,6,7,8,9$. Each panel is the first realisation of its ensemble.
The rows illustrate the typical morphologies of the $B=7$--$10$ regime. The system evolves from noise to stripes and then to spots as $D$ increases. The stripe texture appears at $D=6$ for $B=7,8$ and at $D=7$ for $B=9,10$.}
\label{fig:phase_710}
\end{figure}

Here the morphology passes through all three basic forms as $D$ increases from 5 to 9. The sequence is noise, then stripes, then spots (Fig.~\ref{fig:phase_710}). The middle member is the one the continuous theory knows best. At $B=8$, $D=7$ the equilibrium is a labyrinth of parallel live bands one cell wide, separated by dead channels one cell wide, broken into domains whose direction rotates by ninety degrees from one domain to the next. Its density $\rho=0.45$ is close to the noise's $0.53$, but its live cells are split into $121$ clusters of mean size $24$ against the noise's $10$ clusters of mean size $376$. 
The stripe band itself is narrow in $D$. It covers one or two columns of the scan at each $B$, and what grows with $B$ is not its width but the position of the whole sequence, which slides from $D=6$ at $B=7$ to $D=7$ at $B=9$.

\subsubsection{Orientational symmetry breaking at $B=11$}\label{sec:orient}

\begin{figure}[htbp]
\centering
\includegraphics[width=0.95\linewidth]{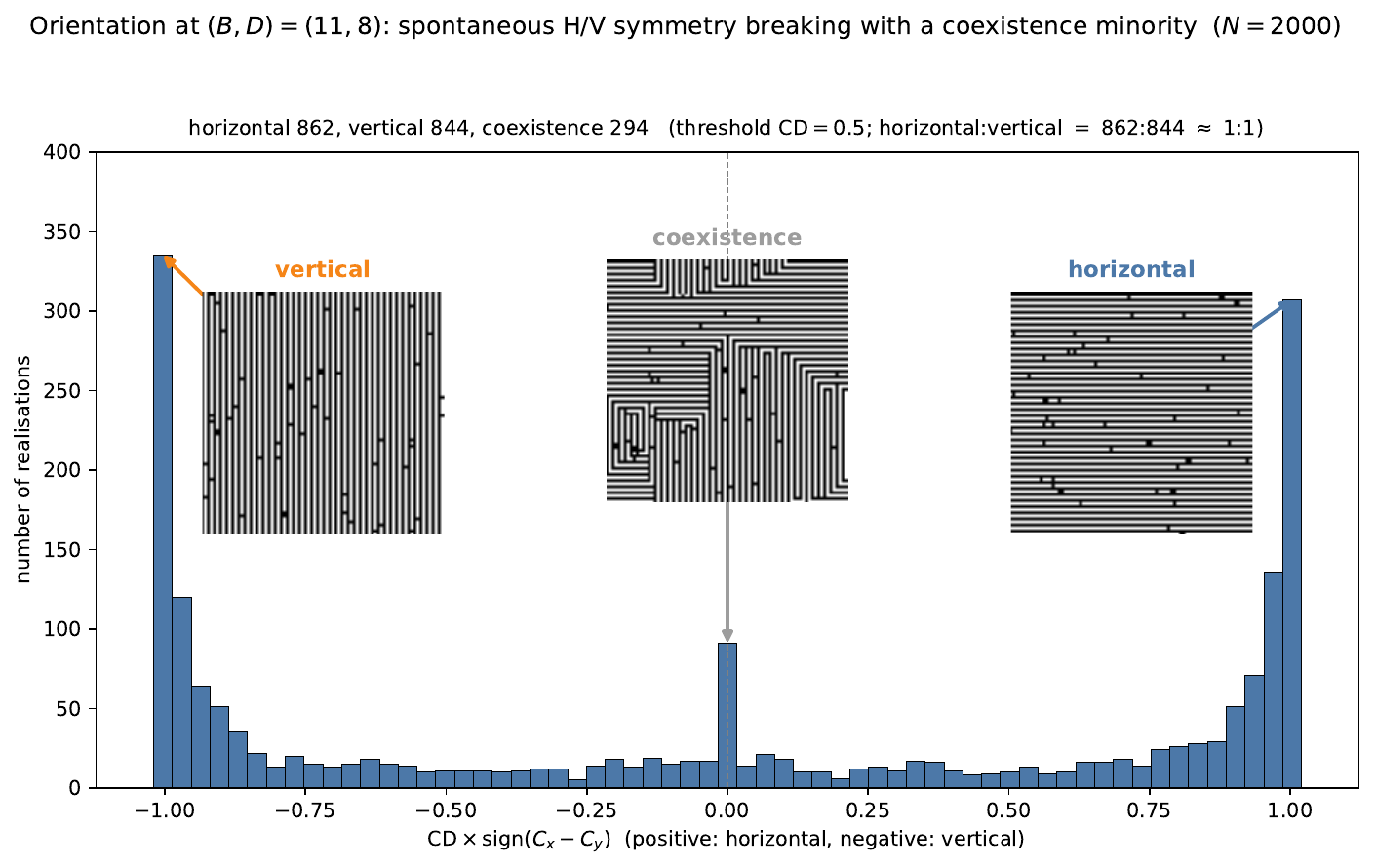}
\caption{Orientational symmetry breaking at $(B,D)=(11,8)$ over the $2000$-realisation ensemble.
Histogram of the signed directional-correlation difference $\mathrm{sign}(C_x-C_y)\,\mathrm{CD}$, where CD is the directional-correlation difference of Sec.~\ref{sec:observables}. A positive value means horizontal stripes, a negative value means vertical stripes, and a value near zero means coexistence of the two orientations.
The oriented realisations split almost evenly between the two axes ($862$ horizontal vs $844$ vertical), so individual oriented realisations break the $90^\circ$ rotation symmetry while the ensemble has no preferred direction. A coexistence minority ($294$, $14.7\%$) forms a patchwork in which horizontal and vertical stripe regions coexist.
The embedded thumbnails show representative realisations of the three classes, each with an arrow pointing to the tallest bar of the class it belongs to.}
\label{fig:phase_11_orient}
\end{figure}

The stripe at $D=8$ has a property the stripes of Fig.~\ref{fig:phase_710} do not. Its bands lock onto one of the two lattice axes. In a labyrinth the bands are straight and axis-aligned inside each domain, but the domains run in both directions and their contributions cancel. At $D=8$ a single domain can span the lattice.
Fig.~\ref{fig:phase_11_orient} quantifies this over the full $2000$-realisation ensemble.
Classifying each realisation by the directional-correlation difference CD~(Sec.~\ref{sec:observables}), $85\%$ form a clean axis-aligned stripe, with a median CD of $0.97$ over the whole ensemble.
Among the oriented realisations the two axes occur in nearly equal numbers. There are $862$ horizontal against $844$ vertical, i.e.\ $1{:}1$ to within sampling noise.
The ensemble therefore selects no preferred axis. Each realisation's choice of orientation is a spontaneous breaking of the lattice $90^\circ$ rotation symmetry, whereas the preferred-direction selection in the anisotropic systems is imposed by their diffusion coefficients~(Figs.~\ref{fig:aniso_ca} and~\ref{fig:aniso_gs}).
The remaining $\sim 15\%$ are not diagonal stripes (a diagonal-correlation check returns near-zero anisotropy for them) but a patchwork of horizontal and vertical stripe regions at the same density ($\rho\simeq 0.49$), in which the two orientations coexist in different parts of the lattice and cancel on average.

This orientational ordering is also dynamically costly.
Forming a globally oriented stripe requires not only growing stripes but coarsening competing orientations into a single axis.
Consistent with this, the recurrence time at $(11,8)$ is exceptionally long. The $2000$-realisation mean is $\sim 1.8\times10^5$ and the median $\sim 1.1\times10^5$, with $82\%$ of realisations taking more than $5000$ steps (Table~\ref{tab:length}).
This makes $(11,8)$ the second-slowest cell in the whole $(B,D)$ plane (only $(10,8)$ is slower) and a sharp spike over its $D$-neighbours ($29$ at $D=7$ and $2.8\times10^3$ at $D=9$).
The oriented-stripe point thus sits at a genuine dynamical slowdown, as expected for a slow symmetry-choice and coarsening process.

The locking is not a property of the $B=11$ row, however. It is a property of the line $D=8$. Classifying every realisation of every cell of the scan by the same criterion used in Fig.~\ref{fig:phase_11_orient}, $\mathrm{CD}>0.5$, oriented realisations occur on that line and nowhere else in the $(B,D)$ plane. The remaining $317$ cells contain none at all. Their number falls off along the line as fast as it does off it. There, $85.3\%$ of the ensemble at $(11,8)$, $14.6\%$ at $(12,8)$, and $4.5\%$ to $7.0\%$ for $B=13$ through $17$. The $D=8$ panel of Fig.~\ref{fig:phase_high} is the intermediate case made visible, a lattice broken into domains of horizontal and of vertical bands whose contributions cancel. Along the whole line the two axes are chosen equally often, so the ensemble selects no direction anywhere on it.

\begin{figure}[htbp]
\centering
\includegraphics[width=0.95\linewidth]{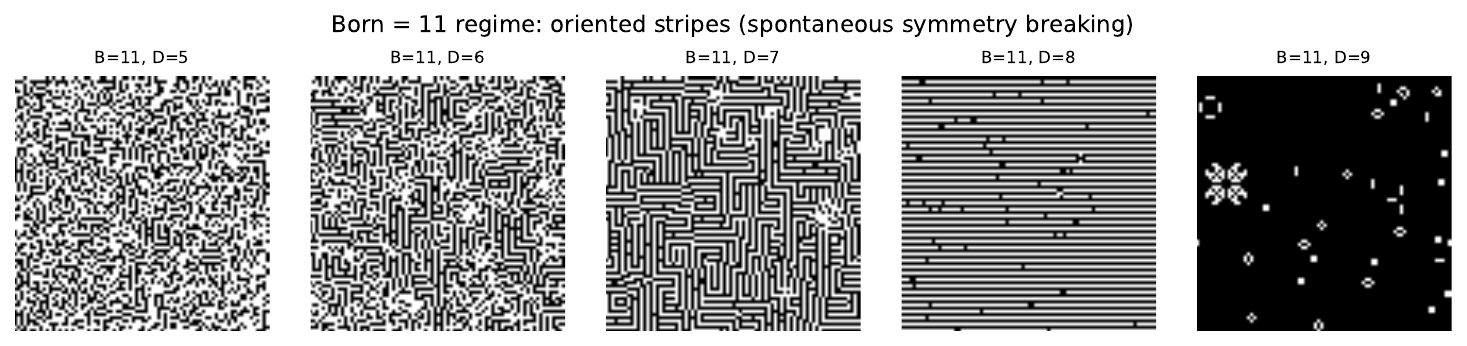}
\caption{Equilibrium configurations for $B=11$, $D=5,6,7,8,9$. Each panel is the first realisation of its ensemble.
The $D=8$ panel is the oriented stripe. The bands lock either vertically or horizontally, and the choice is made by the realisation rather than by the rule. At $D=7$ the same density is reached by a labyrinth in which the two orientations coexist.
The equilibrium is dense up to $D=7$ and dilute from $D=9$ on.}
\label{fig:phase_11}
\end{figure}

\subsubsection{The same sequence with fixed transitions at high $B$}

\begin{figure}[htbp]
\centering
\includegraphics[width=0.95\linewidth]{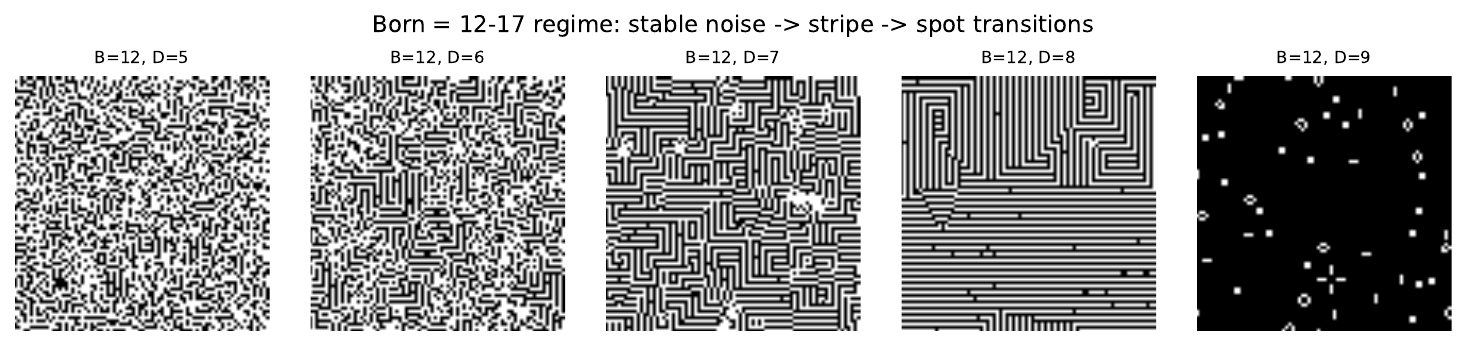}
\caption{Equilibrium configurations for $B=12$, $D=5,6,7,8,9$ in the high-$B$ regime. Each panel is the first realisation of its ensemble.
$B=12$ is representative of the whole $B=12$--$17$ range. The equilibrium is dense through $D=8$ and dilute from $D=9$ on.
The noise$\to$stripes$\to$spots sequence persists, and the $D=8$ panel shows the multi-domain case of the orientation locking of Fig.~\ref{fig:phase_11}.}
\label{fig:phase_high}
\end{figure}

For larger $B$ the same sequence noise$\to$stripes$\to$spots persists, and its position stops moving. From $B=12$ to $B=17$ the stripe band begins at the same place, $D=7$, and the dilute spots begin at the same place, $D=9$ (Fig.~\ref{fig:phase_high}). The $B=12$ row is therefore representative of the whole range $B=12$--$17$, and the whole sequence is already fully developed by $B=12$.
What changes along this part of the scan is not the presence of the locking but its strength. The bands at $D=8$ are still axis-aligned, and still run in domains, but the domains no longer merge into one. The fraction of realisations that reach a single orientation across the whole lattice is $14.6\%$ at $(12,8)$ and a few per cent for larger $B$, against $85.3\%$ at $(11,8)$.

\subsection{Order parameters and the morphology map}\label{sec:observables}

In the tour above we identified the morphologies by eye. To define them we need numbers that take different values in different morphologies, so we measure six observables on each equilibrium configuration.
Let $s_i\in\{0,1\}$ be the state of site $i$, $L$ the linear size, and $\mathcal{C}_k$ the $k$-th eight-connected cluster of live cells on the periodic lattice, of size $|\mathcal{C}_k|$, with $N_c$ clusters in all.
Each observable answers one of the questions the tour had to ask. Density separates a dilute equilibrium from a dense one. The directional-correlation difference asks whether the pattern has a global direction, which is the distinction between the locked stripes of Fig.~\ref{fig:phase_11} and the labyrinths of Fig.~\ref{fig:phase_710}. The cluster count and the mean cluster size ask how the live mass is shared among connected components, which is what tells a dense state whose structure stops at the lattice scale from a labyrinth of the same density, and both from a gas of spots. The cluster-size variance asks whether one configuration holds components of very different sizes at once, the signature of two morphologies meeting inside a single sample. The recurrence time, finally, asks how long the thermalization takes. It is the only one of the six that is not a property of the endpoint geometry, and we use it to mark the boundaries rather than to classify what lies between them. Throughout this subsection the numerical values quoted are ensemble means over the $2000$ realisations at each $(B,D)$. The panels of Figs.~\ref{fig:phase_low}--\ref{fig:phase_high} each show a single representative realisation, and the same means are displayed as heat maps in Fig.~\ref{fig:observables}.

\begin{enumerate}
\item \textit{Density.}
\begin{equation}
\rho = \frac{1}{L^2}\sum_i s_i.
\end{equation}
Density is the most direct order parameter for the dilute--dense transition.

In the low-$B$ regime (Fig.~\ref{fig:phase_low}), for example, the mean $\rho$ falls from $0.50$ at $B=3,D=3$ (dense noise) to $0.43$ at $D=5$ (coexistence) and to $0.05$ at $D=8$ (dilute spots). In the high-$B$ regime (Fig.~\ref{fig:phase_high}) it decreases from $0.60$ at $B=12,D=5$ to $0.03$ at $D=9$.

\item \textit{Directional-correlation difference (CD).}
Define the nearest-neighbour correlations along the two lattice axes, $x$ and $y$
\begin{equation}{
{\begin{aligned}
 C_x&=\frac{1}{L(L-1)}\sum_{i=0}^{L-1}\sum_{j=0}^{L-2}s_{i,j}s_{i,j+1},\\
 C_y&=\frac{1}{L(L-1)}\sum_{i=0}^{L-2}\sum_{j=0}^{L-1}s_{i,j}s_{i+1,j}.
\end{aligned}}
}\end{equation}
Then
\begin{equation}{
{\mathrm{CD} = \begin{cases}|C_x-C_y|/(C_x+C_y),&C_x+C_y\ge10^{-12},\\0,&C_x+C_y<10^{-12}.\end{cases}}
}\end{equation}
These correlations omit the wrap-around pair on each row and column, while the dynamics itself retains periodic boundary conditions.
We also take the maximum of CD computed on the live cells and on the dead cells to avoid sensitivity to which of the two is the minority one.

CD is designed to detect global orientational order. A state whose bands run in one direction over the whole lattice has stronger correlations along the bands than across them, giving CD close to one. It is not a measure of whether bands are present at all. A labyrinth is built of the same axis-aligned bands, but its domains point in both directions and their contributions to $C_x-C_y$ cancel, so a labyrinth scores no differently from spots or from noise.
The numbers say so plainly. At $B=11,D=8$ the mean CD is $0.82$ and the median $0.97$, while at $D=5$ (noise) and $D=9$ (spots) of the same row the mean is below $0.06$. The labyrinth cells are no higher. The largest mean CD anywhere off the $D=8$ line is $0.074$, at the labyrinth cell $(8,7)$, against $0.058$ at the spot cell with the strongest anisotropy, $(10,8)$, and $0.060$ at the most anisotropic of the $111$ cells classified as noise. CD is the order parameter of the orientation locking of Fig.~\ref{fig:phase_11}, and of nothing else in the plane.

\item \textit{cluster count.}
This is the number $N_c$ of eight-connected clusters of live cells on the periodic lattice, i.e.\ the 0th Betti number of the live-cell set. It counts clusters, and takes no account of the holes that a percolating texture encloses.
It separates textures of the same density. A state whose structure stops at the lattice scale is one or a few percolating components, whereas a labyrinth at nearly that same density breaks into hundreds of them. A gas of spots has a count of the same order as a labyrinth's, so $N_c$ is read together with the mean cluster size below, which is what tells the two apart.
For example, in the high-$B$ regime (Fig.~\ref{fig:phase_high}) the mean cluster count is $\approx 2$ at $B=12,D=5$ (a single percolating cluster), rises to $\approx 49$ at $D=7$ (the stripe/labyrinth texture breaks into many clusters) and is $\approx 41$ at $D=9$ (the dilute spots are numerous disconnected clusters, though fewer in absolute number than the labyrinth fragments because the live-cell density is much lower).

\item \textit{Mean cluster size (CSize).}
\begin{equation}
\mathrm{CSize} = \frac{1}{N_c}\sum_k |\mathcal{C}_k|,
\end{equation}
where $N_c$ is the number of live-cell clusters.
For an empty configuration we set CSize and CSV to zero.
CSize complements the cluster count. 
A gas of spots is a small number of tiny clusters, so CSize is of order a few sites. 
A labyrinth is a large number of clusters, each several times bigger. A percolating dense state is a handful of clusters bigger again.
In the $B=7$--$10$ regime (Fig.~\ref{fig:phase_710}), for instance, the mean CSize falls from $376$ at $B=8,D=5$ to $24$ at $D=7$ and to $4$ at $D=9$.

\item \textit{Cluster-size variance (CSV).}
\begin{equation}
\mathrm{CSV} = \frac{1}{N_c}\sum_k \bigl(|\mathcal{C}_k|-\mathrm{CSize}\bigr)^2.
\end{equation}
CSV is large whenever one configuration holds clusters of very different sizes at once, which is what a percolating background carrying debris looks like. It is small where the live mass is shared among comparable components.
In the low-$B$ regime (Fig.~\ref{fig:phase_low}), the mean CSV is $3.5\times10^5$ at $B=3,D=3$ and $7.9\times10^4$ at $D=5$ (coexistence, clusters of very different sizes), whereas among the dilute spots at $D=8$ it drops to $\mathcal{O}(1)$ (here $\approx 4.6$).
Among the dense states it is the one measure that runs the same way in both of the cases the tour distinguishes. There are $1.1\times10^4$ at the labyrinth cell $(8,7)$ against $1.1\times10^6$ at the noise cell $(8,5)$, a percolating cluster and its fragments.

\item \textit{Recurrence time.}
The number of iterations needed for the deterministic dynamics to return to a previously visited configuration, averaged over the 2000 realisations for each $(B,D)$.
This is the one observable that is not a property of the endpoint geometry, and it is what the thermalization reading at the start of this section makes measurable. It counts how long the run takes to settle.
Because the system is finite and deterministic, it is the natural analogue of a relaxation time, and near a boundary, where the dynamics must choose between two morphologies, it is expected to reach a maximum.
It does, and the maximum is narrow and specific. The five slowest cells of the whole scan are $(10,8)$, $(11,8)$, $(6,6)$, $(7,7)$ and $(9,8)$, with mean recurrence times between $6.2\times10^{4}$ and $5.7\times10^{5}$ steps, and at every one of the five the equilibrium one step below, at $D-1$, is a labyrinth. One step further in $D$ the mean is back to $10^{3}$ steps, and at the dense-noise cells $(3,3)$, $(8,5)$ and $(12,5)$ it is $4$, $11$ and $7$.
The slowdown therefore marks one particular boundary rather than boundaries in general. The labyrinth is what the dynamics hesitates over, and it hesitates on the dilute side of it.
Table~\ref{tab:length} gives the values at the cells that bracket each of the changes of the tour, the fast ones included.
\end{enumerate}

\begin{table}[htbp]
\centering
\caption{Recurrence time (mean number of iterations to recurrence) at the cells that bracket the changes of morphology visible in Figs.~\ref{fig:phase_low}--\ref{fig:phase_high}, the fast cells included.
Each entry is the mean over the $2000$ realisations.}
\label{tab:length}
\begin{tabular}{c|ccccccccccc}
$(B,D)$ & (3,5) & (6,5) & (6,6) & (8,7) & (8,8) & (8,9) & (11,7) & (11,8) & (11,9) & (12,8) & (12,9) \\
\hline
$T_{\rm rec}$ & 21 & 46 & 113889 & 2205 & 3998 & 1034 & 29 & 176532 & 2808 & 617 & 3237 \\
\end{tabular}
\end{table}

What matters for what follows is that the five geometric measures describe different properties, with the exact relation $N_c\,\mathrm{CSize}=L^2\rho$ in each nonempty configuration. 
Density separates the dilute states from the dense ones. $N_c$ and CSize together separate the dense states from one another, CSV adds the internal spread of the live mass, and CD is independent of all four, since it responds only to whether the pattern has a direction. The recurrence time is outside this set. It says nothing about the endpoint, and we use it only to locate where the endpoint changes.

\subsubsection{B--D plane maps}

\begin{figure}[p]
\centering
\includegraphics[width=0.95\linewidth]{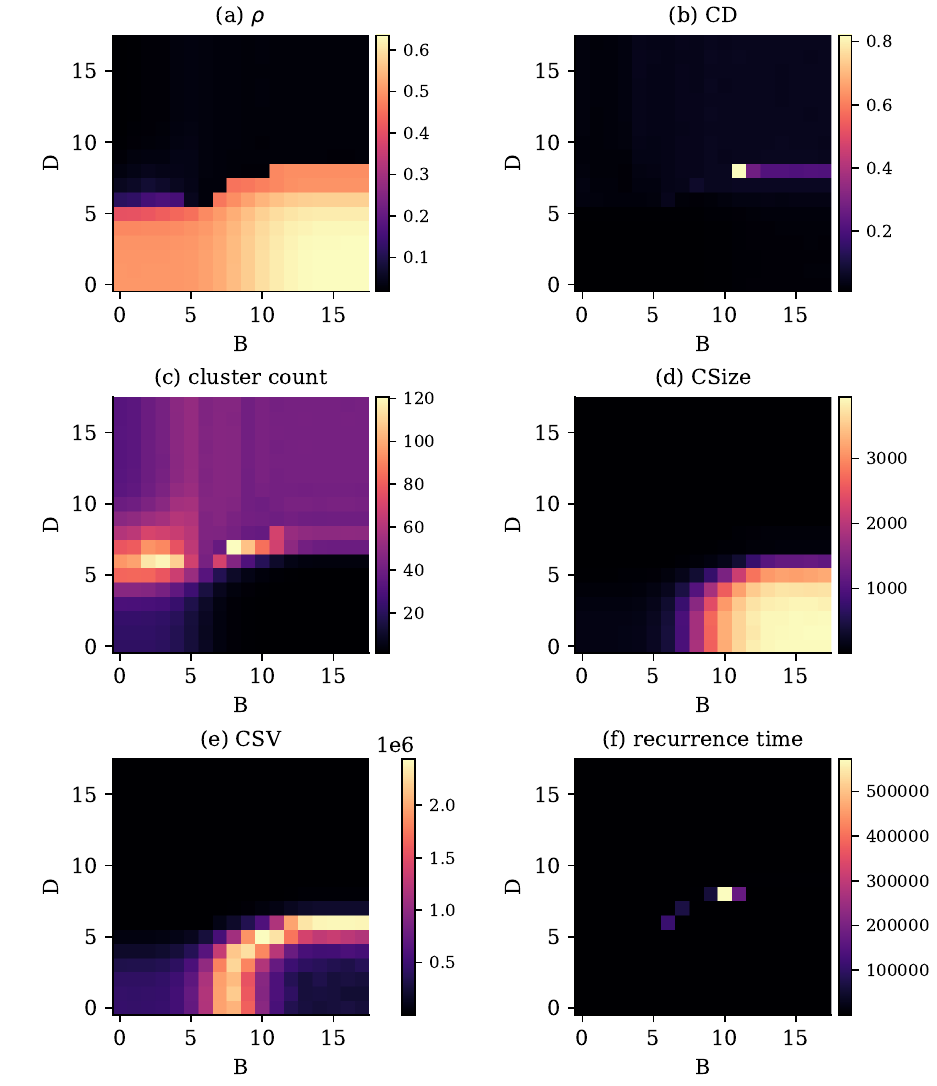}
\caption{Six observables on the $(B,D)$ control plane, averaged over 2000 realisations.
From left to right, top to bottom, the panels show density, CD, cluster count, CSize, CSV, and recurrence time.}
\label{fig:observables}
\end{figure}

\begin{figure}[p]
\centering
\includegraphics[width=0.95\linewidth]{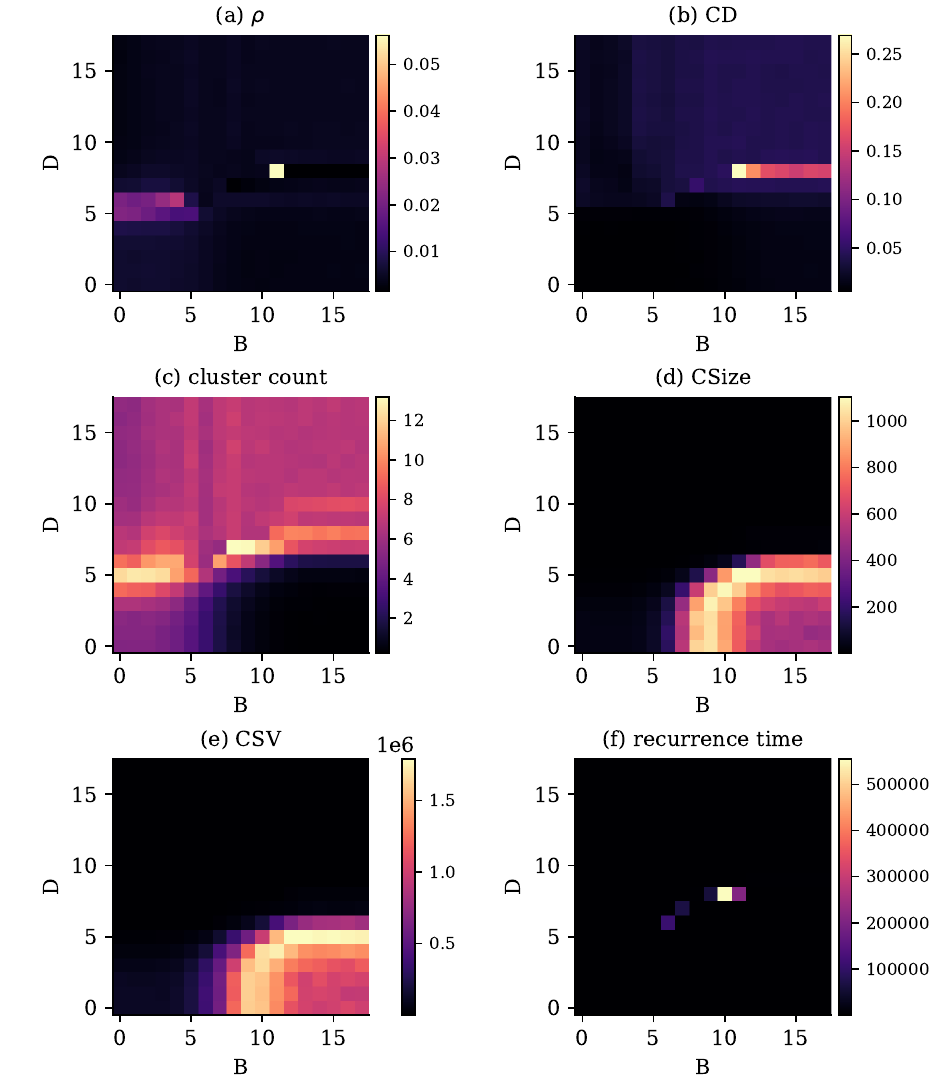}
\caption{Standard deviations of the six observables on the $(B,D)$ control plane, computed from 2000 realisations.
Large values indicate regions where the equilibrium morphology is sensitive to the random initial state.}
\label{fig:observables_std}
\end{figure}

\begin{figure}[p]
\centering
\includegraphics[width=0.95\linewidth]{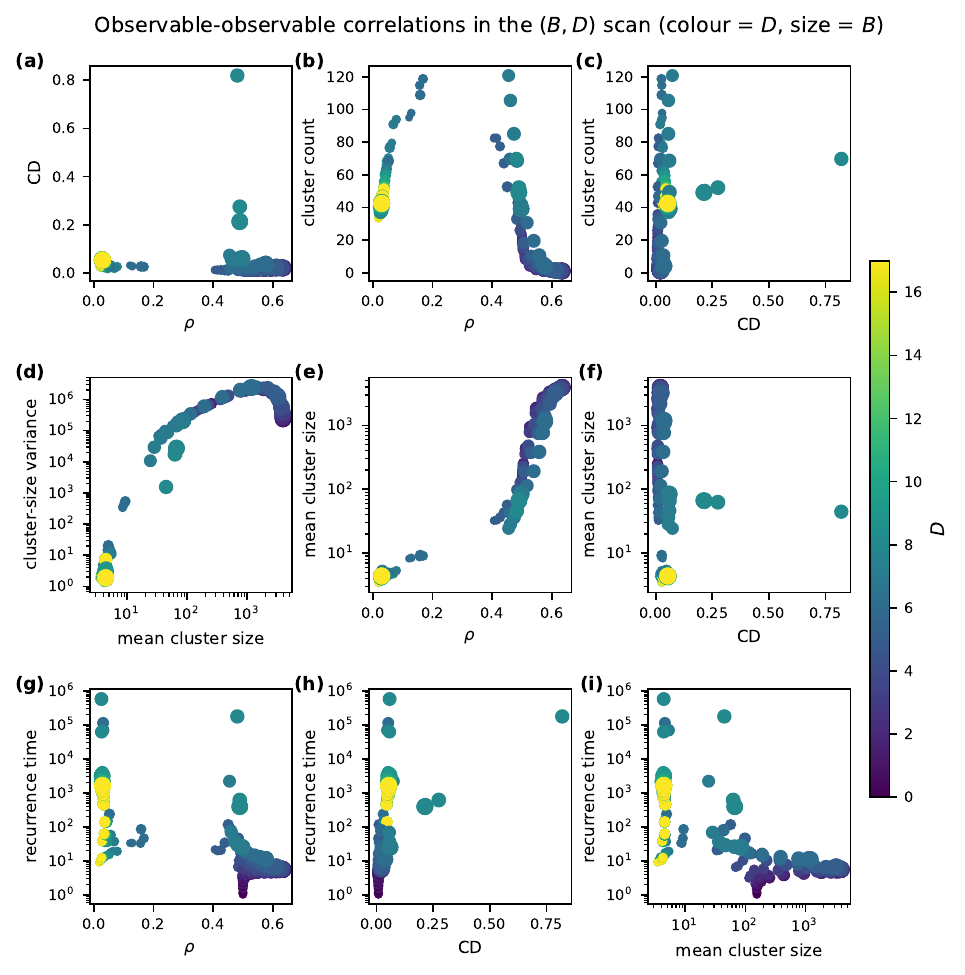}
\caption{Observable-observable scatter plots of the $(B,D)$ scan.
Each point represents the mean values of two observables for one $(B,D)$ cell, averaged over 2000 realisations.
Colour encodes $D$ and marker size encodes $B$ (larger symbols correspond to larger $B$).
The panels are (a) CD vs density, (b) cluster count vs density, (c) cluster count vs CD, (d) cluster-size variance vs mean cluster size, (e) mean cluster size vs density, (f) mean cluster size vs CD, (g) recurrence time vs density, (h) recurrence time vs CD, and (i) recurrence time vs mean cluster size.
Logarithmic axes are used whenever a quantity spans several orders of magnitude.}
\label{fig:observables_scatter}
\end{figure}

\begin{figure}[p]
\centering
\includegraphics[width=0.7\linewidth]{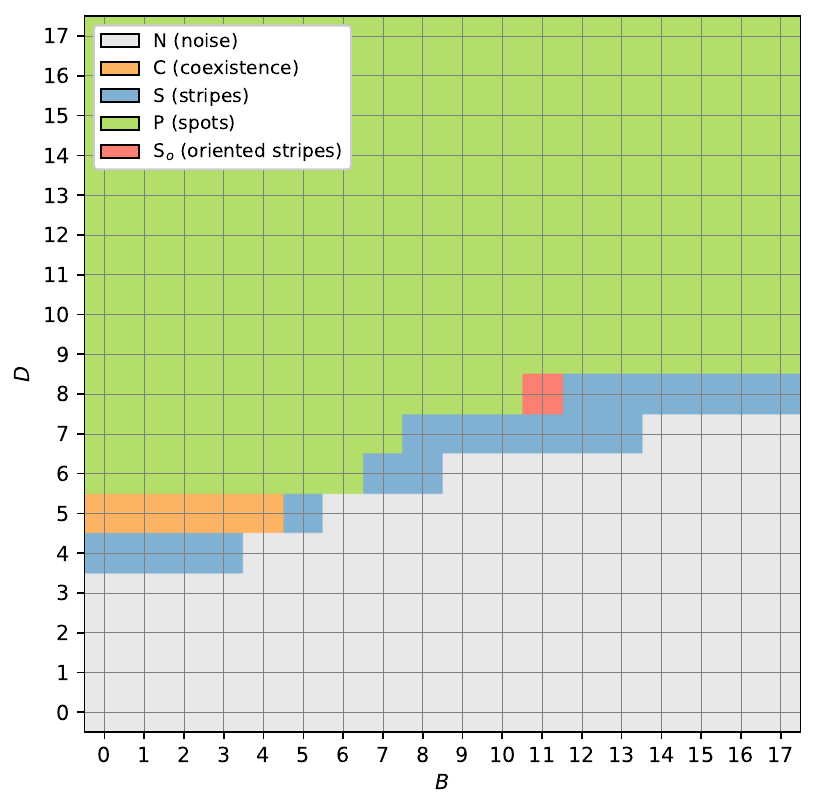}
\caption{Morphology map of the $(B,D)$ plane. The classification is motivated by the visual inspection of Figs.~\ref{fig:phase_low}--\ref{fig:phase_high} and reproduces the noise (N), coexistence (C), stripes (S), oriented stripes (S$_o$) and spots (P) regions. The criteria and the order in which they are applied are given in Table~\ref{tab:criteria}. The boundaries are drawn on the grid of the scan, so they are one cell wide by construction, and several thresholded observables depend on system size. The map is a statement about this $80\times80$ lattice and would have to be rescaled to be read at another size.}
\label{fig:phase_diagram}
\end{figure}

Fig.~\ref{fig:observables} shows the mean values of the six observables on the $(B,D)$ plane, each averaged over 2000 realisations.
The density map separates the dilute from the dense regions. The CD map is dark over the whole plane except along the line $D=8$, where the orientation locking of Fig.~\ref{fig:phase_11} lives. The cluster count and cluster-size statistics distinguish spots from stripes and both from the percolating dense states. The recurrence time is largest in the cells immediately above the stripe band, where the dynamics struggles to settle.
These maps, together with the observable-observable correlations shown in Fig.~\ref{fig:observables_scatter}, are used to build the morphology map in Fig.~\ref{fig:phase_diagram}.
The corresponding standard deviations are shown in Fig.~\ref{fig:observables_std}. They are small. The median spread of the density across the whole plane is $0.005$, and the largest value anywhere is $0.056$, at $(11,8)$, the one cell of the scan whose ensemble splits at all. $98.5\%$ of its realisations remain dense, $1.5\%$ fall to a dilute state, and $85.3\%$ of the full ensemble satisfy $\mathrm{CD}>0.5$. 
The endpoint is therefore set by $(B,D)$ and not by the particular realisation almost everywhere, and the single exception is the oriented cell itself.

The correlations between the different observables are made explicit in Fig.~\ref{fig:observables_scatter}. Each point is one $(B,D)$ cell, coloured by $D$ and sized by $B$.
Different morphologies occupy different regions of the observable-observable planes. The oriented cells of the $D=8$ line are separated from everything else in every panel that contains CD, since the seven of them are the only cells of the scan with $\mathrm{CD}>0.2$. The dense-noise cells are the outliers of the CSize--CSV plane, reaching mean cluster sizes of a few thousand cells with variances of $10^{6}$, where a labyrinth of nearly the same density sits at tens of cells and $10^{3}$ to $10^{5}$.
These scatter plots therefore provide an independent, low-dimensional view of the same structure that is visible in the heat maps.

Applying the criteria of Table~\ref{tab:criteria} gives $111$ N cells, $5$ C cells, $19$ S cells, $188$ P cells and one S$_o$ cell. The S cells are at $D=4$ for $B=0$--$3$, $D=5$ for $B=5$, $D=6$ for $B=7$, $D=6,7$ for $B=8$, $D=7$ for $B=9$--$11$, $D=7,8$ for $B=12,13$, and $D=8$ for $B=14$--$17$. There is no S cell at $B=4$ or $6$. The C cells remain at $D=5$ for $B\le4$.
Where an S band is assigned, it is one or two cells wide. Nowhere in the scan does a stripe state survive three steps in $D$.
The oriented stripes (S$_o$) occupy the single cell $(11,8)$. The weakening locking that continues along $D=8$ up to $B=17$ is not caught by the threshold below, and is reported in Sec.~\ref{sec:orient} from the per-realisation counts instead.

These qualitative distinctions can be turned into a quantitative morphology map by thresholding the observables (composite order parameters that contain no explicit dependence on the control parameters $B$ and $D$). The rules are collected in Table~\ref{tab:criteria}. These numbers are simply a way to encode the visual distinctions, and none of them is a parameter of the model.
Two things have to be said for the rules to define the map. Several of them can be satisfied by the same cell, so the assignment is a procedure rather than a set of conditions. The criteria are tested in the order P, S, C, N, S$_o$, each overwriting whatever the previous ones left, and a cell that satisfies none of the five is called N if $\rho>0.35$ and P otherwise. Some of these observables depend on system size. $N_c>60$ and $\mathrm{CSV}>10^5$ are statements about a lattice of $80\times80$ sites, and do not transfer to another size without rescaling. That is why the map is drawn at one $L$ and why its boundaries are approximate.

\begin{table}[htbp]
\centering
\caption{Order-parameter criteria used to assign each $(B,D)$ cell to one of the five morphologies, applied in the order P, S, C, N, S$_o$ as described in the text.}
\label{tab:criteria}
\begin{tabular}{lp{0.60\linewidth}}
\hline
Morphology & Criterion \\
\hline
N (noise) & $\rho>0.45$ and ($N_c<30$ or (CSize$>80$ and CSV$>10^5$)) \\
C (coexistence) & $0.35<\rho<0.45$, CSV$>2\times10^4$, $N_c>60$ \\
S (stripes) & $\rho>0.35$ and $N_c>10$ \\
P (spots) & $\rho<0.2$ or (CSize$<10$ and CSV$<100$) \\
S$_o$ (oriented stripes) & CD$>0.5$ and $\rho>0.3$ \\
\hline
\end{tabular}
\end{table}

The resulting morphology map is shown in Fig.~\ref{fig:phase_diagram}. It reproduces the bulk of the tour. It gives the spots at large $D$, the noise at small $D$, the coexistence band at low $B$, the stripe band in between, and the oriented cell at $(11,8)$.
The numerical classification and visual morphologies differ at specific cells. Four dense-noise cells, $(B,4)$ for $B=0$--$3$, are labelled S. Conversely, the four labyrinth cells $(B,7)$ for $B=14$--$17$ are labelled N because their cluster sizes and variances satisfy the N criterion. We retain the stated thresholds and these assignments in Fig.~\ref{fig:phase_diagram}.

\subsection{Dependence on the initial density}\label{sec:density}

\begin{figure}[htbp]
\centering
\includegraphics[width=0.85\linewidth]{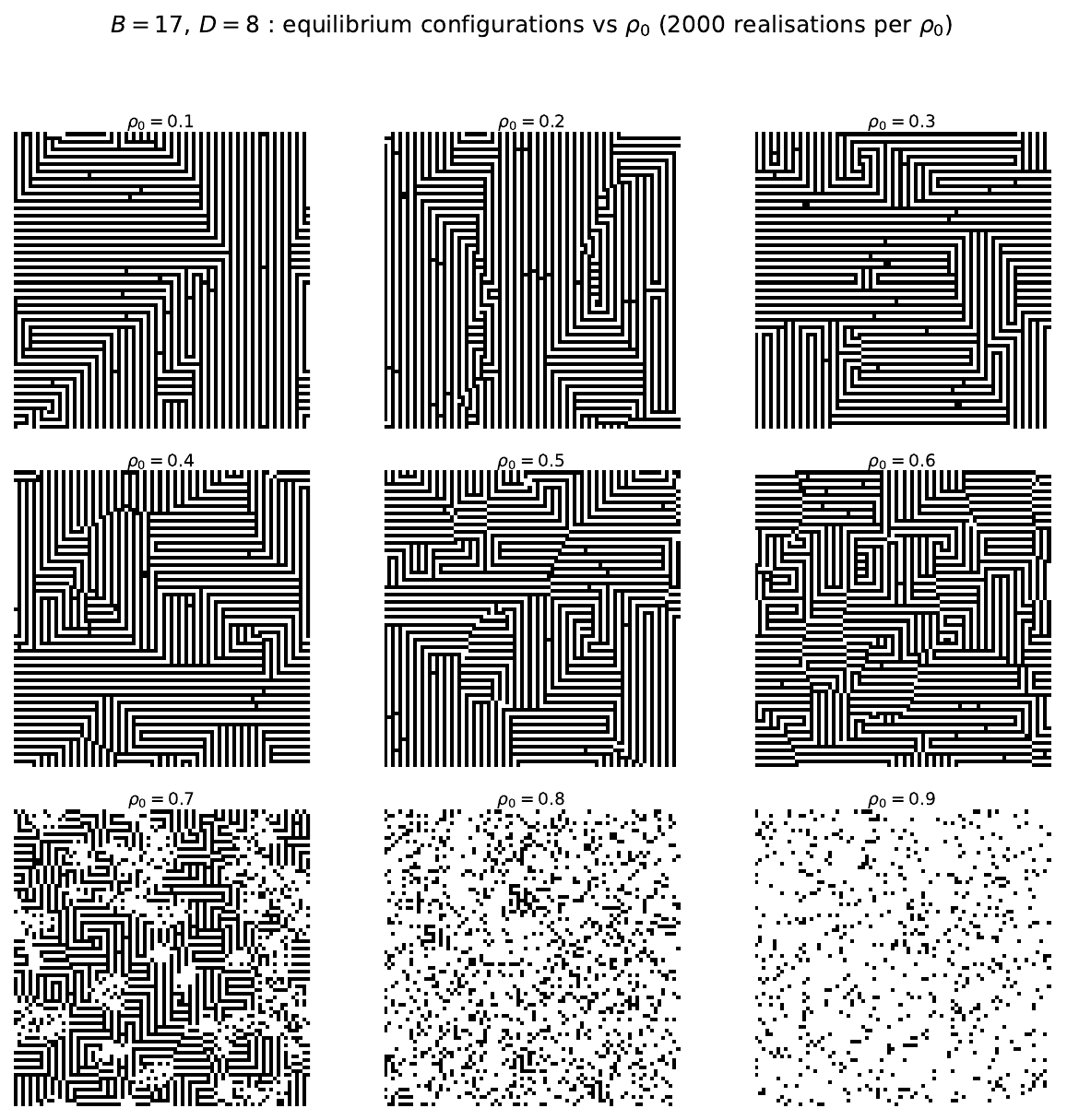}
\caption{Equilibrium configurations for $(B,D)=(17,8)$ as the initial density $\rho_0$ is varied from $0.1$ to $0.9$ ($3\times3$, left to right, top to bottom).
Each panel is one of the $2000$ realisations run at that $\rho_0$.
Over $\rho_0=0.2$--$0.6$ the mean density stays near $0.49$ with standard deviation below $0.002$, while the fraction of globally oriented realisations decreases. At $\rho_0=0.1$, the mean and standard deviation are $0.48321$ and $0.06640$. At $\rho_0=0.8$ the dense configuration is reached after evolution, while the displayed $\rho_0=0.9$ configuration is unchanged from its initial state.}
\label{fig:density_178}
\end{figure}

\begin{figure}[htbp]
\centering
\includegraphics[width=0.85\linewidth]{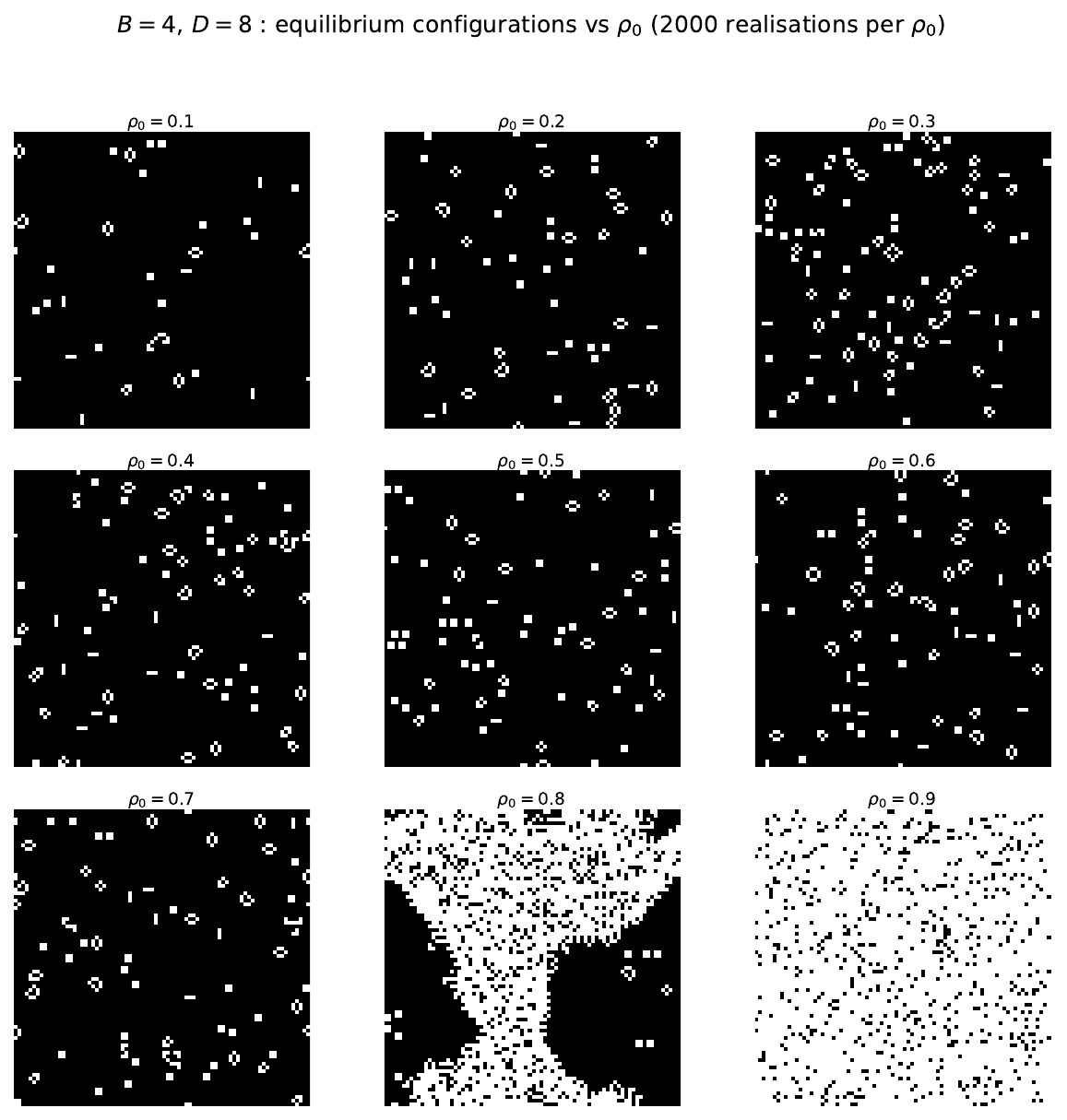}
\caption{As in Fig.~\ref{fig:density_178} for $(B,D)=(4,8)$. A dilute spot attractor pinned at $\rho_{\rm eq}\simeq0.05$ for $0.2\le\rho_0\le0.7$, a split ensemble at $\rho_0=0.8$, and a frozen initial configuration at $\rho_0=0.9$.
The $\rho_0=0.8$ panel shows a typical coexistence (dense$+$void) realisation. At this strongly fluctuating density realisation $0$ is a sparse outlier (see the discussion of ensemble fluctuations in Sec.~\ref{sec:observables}), so a typical realisation is displayed instead. Every other panel of Figs.~\ref{fig:density_178} and~\ref{fig:density_48} shows realisation $0$.}
\label{fig:density_48}
\end{figure}

The morphology map of Fig.~\ref{fig:phase_diagram} is obtained at fixed initial density $\rho_0=0.5$.
This subsection is the test of the assumption the section rests on. If the equilibrium is selected by the rule, then how the random state was prepared should not matter to it, or should matter only over a window of $\rho_0$ wide enough for the scan above to be read as a property of the rule and not of the initial condition. To find out, we scanned $\rho_0$ from $0.1$ to $0.9$ for several representative $(B,D)$ rules, with $2000$ realisations at each value of $\rho_0$ for each rule, each evolved to its first recurrent state.
Four rules were swept, chosen to bracket the map. These are $(B,D)=(0,1)$, where no cell can ever be born, $(4,6)$ and $(4,8)$ from the spots, and $(17,8)$ from the stripe band. Figs.~\ref{fig:density_178} and~\ref{fig:density_48} show the two that carry the argument. The other two are quoted below.

The rules answer in different ways, and the differences are the content of the sweep.

\textit{The reference case.} At $(0,1)$ the birth clause cannot fire, since it asks for $n_5<0$, but live cells can still die if both survival conditions fail. At $\rho_0=0.5$, the mean recurrence time is $1.4215$ steps and $58.1\%$ of the initial configurations are already fixed points. At $\rho_0=0.9$, every sampled configuration is fixed from the start.

\textit{A dilute attractor, and its coexistence.} At $(4,8)$ (Fig.~\ref{fig:density_48}) the equilibrium is a dilute spot state with $\rho_{\rm eq}$ between $0.042$ and $0.053$ for every $\rho_0$ from $0.2$ to $0.7$, whether the initial configuration held $20\%$ or $70\%$ of the lattice live. Only at $\rho_0=0.1$ is there too little live mass to reach even this attractor, and $\rho_{\rm eq}=0.018$. The rule selects the density and the preparation does not. At $\rho_0=0.8$ this stops, and it stops by splitting. There, $88\%$ of the realisations stay dense, $11\%$ fall to the dilute attractor, and the ensemble mean of $0.67$ sits well below its median of $0.78$. This is coexistence across realisations. The criterion C in Table~\ref{tab:criteria} instead identifies dense and dilute regions within a configuration. At $\rho_0=0.9$, $\rho_{\rm eq}=0.900$, the mean recurrence time is $1.3215$ steps, and $71.1\%$ of the initial configurations are already fixed points.

\textit{A crossover without a jump.} At $(4,6)$ the dilute attractor is weaker and the approach to it is smooth. There $\rho_{\rm eq}$ rises $0.018$, $0.043$, $0.053$, $0.066$, $0.159$, $0.527$, $0.694$ as $\rho_0$ runs from $0.1$ to $0.7$, and then tracks $\rho_0$. The ensemble stays narrow at every point (standard deviation at most $0.030$), so no two attractors are in competition here. Changing the threshold from $D=8$ to $D=6$ replaces the pinned-density branch and discontinuity by a smooth crossover.

\textit{The stripe band, where the two order parameters thermalize at different rates.} Fig.~\ref{fig:density_178} shows $(17,8)$. Over $\rho_0=0.2$--$0.6$ the mean equilibrium density ranges from $0.4853$ to $0.4925$, with an ensemble standard deviation below $0.002$. At $\rho_0=0.1$, the mean is $0.48321$ and the standard deviation is $0.06640$. The initial density is forgotten as completely as the scan of Sec.~\ref{sec:observables} assumes. The orientational order is not forgotten at all. The fraction of realisations that reach a single global orientation, measured by the same $\mathrm{CD}>0.5$ criterion as Fig.~\ref{fig:phase_11_orient}, falls monotonically over $0.1\le\rho_0\le0.6$. It is $54\%$ at $\rho_0=0.1$, $51\%$ at $0.2$, $33\%$ at $0.3$, $18\%$ at $0.4$, $6\%$ at $0.5$ and $0.1\%$ at $0.6$. Two properties of one and the same equilibrium thus thermalize on different scales. The density is set by the rule, the orientation is inherited from the initial condition. The distinction is the one the tour already forced at $(11,8)$, where the orientation is chosen afresh in every realisation, and it says that at $\rho_0\ne0.5$ the stripe band remembers how it was filled.

Above $\rho_0=0.7$, the equilibrium density approaches the initial density, with $\rho_{\rm eq}=0.794$ at $\rho_0=0.8$ and $0.900$ at $\rho_0=0.9$. The respective mean recurrence times are $9.941$ and $1.469$ steps. No sampled initial configuration is fixed at $\rho_0=0.8$, whereas $59.4\%$ are fixed at $\rho_0=0.9$. Birth still requires exactly three live near neighbours, while survival can be sustained by more than eight live sites in the distant ring.

Here the analogy with the reaction--diffusion systems of Sec.~\ref{sec:turing} ends. Reaction--diffusion equations can have stationary inhomogeneous solutions, as in Sec.~\ref{sec:turing}. In the dense lattice runs, some initial configurations satisfy the update rule at every site and are therefore fixed points. The morphology map of Fig.~\ref{fig:phase_diagram} is drawn at $\rho_0=0.5$, where the rule selects the equilibrium from the density alone, and the high-density recurrence statistics are given above.

\begin{figure}[htbp]
\centering
\includegraphics[width=0.85\linewidth]{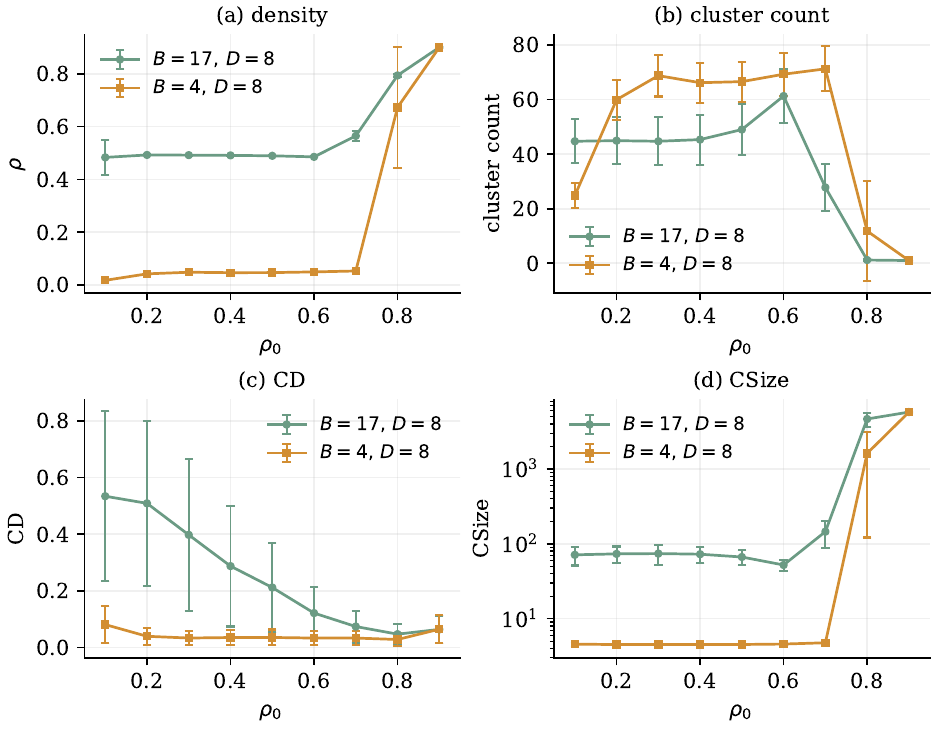}
\caption{Mean equilibrium density $\rho$, cluster count, directional-correlation difference CD and mean cluster size CSize versus initial density $\rho_0$ for $(B,D)=(17,8)$ and $(4,8)$. The points are means with standard-deviation error bars over the $2000$ realisations.
Same data as the configurations of Figs.~\ref{fig:density_178} and~\ref{fig:density_48}.}
\label{fig:density_obs}
\end{figure}

Fig.~\ref{fig:density_obs} collects the mean equilibrium density, cluster count, CD and mean cluster size behind the two cases.

\subsection{Intrinsic dimension of the equilibrium configurations}\label{sec:id}

A complementary question about the endpoints, which needs no thresholds, is how much configuration space each ensemble actually fills. An ordered morphology is confined to a narrow manifold of binary configurations, whereas a disordered one spreads over a much larger volume.
We therefore estimate the intrinsic dimension of the set of equilibrium configurations\cite{Facco2017}. Intrinsic-dimension diagnostics have also been used to distinguish multiple transitions in lattice spin and gauge systems.~\cite{Mei:2026vay}
It will not separate the ordered morphologies from one another, but it separates the ordered from the disordered without invoking a threshold, and it isolates a structure that none of the six observables of Sec.~\ref{sec:observables} contains.
For each $(B,D)$ we collect the $2000$ equilibrium $80\times80$ configurations of that cell, flatten each into a binary vector $\mathbf{x}_i\in\{0,1\}^{L^2}$ with $L=80$, and remove duplicates, since identical configurations would produce zero distances and make the estimator ill-defined.
In practice no cell of the scan contains a repeated configuration, so all $2000$ realisations of every cell are available to the estimate.
The distance between two configurations is the Euclidean distance normalised by the linear size,

\begin{equation}
d_{ij}=\left[\frac{1}{L^2}\sum_{k=1}^{L^2}\bigl|x_{i,k}-x_{j,k}\bigr|\right]^{1/2}.
\end{equation}
Because the components are binary, $|x_{i,k}-x_{j,k}|=(x_{i,k}-x_{j,k})^2$, so $d_{ij}=\|\mathbf x_i-\mathbf x_j\|_2/L$.
The overall factor $1/L$ drops out of the neighbour-distance ratios $r_{i,2k}/r_{i,k}$ used below, but it keeps the absolute distances of order unity and avoids finite-precision issues.

\begin{figure}[p]
\centering
\includegraphics[width=\linewidth,height=0.68\textheight,keepaspectratio]{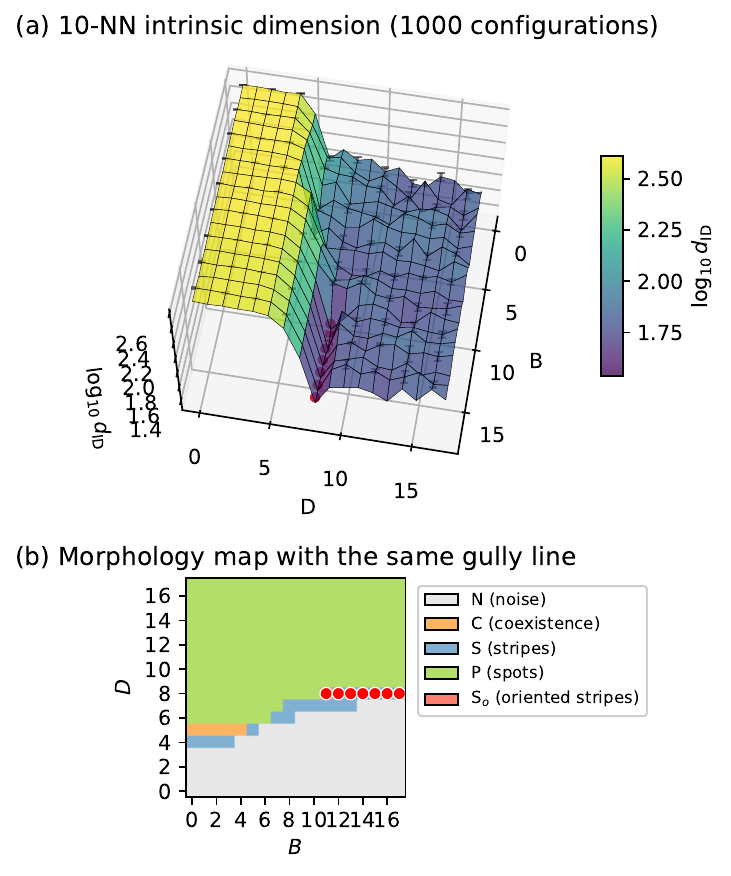}
\caption{(a) Three-dimensional view of the intrinsic dimension of the equilibrium-state manifold in the $(B,D)$ plane, estimated with the 10-nearest-neighbours ($k=10$) method on $1000$ configurations drawn from the $2000$ realisations of each cell.
The vertical axis is $\log_{10}d_{\rm ID}$ so that the gully of low $d_{\rm ID}$ remains visible next to the plateau of high $d_{\rm ID}$.
Vertical error bars show the bootstrap standard deviation obtained from four resamples of $400$ realisations.
The red dashed curve marks the primary low-$d_{\rm ID}$ gully. The tracking procedure returns the line $D=8$ for $B=11$ through $17$ and no point elsewhere (see text).
(b) The same $(B,D)$ morphology map as Fig.~\ref{fig:phase_diagram}, with that gully line overlaid in red for direct comparison.
Low values correspond to the ordered stripe and spot morphologies. High values correspond to the disordered ones.}
\label{fig:id}
\end{figure}

We estimate the intrinsic dimension with the $k$-nearest-neighbours ($k$-NN) generalisation of the two-NN estimator~\cite{Facco2017,Denti2022}.
For every configuration $\mathbf{x}_i$ we denote by $r_{i,k}$ and $r_{i,2k}$ its distances to the $k$-th and $2k$-th nearest neighbours, and define $\mu_i=r_{i,2k}/r_{i,k}$.
For locally uniform sampling in $d$ dimensions, the volume inside radius $r$ is proportional to $r^d$. Conditional on the $2k$-th neighbour radius, the $2k-1$ interior points are uniform in that volume. Thus $z=\mu^{-d}$, the fractional volume at the $k$-th neighbour, has density $z^{k-1}(1-z)^{k-1}/B(k,k)$ on $0<z<1$, where $B(k,k)=\int_0^1z^{k-1}(1-z)^{k-1}\,\mathrm dz$. Its logarithmic moment is $\langle\ln z\rangle=-\sum_{j=k}^{2k-1}j^{-1}$. Using $\ln z=-d\ln\mu$ therefore gives

\begin{equation}
d_{\rm ID}=\frac{\displaystyle\sum_{j=k}^{2k-1}j^{-1}}{\langle\ln\mu_i\rangle},\qquad \sum_{j=10}^{19}j^{-1}=0.7187714\ldots .
\label{eq:knn}
\end{equation}
Larger $k$ smooths out the ratio and reduces the bootstrap variance.
For the results shown in Fig.~\ref{fig:id} we use $k=10$.
The distance matrix grows as the square of the number of configurations, so the estimate is formed from $1000$ of the $2000$ configurations of the cell, drawn without replacement.
The uncertainty is estimated by bootstrap. The dimension is recomputed on four resamples of $400$ configurations drawn with replacement from the $2000$ of the cell, duplicates are removed in each resample, and the standard deviation of the four estimates is reported.

A low $d_{\rm ID}$ means that the endpoints of a cell are confined to a narrow manifold of configurations. A high value means that they are not.
Two cautions belong with the number before it is used.
The first concerns its scale. The estimate is capped at the embedding dimension, $d_{\rm ID}\le L^2=6400$, and the largest value obtained anywhere in the scan is $413$, so the cap never binds and only the ordering that the estimator induces will be read off the map.
The second is that the dimension is not independent of the density.
Over the whole plane $d_{\rm ID}$ and the equilibrium density $\rho$ have a Pearson correlation of $0.90$. That is unsurprising, since a configuration with fewer live cells has fewer ways of differing from its neighbours. The map taken by itself would say little more than that the dilute regions are dilute.
The comparisons below are therefore made either across the gap in the distribution of $d_{\rm ID}$, or between cells at the same density.

To locate the sharpest structure of the surface we detect a gully of low $d_{\rm ID}$ in the surface of $\log_{10}d_{\rm ID}$. The dimension enters none of the criteria of Table~\ref{tab:criteria}, so the two routes are independent and can be compared.
For each $B$ we find the local minima of $\log_{10}d_{\rm ID}$ along $D$ inside the window $5\le D\le16$ whose prominence, how far a minimum stands below the ridge on either side, exceeds $0.12$ in $\log_{10}$. The valley is anchored at the most prominent of these minima and followed in both directions across neighbouring $B$ values, each step being allowed to move by at most $|\Delta D|\le2$. A three-point median filter removes isolated jitter.
The procedure returns one straight segment and nothing else. The line is $D=8$, traced from $B=11$ to $B=17$.
It does not trace the boundary between the plateau and the ordered region, so the red dashed curve of Fig.~\ref{fig:id}(a) and the red dots of panel (b) mark that line and no other.
Why the segment stops at $B=10$ is itself informative. The minima it chains together have prominences of $0.40$ to $0.57$ in $\log_{10}$, a factor $2.5$ to $3.7$ in the dimension, whereas no minimum anywhere else in the plane exceeds $0.26$, and at $B=10$ there is no minimum in the window. The weaker minimum at $(9,8)$, of prominence $0.120$, is disconnected from the segment by that gap.

Fig.~\ref{fig:id} shows the resulting map.
Sorted, the $324$ values split into two groups with an empty interval between them. $205$ cells lie between $19.6$ and $138.8$, and $119$ cells between $247.1$ and $412.9$.
The split is a split by column, with one column in dispute. The high group is every cell with $D\le5$ together with the cells at $D=6$ and $B\ge7$, and the low group is every cell with $D\ge7$ together with the cells at $D=6$ and $B\le6$.

The high group is the dense, disordered part of the plane.
For $D\le4$ the dimension lies between $361.5$ and $412.9$ at every $B$, the largest value being $412.9$ at $(B,D)=(3,0)$.
The remaining $29$ cells of the high-dimensional group belong to $D=5$ and $D=6$, the two columns adjacent to the low-dimensional region.

The low group is the ordered region, and within it the dimension falls along the sequence the tour describes rather than rising.
For $B=8$--$17$ at $D=7$, the values lie between $108.7$ and $138.8$. The spot cells lie between $39.2$ and $113.1$, with median $65.0$. The seven cells at $D=8$ with $B\ge11$ lie between $19.6$ and $27.2$.
The lowest-dimensional ensemble in the whole plane is therefore not the dilute spots but the dense oriented state, and that is the content of the gully.
The comparison has to be made at fixed density, and it can be. At $B=11$ the labyrinth cell $(11,7)$ and the oriented cell $(11,8)$ have densities $0.483$ and $0.481$ and $68.6$ and $69.7$ live clusters, so they differ in neither the amount of live matter nor the number of pieces it is broken into, yet their dimensions are $114.1$ and $25.6$.
The same holds along the whole line, $108.7$ to $114.1$ against $19.6$ to $27.2$ at $\rho\simeq0.49$ throughout.
The comparison concerns the estimated number of independent directions in configuration space, not the number of distinct configurations. Both cells contain $2000$ distinct sampled configurations.
CD and $d_{\rm ID}$ are computed from the same ensemble, with $2000$ configurations used for CD and $1000$ for $d_{\rm ID}$ by unrelated formulas (one from two-point correlations along the lattice axes, the other from the geometry of the ensemble as a whole), and the seven cells with $d_{\rm ID}<28$ are exactly the seven cells of Sec.~\ref{sec:orient} on which some realisations are oriented.
The orientational locking found there is therefore not an artefact of the threshold built into CD.

The two classifications disagree at eleven cells when comparing the S/N labels with the two dimension groups. Seven S-labelled cells lie in the high-dimensional group, namely $(B,4)$ for $B=0$--$3$, $(5,5)$, $(7,6)$, and $(8,6)$, with $d_{\rm ID}=299.8$--$405.4$. Four N-labelled cells, $(B,7)$ for $B=14$--$17$, lie in the low-dimensional group. The numerical morphology thresholds and the nearest-neighbour dimension therefore give different assignments at these boundaries.

\section{\label{sec:5}A Turing-type density equation for a self-gravitating toy universe}

The examples in the previous sections show that Turing-like patterns can form when a local activation is combined with a longer-range inhibition. 
Here we ask whether these competing effects are present in an expanding box of particles held together by their own gravity, and whether the structures such a system builds can be described, approximately, by a density equation of the activation--inhibition type.
Gravity only attracts, yet mass conservation gives the two competing effects. 
A region denser than its surroundings pulls matter toward itself and so grows denser still. This is self-amplification, the activation. 
The matter it pulls comes from around it, so the surrounding region loses density. This is depletion, the inhibition. These are the two effects that the spherical infall model of a density peak describes, growth of the core and depletion of the shell around it~\cite{Gunn1972,Bertschinger1985}.  The two effects are amplification and depletion, and the third ingredient concerns their relative spatial ranges. Both effects arise from the same gravitational flow, so the long-range force alone does not establish a longer inhibition range.

\subsection{\label{sec:5A}Particle simulation of the toy universe}

The toy universe is an expanding cube with periodic boundary conditions, the setting of the cosmological $N$-body problem \cite{Springel:2006vs}. 
It holds $N$ particles of equal mass $m$, and gravity is the only interaction between them. The structure they build is the cosmic web \cite{Bond:1995yt}. 
Positions are given in comoving coordinates $\mathbf r$, which the expansion does not stretch, i.e., $\mathbf r$ is dimensionless and runs from $0$ to $1$, so the box has unit side, and two points whose coordinates differ by $\Delta \mathbf r$ are a physical distance $a(t)|\Delta \mathbf r|$ apart. 
Here $t$ is the cosmic time, $a(t)$ the scale factor, which grows with time, and $H=\dot a/a$ the expansion rate. 
The evolution is advanced in a sequence of discrete steps $t_0<t_1<\cdots$, and the position of particle $i$ at the $j$th of them is written $\mathbf x_{i,j}$.

The box is divided into $M^{3}$ cells, and we use $N=M^{3}$. 
The density is defined at the sites, the corners the cells share, written $\rho_{\mathbf l}$ with $\mathbf l$ an integer triple running from $0$ to $M-1$ along each axis. 
Particle $i$, at $\mathbf x_i$, gives its mass to the eight sites of the cell containing it, with a weight $W_{\mathbf l}({\mathbf x} _i)=M^{-3}W(\mathbf l/M-\mathbf x_i)$, i.e. the cloud-in-cell~(CIC) assignment \cite{Hockney1988}, 
\begin{equation}
\begin{aligned}
&W(\mathbf r-\mathbf x_i)\equiv M^{3}\prod_{\mu}\max\bigl(0,\,1-M|r^{\mu}-x_i^{\mu}|\bigr),
\end{aligned}
\label{eq:density_cicweight}
\end{equation}
with periodic separations $r^\mu-x_i^\mu$ taken in $[-1/2,1/2)$, and $\int_{[0,1)^3} W\,\mathrm d^{3}r=1$, so that
\begin{equation}
\rho_{\mathbf l}=\frac{Nm}{a^{3}}\sum_{i}W_{\mathbf l}(\mathbf x_i).\\
\label{eq:density_def}
\end{equation}
The $\max$ in Eq.~\eqref{eq:density_cicweight} merely keeps the eight corners of the cell that contains the particle, so a particle enters the sum over $i$ in Eq.~\eqref{eq:density_def} only if it lies in one of the eight cells that have the site $\mathbf l$ as a corner.
The density contrast is $\delta_{\mathbf l}\equiv\bigl[\rho_{\mathbf l}-\bar\rho\bigr]/\bar\rho$, with $\bar{\rho}=Nm/a^3$.

The equation of motion particles obey, and the equation that fixes the potential $\Phi$, have the following continuous versions \cite{Efstathiou:1985re},
\begin{align}
\ddot{\mathbf x}_i+2H\dot{\mathbf x}_i
&=-\frac{1}{a^{2}}\nabla\Phi ({\bf x}_i),
\label{eq:comoving_eom}\\
\nabla^{2}\Phi
&=4\pi G a^{2}\bigl(\rho-\bar\rho\bigr),
\label{eq:poisson}
\end{align}
where $\mathbf x_i$ is the position of particle $i$ at time $t$, $\dot{\mathbf x}_i$ its velocity in these coordinates and a dot denotes $\mathrm d/\mathrm dt$, and $G$ Newton's constant. 
The gradient $\nabla$ is taken with respect to position $\mathbf r$. Where it acts on a field, as in $\nabla\Phi$, the field is a function of that coordinate. Where the result is needed at a particle, as in $\nabla\Phi(\mathbf x_i)$, it is read at the position of that particle.
The mean density drives the expansion, which Eq.~\eqref{eq:comoving_eom} already carries in $H$, so only the excess $\rho-\bar\rho$ sources $\Phi$. 
Eq.~\eqref{eq:poisson} is solved mode by mode~\cite{Hockney1988,Efstathiou:1985re}. 
Writing $\mathbf k=2\pi\mathbf n$ for the modes of the unit box, $\mathbf n$ an integer triple, $k=|\mathbf k|$, and $\delta_{\mathbf k}$, $\Phi_{\mathbf k}$ for the Fourier components of $\delta$ and $\Phi$, transforming Eq.~\eqref{eq:poisson} gives
\begin{equation}
\begin{aligned}
\delta_{\mathbf k}&=\sum_{\mathbf l}\delta_{\mathbf l}\,\mathrm{e}^{-\mathrm{i}\mathbf k\cdot\mathbf l/M},\\
\Phi_{\mathbf k}&=-\frac{3}{2a}\,\frac{\delta_{\mathbf k}}{k^{2}},
\qquad
\bigl(\nabla\Phi\bigr)_{\mathbf k}=\mathrm{i}\mathbf k\,\Phi_{\mathbf k},
\qquad \mathbf k\ne\mathbf 0 ,\\
\nabla\Phi(\mathbf r)&=\frac{1}{M^{3}}\sum_{\mathbf k}\bigl(\nabla\Phi\bigr)_{\mathbf k}\,\mathrm{e}^{\mathrm{i}\mathbf k\cdot {\mathbf r}} ,
\end{aligned}
\label{eq:poisson_k}
\end{equation}

The kick--drift--kick organisation follows cosmological particle-integration methods.~\cite{Quinn:1997iy} Under the discretisation of time above, Eq.~\eqref{eq:comoving_eom} is solved by the following kick--drift--kick update with explicit Hubble drag,
\begin{equation}
\begin{aligned}
\mathbf v_{i,j+\frac12}&=\mathbf v_{i,j}+\frac{\Delta t_j}{2}\Bigl[-\frac{1}{a_{j}^{2}}\bigl(\nabla\Phi({\bf x}_{i,j})\bigr)_{j}-2H_j\,\mathbf v_{i,j}\Bigr],\\
\mathbf x_{i,j+1}&=\mathbf x_{i,j}+\Delta t_j\,\mathbf v_{i,j+\frac12},\\
\mathbf v_{i,j+1}&=\mathbf v_{i,j+\frac12}+\frac{\Delta t_j}{2}\Bigl[-\frac{1}{a_{j+1}^{2}}\bigl(\nabla\Phi({\bf x}_{i,j+1})\bigr)_{j+1}-2H_{j+1}\,\mathbf v_{i,j+\frac12}\Bigr],
\end{aligned}
\label{eq:leapfrog}
\end{equation}
where $\mathbf v_{i,j}=\dot{\mathbf x}_{i,j}$ is the velocity of particle $i$ at step $j$. The first and third lines carry it forward by half a step at a time, and each gradient is interpolated to the particle from the mesh.
By interpolation we mean that $\nabla\Phi$ at the position of a particle is the sum of its values at the eight sites of the cell containing that particle, each weighted by $W_{\mathbf l}(\mathbf x_{i,j})$ of Eq.~\eqref{eq:density_cicweight}.
The background is Einstein--de~Sitter, so $a(t)=\left((3/2)\,t\right)^{2/3}$, and $G=3/(8\pi)$ in the units chosen here~\cite{Peebles:1980yev,Bernardeau:2001qr}.
For simplicity, we set $m=1/N$.
The integration uses $a_0=0.05$ and $a_{j+1}=a_j+\Delta a_j$, with $\Delta a_j\le0.002$ shortened so that $\Delta t_j\max_{i,\mu}|v_{i,j}^{\mu}|\le0.2/M$ and each output epoch is reached exactly. Here $t_j=(2/3)a_j^{3/2}$ and $H_j=a_j^{-3/2}=2/(3t_j)$. The update uses $\Delta t_j=a_j^{1/2}\Delta a_j$, the first-order expansion of the exact interval $t_{j+1}-t_j=(2/3)[(a_j+\Delta a_j)^{3/2}-a_j^{3/2}]$.

The particles carry no attribute but their position, so the initial contrast is carried by a displacement of a regular array. 
With one particle at the centre of each cell that array deposits a strictly uniform density, and displacing it by $\mathbf s_{\mathbf l}$ produces the contrast. 
Labelling each of them by the integer triple $\mathbf l$ of the cell it starts from, and writing $\mathbf s_{\mathbf l}$ for the displacement it is given, at the initial step $j=0$
\begin{align}
\mathbf x_{\mathbf l,0}&=\frac{\mathbf l+\tfrac12}{M}+\mathbf s_{\mathbf l},
\label{eq:zeldovich_x}\\
\mathbf v_{\mathbf l,0}&=H_0\,\mathbf s_{\mathbf l},
\label{eq:zeldovich_v}
\end{align}
where $a_0=0.05$ is the instant Eq.~\eqref{eq:leapfrog} starts from, $H_0=H(a_0)=a_0^{-3/2}$, and $\mathbf s_{\mathbf l}=\mathbf0$ throughout leaves the uniform density unchanged. 
Every particle run reported here starts at $a_0$ and ends at $a=1$.
The first-order relation $\delta=-\nabla\cdot\mathbf s$ implies that a displacement proportional to $a$ gives a linear density contrast proportional to $a$. The run starts on the growing mode of the Einstein--de~Sitter background, and $\mathbf v_{\mathbf l,0}=H_0\mathbf s_{\mathbf l}$ is the rate of change of that displacement at $a_0$. Both initial data follow from the single field $\mathbf s$. This is the Zel'dovich initial condition \cite{Zeldovich:1969sb}.

The displacement follows from the initial contrast, namely the $\delta_{\mathbf l}$ of the next paragraph and its mode $\delta_{\mathbf k}$, through $\delta=-\nabla\!\cdot\mathbf s$. Over the modes of Eq.~\eqref{eq:poisson_k} that condition has the solution $\mathbf s_{\mathbf k}=\mathrm{i}\mathbf k\,\delta_{\mathbf k}/k^{2}$ at $\mathbf k\ne\mathbf0$, the displacement being curl-free, and $\mathbf s_{\mathbf0}=\mathbf0$. A curl-free displacement is what makes the map of Eq.~\eqref{eq:zeldovich_x} a gradient flow, and it is what the approximation rests on~\cite{Schandarin:1989sr,Gurbatov:1989az}.

The displacement is a Gaussian random field~\cite{Bertschinger:1995er} over the whole box, which puts many scales in at once. With $k$ the wave number of Eq.~\eqref{eq:poisson_k}, the coefficients $c_{\mathbf k}$ supplied on the stored half-grid to the inverse real Fourier transform are

\begin{equation}
c_{\mathbf k}\propto\Bigl[\frac{k}{8\pi}\Bigl(1+\frac{k^{2}}{64\pi^{2}}\Bigr)^{-2}\exp\Bigl(-\frac{k^{4}}{32^{4}\pi^{4}}\Bigr)\Bigr]^{1/2}g_{\mathbf k},
\label{eq:web_field}
\end{equation}
where $g_{\mathbf k}$ has independent zero-mean Gaussian real and imaginary parts on the stored Fourier half-grid. A real inverse Fourier transform produces the real density field, whose full Fourier coefficients are denoted by $\delta_{\mathbf k}$. The zero mode is set to zero.
The resulting $\delta_{\mathbf k}$ are fixed up to one overall normalization factor. 
To fix it, multiply $\delta_{\mathbf k}$ by $\exp\bigl[-k^{2}(2/M)^{2}/2\bigr]$, that is, smooth the field with a Gaussian of width two cells (a Gaussian window is written $\exp(-k^{2}\ell^{2}/2)$ throughout this paper), transform the result back to coordinate space, and divide $\delta_{\mathbf k}$ by the standard deviation of that coordinate-space field. 
The smoothing is what makes the factor meaningful. It removes the modes closest to the grid cutoff, whose amplitude is set by the discretisation rather than by the spectrum, so that the normalisation refers to the amplitude on the scales the mesh resolves. 
The smoothed field then has unit standard deviation at $a=1$, so that the growing mode brings it to $0.05$ at $a_0$.

\subsection{\label{sec:5B}From particle motion to two fields}

The mechanism suggested by the particle configurations is simple. A density enhancement attracts more matter, while the region supplying that matter is depleted. 
To express these two effects in one density equation, we start from the equations of motion of the point masses, Eqs.~\eqref{eq:comoving_eom} and~\eqref{eq:poisson}, rather than from a postulated continuous field, so that every coefficient below is inherited from them.

We use the scale factor $a$ as time and write $\mathbf x_i'=\mathrm d\mathbf x_i/\mathrm da$.
Since $\mathrm d/\mathrm dt=aH\,\mathrm d/\mathrm da$ and $H=a^{-3/2}$, this velocity is related to the velocity in Eq.~\eqref{eq:leapfrog} by $\mathbf x_i'=a^{1/2}\dot{\mathbf x}_i$.
Let $\bar\rho_c=Nm$ be the constant mean comoving density.
The normalised density and its mass-weighted velocity with respect to $a$, with $W$ the weight of Eq.~\eqref{eq:density_cicweight}, are \cite{Dominguez:2000rm}
\begin{equation}
\begin{aligned}
\frac{\rho(\mathbf r,a)}{\bar\rho}&=1+\delta(\mathbf r,a)=\frac{1}{\bar\rho_c}\sum_i mW(\mathbf r-\mathbf x_i).\\
\mathbf w(\mathbf r,a)&\equiv\langle\mathbf x'\rangle_W =\frac{\sum_i m\mathbf x_i'W(\mathbf r-\mathbf x_i)}{\sum_i mW(\mathbf r-\mathbf x_i)} .
\end{aligned}
\label{eq:particle_fields}
\end{equation}
Multiplying the average $\langle\,\ldots\,\rangle_W$ of any particle quantity by $1+\delta$ removes the denominator, since $1+\delta$ is that same sum,
\begin{equation}{
(1+\delta)\bigl\langle\,\ldots\,\bigr\rangle_W=\frac{1}{\bar\rho_c}\sum_i m\,(\,\ldots\,)\,W(\mathbf r-\mathbf x_i),
}\label{eq:mean_identity}
\end{equation}
Thus $\delta$ is the contrast of Eq.~\eqref{eq:density_def}, its normalisation now taken by the comoving mean, the factor of $a^{3}$ cancelling in the ratio, and $\mathbf w$ is the mass-weighted average of $\mathrm d\mathbf x/\mathrm d a$, a velocity in the $a$ clock. 
Differentiating the assignment weight at a fixed particle gives $\partial_aW=-\mathbf x_i'\!\cdot\nabla W$. 
Since $\mathbf x_i'$ does not depend on the field point $\mathbf r$, $\mathbf x_i'\!\cdot\nabla W=\nabla\!\cdot(\mathbf x_i'W)$. 
Summing over particles, the left-hand side is $\partial_a\delta$ and the right-hand side is $-\nabla\cdot\bigl[\bar\rho_c^{-1}\sum_i m\mathbf x_i'W\bigr]$, where the bracket is $(1+\delta)\mathbf w$ by the definition above.
Hence
\begin{equation}
\partial_a\delta=-\nabla\!\cdot\bigl[(1+\delta)\mathbf w\bigr].
\label{eq:continuity}
\end{equation}
This relation is mass conservation, i.e., every change of $\delta$ comes from the transport of mass by the particles.

The particle equation of motion, Eq.~\eqref{eq:comoving_eom}, and Poisson's equation, in the same clock, are
\begin{subequations}\label{eq:particle_a}
\begin{align}
\mathbf x_i''+\frac{3}{2a}\mathbf x_i'&=-\frac{1}{a}\nabla\Phi(\mathbf x_i),\label{eq:particle_a_motion}\\
\nabla^2\Phi&=\frac{3}{2a}\delta .\label{eq:particle_a_poisson}
\end{align}
\end{subequations}
We also read the force on a cell from the mesh at that point, rather than averaging the forces on the particles that deposit there. 
Write $W_i=W(\mathbf r-\mathbf x_i)$ and $\partial_\mu=\partial/\partial r^\mu$, with repeated spatial indices $\mu,\nu\in\{1,2,3\}$ summed. Taking the ellipsis in Eq.~\eqref{eq:mean_identity} to be $x'_\mu$ gives $(1+\delta)w_\mu=\bar\rho_c^{-1}\sum_i mW_ix'_{i\mu}$. Differentiating it at fixed $\mathbf r$, and using the derivative of the weight,
\begin{equation}{
\begin{aligned}
\partial_a\bigl[(1+\delta)w_\mu\bigr]
&=\frac{1}{\bar\rho_c}\sum_i m\bigl[W_ix''_{i\mu}+x'_{i\mu}\partial_aW_i\bigr]\\
&=\frac{1}{\bar\rho_c}\sum_i mW_ix''_{i\mu}
 -\frac{1}{\bar\rho_c}\sum_i mx'_{i\mu}\,\mathbf x_i'\!\cdot\nabla W_i\\
&=\frac{1}{\bar\rho_c}\sum_i mW_ix''_{i\mu}
 -\partial_\nu\Bigl[\frac{1}{\bar\rho_c}\sum_i mx'_{i\mu}x'_{i\nu}W_i\Bigr]\\
&=\frac{1}{\bar\rho_c}\sum_i mW_ix''_{i\mu}
 -\partial_\nu\bigl[(1+\delta)\langle x'_\mu x'_\nu\rangle_W\bigr],
\end{aligned}
}\label{eq:dil}
\end{equation}
where the second line uses $\partial_aW_i=-\mathbf x_i'\!\cdot\nabla W_i$, the third $\partial_\nu(x'_{i\mu}x'_{i\nu}W_i)=x'_{i\mu}x'_{i\nu}\partial_\nu W_i=x'_{i\mu}\,\mathbf x_i'\!\cdot\nabla W_i$, which holds because $\mathbf x_i'$ carries no dependence on $\mathbf r$ and the derivative therefore passes through the sum, and the fourth the ellipsis $x'_\mu x'_\nu$ in Eq.~\eqref{eq:mean_identity}. 
Writing $\langle x'_\mu x'_\nu\rangle_W=w_\mu w_\nu+\sigma_{\mu\nu}$ with $\sigma_{\mu\nu}\equiv\langle(x'_\mu-w_\mu)(x'_\nu-w_\nu)\rangle_W$, the spread of velocities in a cell, and differentiating the product in the first term with the continuity equation $\partial_a(1+\delta)=-\partial_\nu[(1+\delta)w_\nu]$ of Eq.~\eqref{eq:continuity}, Eq.~\eqref{eq:dil} reads \cite{Dominguez:2000rm,Pueblas:2008uv}
\begin{equation}{
\begin{aligned}
&\frac{1}{\bar\rho_c}\sum_i mW_ix''_{i\mu}=\partial_a\bigl[(1+\delta)w_\mu\bigr]+\partial_\nu\bigl[(1+\delta)(w_\mu w_\nu+\sigma_{\mu\nu})\bigr]\\
&\quad=(1+\delta)\partial_aw_\mu-w_\mu\partial_\nu\bigl[(1+\delta)w_\nu\bigr]
 +w_\mu\partial_\nu\bigl[(1+\delta)w_\nu\bigr]+(1+\delta)w_\nu\partial_\nu w_\mu+\partial_\nu\bigl[(1+\delta)\sigma_{\mu\nu}\bigr]\\
&\quad=(1+\delta)\bigl[\partial_aw_\mu+w_\nu\partial_\nu w_\mu\bigr]+\partial_\nu\bigl[(1+\delta)\sigma_{\mu\nu}\bigr].
\end{aligned}
}\label{eq:momentum_exacta}
\end{equation}
Averaging Eq.~\eqref{eq:particle_a_motion} with the same weight, with the ellipsis filled by $\partial_\mu\Phi(\mathbf x_i)$, gives
\begin{equation}{
\frac{1}{\bar\rho_c}\sum_i mW_ix''_{i\mu}
=-\frac{3}{2a}(1+\delta)w_\mu-\frac{1}{a}(1+\delta)\bigl\langle\partial_\mu\Phi(\mathbf x_i)\bigr\rangle_W.
}\label{eq:accmean}
\end{equation}
Using Eqs.~\eqref{eq:momentum_exacta} and \eqref{eq:accmean},
\begin{equation}{
\begin{aligned}
&(1+\delta)\bigl[\partial_aw_\mu+w_\nu\partial_\nu w_\mu\bigr]+\partial_\nu\bigl[(1+\delta)\sigma_{\mu\nu}\bigr]=-\frac{3}{2a}(1+\delta)w_\mu-\frac{1}{a}(1+\delta)\bigl\langle\partial_\mu\Phi(\mathbf x_i)\bigr\rangle_W ,
\end{aligned}
}\label{eq:momentum_exact}
\end{equation}
Dividing by $1+\delta$, dropping the $\sigma_{\mu\nu}$ term which is the flux of the spread, and reading the force at the cell, $\langle\partial_\mu\Phi(\mathbf x_i)\rangle_W\to\partial_\mu\Phi(\mathbf r)$, gives \cite{Bernardeau:2001qr,Pueblas:2008uv}
\begin{equation}
\partial_a\mathbf w+(\mathbf w\!\cdot\nabla)\mathbf w+\frac{3}{2a}\mathbf w=-\frac{1}{a}\nabla\Phi.
\label{eq:resolved_motion}
\end{equation}
To obtain a model with only two fields, we assume the mesh velocity is curl-free with zero spatial mean and write it through one periodic potential, $\mathbf w=\nabla\chi$, and its initial value follows from the gradient displacement in Sec.~\ref{sec:5A}~\cite{Sahni:1995rm,Bernardeau:2001qr}.
For a gradient velocity, $(\mathbf w\!\cdot\nabla)\mathbf w=\nabla|\nabla \chi|^2/2$, since $w_\nu\partial_\nu w_\mu=\partial_\nu\chi\,\partial_\mu\partial_\nu\chi=\partial_\mu|\nabla \chi|^2/2$, and $3\mathbf w/(2a)=\nabla\bigl(3\chi/(2a)\bigr)$, so Eq.~\eqref{eq:resolved_motion} reads
\begin{equation*}
\nabla\Bigl[\partial_a\chi+\frac12|\nabla\chi|^2+\frac{3}{2a}\chi+\frac{1}{a}\Phi\Bigr]=0.
\end{equation*}
The bracket is therefore a function of $a$ alone, $C(a)$. Poisson's equation \eqref{eq:particle_a_poisson} gives $\Phi=3\nabla^{-2}\delta/(2a)$ up to a function of $a$, which is absorbed in $C(a)$. 
Solving the bracket for $\partial_a\chi$, and writing the continuity equation \eqref{eq:continuity} for $\mathbf w=\nabla\chi$, gives the two equations for $\delta$ and $\chi$,
\begin{align}
 \partial_a \delta&=-\nabla\!\cdot[(1+\delta)\nabla \chi],
 \label{eq:delta_eq}\\
 \partial_a \chi&=-\frac{3}{2a}\chi-\frac{3}{2a^2}\nabla^{-2}\delta-\frac12|\nabla \chi|^2+C(a).
 \label{eq:chi_eq}
\end{align}
Here $\nabla^{-2}$ is the inverse Laplacian on the nonzero modes of the periodic box, the mean of $\delta$ being zero by mass conservation. 
The spatially constant $C(a)$ is the function of $a$ left undetermined by that integration. 
It shifts $\chi$ by a function of $a$ alone, which leaves $\mathbf w=\nabla \chi$, and with it the density, unchanged. 
The choice $\int\chi\,\mathrm d^3r=0$ at every $a$ fixes it. 
Volume-averaging Eq.~\eqref{eq:chi_eq} over the box, whose volume is one, gives
\begin{equation*}
\partial_a\int\chi\,\mathrm d^3r
=-\frac{3}{2a}\underbrace{\int\chi\,\mathrm d^3r}_{=0}
-\frac{3}{2a^2}\underbrace{\int\nabla^{-2}\delta\,\mathrm d^3r}_{=0}
-\frac{1}{2}\int|\nabla\chi|^2\,\mathrm d^3r
+C(a)\underbrace{\int1\,\mathrm d^3r}_{=1},
\end{equation*}
where $\int\chi\,\mathrm d^3r$ vanishes because the mean of $\chi$ is zero at every $a$, so the left-hand side vanishes with it, and $\int\nabla^{-2}\delta\,\mathrm d^3r$ vanishes because the inverse Laplacian has no zero mode. 
What remains nonzero gives $C(a)=(1/2)\int|\nabla\chi|^2\,\mathrm d^3r$.

Every coefficient in Eqs.~\eqref{eq:delta_eq} and \eqref{eq:chi_eq} has a particle origin. 
The damping $3/(2a)$ comes from the expansion, the coefficient $3/(2a^2)$ is that acceleration with Poisson's equation already used to remove $\Phi$, and the $1/2$ comes from following a gradient velocity along its own trajectory. 
The same $\chi$ determines both the convergence and the transport of density~\cite{Sahni:1995rm}.

The particle initial condition is the Zel'dovich growing mode of Sec.~\ref{sec:5A}. 
The particle velocities are tied to the same displacement $\mathbf s_{\mathbf l}$ that sets the density.
Write $\delta_0^{\rm lin}$ for the contrast specified by Eq.~\eqref{eq:web_field} at $a_0$, whose Gaussian-smoothed standard deviation at that time is $0.05$, before the displaced particles are assigned to the mesh.
It is related to the initial displacement by $\delta_0^{\rm lin}=-\nabla\cdot\mathbf s$.
For the particles labelled by their initial grid sites, Eq.~\eqref{eq:zeldovich_v} and the change of clock give $\mathbf x'_{\mathbf l,0}=a_0^{1/2}\mathbf v_{\mathbf l,0}=\mathbf s_{\mathbf l}/a_0$.
Setting $\nabla\chi_0=\mathbf s/a_0$ and taking its divergence, with $\delta_0^{\rm lin}=-\nabla\cdot\mathbf s$, gives $\nabla^2\chi_0=-\delta_0^{\rm lin}/a_0$.

Let $h=1/M$ be the grid spacing and $\mathbf e_\mu$ the integer unit vector in direction $\mu$. The nearest-neighbour Laplacian of a scalar field $f$ sampled at $\mathbf r=\mathbf l/M$ is~\cite{Hockney1988}
\begin{equation}
(\Delta_h f)_{\mathbf l}=\frac{1}{h^2}\sum_{\mu=1}^3\bigl(f_{\mathbf l+\mathbf e_\mu}-2f_{\mathbf l}+f_{\mathbf l-\mathbf e_\mu}\bigr).
 \label{eq:field_laplacian}
\end{equation}
Applying it to $\mathrm{e}^{\mathrm i\mathbf k\cdot\mathbf r}$ gives the eigenvalue $-\kappa_{\rm grid}^2(\mathbf k)$, where
\begin{equation}
 \kappa_{\rm grid}^2(\mathbf k)=\frac{4}{h^2}\sum_{\mu=1}^3\sin^2\!\left(\frac{k_\mu h}{2}\right),
 \qquad
 \chi_{0,\mathbf k}=\frac{\delta^{\rm lin}_{0,\mathbf k}}{a_0\kappa_{\rm grid}^2(\mathbf k)}
 \quad(\mathbf k\ne\mathbf0),\qquad \chi_{0,\mathbf0}=0.
 \label{eq:field_initial_potential}
\end{equation}
This is the initial potential used in the calculations.

The field calculation uses the CIC deposit of the initial particle positions for $\delta_0$. Its initial potential is obtained from the linear density template by Eq.~\eqref{eq:field_initial_potential}, rather than by depositing the particle velocities. Thus the initial densities coincide, while the initial mesh velocities are constructed by the two different prescriptions.

\subsection{\label{sec:5C}Local amplification and depletion of the surroundings}

Expanding the divergence in Eq.~\eqref{eq:continuity} with $\mathbf w=\nabla\chi$ gives
\begin{equation*}
\nabla\cdot\bigl[(1+\delta)\nabla\chi\bigr]=(1+\delta)\nabla^2\chi+\nabla\chi\cdot\nabla\delta,
\end{equation*}
so that the density change splits into an activation $\mathcal A$ from the convergence of the flow and an inhibition $\mathcal I$ from the density the flow carries,
\begin{equation}{
\partial_a\delta=\underbrace{-(1+\delta)\nabla^2\chi}_{\mathcal A}\;\underbrace{-\nabla\chi\!\cdot\nabla\delta}_{\mathcal I}.
}\label{eq:AI}
\end{equation}
The convergence term $\mathcal A$ increases the density where the flow contracts.
Write $\delta=\epsilon\delta_1+O(\epsilon^2)$ and $\chi=\epsilon\chi_1+O(\epsilon^2)$ with $\epsilon\ll1$. The profiles $\delta_1,\chi_1$ and their spatial derivatives through second order remain bounded as $\epsilon\to0$, so the products of perturbations and gradients below are $O(\epsilon^2)$.
A product of two perturbations is then negligible against a single one, and the two equations read
\begin{subequations}
\begin{align}
\partial_a\delta&=-(1+\delta)\nabla^2\chi-\nabla\chi\cdot\nabla\delta
=-\nabla^2\chi\;\underbrace{-\delta\nabla^2\chi}_{O(\epsilon^2)}\;\underbrace{-\nabla\chi\cdot\nabla\delta}_{O(\epsilon^2)},
\label{eq:delta_lin}\\
\partial_a\chi&=-\frac{3}{2a}\chi-\frac{3}{2a^2}\nabla^{-2}\delta
\underbrace{-\frac{1}{2}|\nabla\chi|^2}_{O(\epsilon^2)}+C(a),
\label{eq:chi_lin}
\end{align}
\end{subequations}
so dropping the terms marked $O(\epsilon^2)$, applying $\nabla^2$ to the second equation, which removes $C(a)$ because that function depends on $a$ alone, and using the first, i.e., replacing $\nabla^2\chi$ with $-\partial_a\delta$, gives
\begin{equation}
 \partial_a^2\delta+\frac{3}{2a}\partial_a\delta-\frac{3}{2a^2}\delta=0.
\label{eq:growth}
\end{equation}
At each position, Eq.~\eqref{eq:growth} has the solution $\delta(\mathbf r,a)=C_1(\mathbf r)a+C_2(\mathbf r)a^{-3/2}$, where $C_1$ and $C_2$ are fixed by the initial density and its derivative~\cite{Peebles:1980yev,Bernardeau:2001qr}.
The Zel'dovich initial data of Sec.~\ref{sec:5A} tie the velocities to the displacement so that $\partial_a\delta=\delta/a$ at $a_0$, which sets $C_2=0$, and then $\partial_a\delta=\delta/a$ at every $a$. 
Eq.~\eqref{eq:delta_lin} then gives $\nabla^2\chi=-\delta/a$. The initial potential is calculated from the linear template $\delta_0^{\rm lin}$ through the discrete version of this relation, Eq.~\eqref{eq:field_initial_potential}. 
At the same order, $\mathcal A=-\nabla^2\chi+O(\epsilon^2)=\delta/a+O(\epsilon^2)$, so a positive perturbation grows. The discarded second-order terms cannot be determined from this linear relation.

The same flow must obtain the accumulating mass from somewhere. 
The transport term $\mathcal I$ is the advective part of the same flux. It is the change of density that the flow $\mathbf w=\nabla \chi$ produces as it moves material across the spatial slopes of $\delta$. 
The density drives $\chi$ through the zero-mean periodic inverse Laplacian. On the unit box $V=[0,1)^3$, its kernel representation is

\begin{equation*}
\begin{aligned}
 \nabla^{-2}\delta(\mathbf r)&=\int_VG_{\rm per}(\mathbf r-\mathbf r')\delta(\mathbf r')\,\mathrm d^3r',\\
 G_{\rm per}(\mathbf s)&=-\sum_{\mathbf k\ne\mathbf0}\frac{\mathrm{e}^{\mathrm i\mathbf k\cdot\mathbf s}}{k^2},\qquad\mathbf k=2\pi\mathbf n.
\end{aligned}
\end{equation*}
where $\mathbf n\in\mathbb Z^3$. The sum defines the periodic kernel through its Fourier coefficients. Applying $\nabla^2$ multiplies each nonzero mode by $-k^2$ and returns $\delta$, whose zero mode vanishes. The integral couples each point to density throughout the box.
Local amplification is therefore coupled to depletion over a region of the size of the box, which is the long-range inhibitory action in the model.

The two effects are tied by conservation. 
Integrating Eq.~\eqref{eq:AI} over the box gives $\partial_a\int\delta\,\mathrm d^3r=\int\mathcal A\,\mathrm d^3r+\int\mathcal I\,\mathrm d^3r$, and the box holds unit mass, $\int(1+\delta)\,\mathrm d^3r=1$ by Eq.~\eqref{eq:particle_fields}, so the left-hand side vanishes,
\begin{equation}
\int\mathcal I\,\mathrm d^3r=-\int\mathcal A\,\mathrm d^3r.
\label{eq:AI_mass}
\end{equation}
Eq.~\eqref{eq:AI_mass} fixes only the total. 
Whether moderate overdensities keep growing across the web, or most of the mass collects in a few condensations, depends on that spatial distribution. 
And because $\mathcal A$ and $\mathcal I$ are both built from the same $\mathbf w$, a change of the motion changes both.
Eq.~\eqref{eq:growth} contains no wave number, so every scale grows at the same rate. 
The link to Turing patterns is therefore the nonlinear coupling of an activation and a long-range inhibition, not the finite-wavelength linear instability of a reaction--diffusion system.

\subsection{\label{sec:5D}The two-field equations and the gravitational web}

The activation $\mathcal A$ and the inhibition $\mathcal I$ depend on two evolving fields, the density contrast $\delta$ and the velocity potential $\chi$.
In the generic notation $(u,v)$ of Sec.~\ref{sec:turing}, set $u\equiv\delta$ and $v\equiv\chi$.
Eqs.~\eqref{eq:AI} and \eqref{eq:chi_eq} then give
\begin{subequations}\label{eq:uv_toy}
\begin{align}
\partial_a u&=-(1+u)\nabla^2v-\nabla v\cdot\nabla u,\label{eq:uv_toy_density}\\
\partial_a v&=-\frac{3}{2a}v-\frac{3}{2a^2}\nabla^{-2}u-\frac12|\nabla v|^2+\frac12\int|\nabla v|^2\,\mathrm d^3r.
\label{eq:uv_toy_potential}
\end{align}
\end{subequations}
The last term is the $C(a)$ already fixed in Sec.~\ref{sec:5B} by $\int v\,\mathrm d^3r=0$.
Both time derivatives are therefore determined by the current $u,v$.
This is the two-field representation of the competition.
We return to the physical notation $\delta,\chi$ below, with $\mathbf w=\nabla\chi$ the velocity with respect to $a$.

We now compare the evolution of these two fields with that of the particles under the original force law.
Both calculations use the unit periodic box, $M=64$, $a_0=0.05$, and the same random realisation of Eq.~\eqref{eq:web_field}.
The particle number is $N=M^3$, each mass is $m=1/N$, and the largest step is $\Delta a=0.002$.

The density is stored at the sites, $\delta_{\mathbf l}$.
The velocity is stored on the faces of the mesh, i.e., we write $\mathbf l+\mathbf e_\mu/2$ for the face that separates the sites $\mathbf l$ and $\mathbf l+\mathbf e_\mu$, and $w_{\mu,\mathbf l+\mathbf e_\mu/2}$ for the velocity through that face along the axis $\mu$.
That velocity is the potential difference across the face,
\begin{equation*}
w_{\mu,\mathbf l+\frac12\mathbf e_\mu}=\frac{\chi_{\mathbf l+\mathbf e_\mu}-\chi_{\mathbf l}}{h},
\end{equation*}
To calculate mass flux, the density on a face is used, which is the value in the cell the flow comes from, extrapolated to the face by half the slope of that cell, \cite{VanLeer1979,LeVeque2002}
\begin{equation*}
\delta_{\mathbf l+\frac12\mathbf e_\mu}=
 \begin{cases}
  \delta_{\mathbf l}+\tfrac12 \theta_{\mu,\mathbf l},
   & w_{\mu,\mathbf l+\frac12\mathbf e_\mu}>0,\\[2pt]
  \delta_{\mathbf l+\mathbf e_\mu}-\tfrac12 \theta_{\mu,\mathbf l+\mathbf e_\mu},
   & w_{\mu,\mathbf l+\frac12\mathbf e_\mu}<0,
 \end{cases}
\end{equation*}
where the slope at a site is the smaller of its two one-sided differences, taken with that sign, and zero where the two differ in sign, \cite{LeVeque2002}
\begin{equation*}
\theta_{\mu,\mathbf l}=\mathrm{minmod}\bigl(\delta_{\mathbf l}-\delta_{\mathbf l-\mathbf e_\mu},\;\delta_{\mathbf l+\mathbf e_\mu}-\delta_{\mathbf l}\bigr),
\qquad
\mathrm{minmod}(x,y)=
\begin{cases}
\mathrm{sgn}(x)\min(|x|,|y|), & xy>0,\\
0, & xy\le0.
\end{cases}
\end{equation*}
The mass flux through the face is
\begin{equation*}
F_{\mu,\mathbf l+\frac12\mathbf e_\mu}=\bigl(1+\delta_{\mathbf l+\frac12\mathbf e_\mu}\bigr)\,w_{\mu,\mathbf l+\frac12\mathbf e_\mu},
\end{equation*}
and each face enters the two cells it separates with opposite signs,
\begin{equation}
 \partial_a\delta_{\mathbf l}=-\frac{1}{h}\sum_{\mu=1}^3\bigl(F_{\mu,\mathbf l+\frac12\mathbf e_\mu}-F_{\mu,\mathbf l-\frac12\mathbf e_\mu}\bigr),
 \label{eq:flux_update}
\end{equation}
so the sum of $1+\delta$ over the box is unchanged.
Where $\delta$ is smooth the slope makes the face value second-order accurate. 

The quadratic term of Eq.~\eqref{eq:uv_toy_potential} needs $\partial_\mu\chi$ at a site, where two one-sided differences are available, \cite{OsherShu1991}
\begin{equation*}
 \Bigl[\tfrac12|\nabla\chi|^2\Bigr]_{\mathbf l}
 =\frac12\sum_{\mu=1}^3
  \max\Bigl\{
   \Bigl[\max\Bigl(\frac{\chi_{\mathbf l}-\chi_{\mathbf l-\mathbf e_\mu}}{h},0\Bigr)\Bigr]^2,\;
   \Bigl[\min\Bigl(\frac{\chi_{\mathbf l+\mathbf e_\mu}-\chi_{\mathbf l}}{h},0\Bigr)\Bigr]^2
  \Bigr\},
\end{equation*}
which uses the backward difference when both are positive and the forward difference when both are negative. When their signs differ, it returns zero at a local minimum and half the larger squared one-sided slope at a local maximum, for each coordinate direction.

Time stepping is second-order Runge--Kutta, with $f$ the right-hand sides of Eqs.~\eqref{eq:uv_toy_density} and \eqref{eq:uv_toy_potential}, \cite{ShuOsher1988}
\begin{equation*}
 \begin{pmatrix}\delta^{n+1}\\ \chi^{n+1}\end{pmatrix}
 =\begin{pmatrix}\delta^{n}\\ \chi^{n}\end{pmatrix}
 +\frac{\Delta a}{2}\left[
  f\left(a,\begin{pmatrix}\delta^n\\\chi^n\end{pmatrix}\right)
  +f\left(a+\Delta a,\begin{pmatrix}\delta^n\\ \chi^n\end{pmatrix}
 +\Delta af\left(a,\begin{pmatrix}\delta^n\\\chi^n\end{pmatrix}\right)\right)\right].
\end{equation*}
The step is at most $\Delta a=0.002$ and is shortened so that \cite{Courant1967,LeVeque2002}

\begin{equation*}
 \Delta a\sum_{\mu=1}^3\max_{\mathbf l}
 \bigl|w_{\mu,\mathbf l+\frac12\mathbf e_\mu}\bigr|\le\frac{h}{5},
\end{equation*}
and it is halved, at most sixteen times, whenever a trial step would make $1+\delta$ negative. If sixteen halvings do not suffice, the run is stopped.
In the runs reported here $1+\delta$ stays positive and its spatial mean stays one, equivalently $h^3\sum_{\mathbf l}(1+\delta_{\mathbf l})=1$.

For the density comparison we smooth both calculations with a Gaussian of width $\ell_s=1.5h$ and write the result as $\rho_s$.
More generally, $\mathcal S_b$ will denote Gaussian smoothing of width $b$, defined in Fourier space by~\cite{Bernardeau:2001qr}
\begin{equation}
 (\mathcal S_b f)_{\mathbf k}=\mathrm{e}^{-b^2k^2/2}f_{\mathbf k},
 \qquad
 \frac{\rho_s}{\bar\rho}=1+\delta_s,
 \qquad \delta_s=\mathcal S_{\ell_s}\delta.
 \label{eq:density_smoothing}
\end{equation}

\begin{figure}[htbp]
\centering
\includegraphics[width=0.98\linewidth]{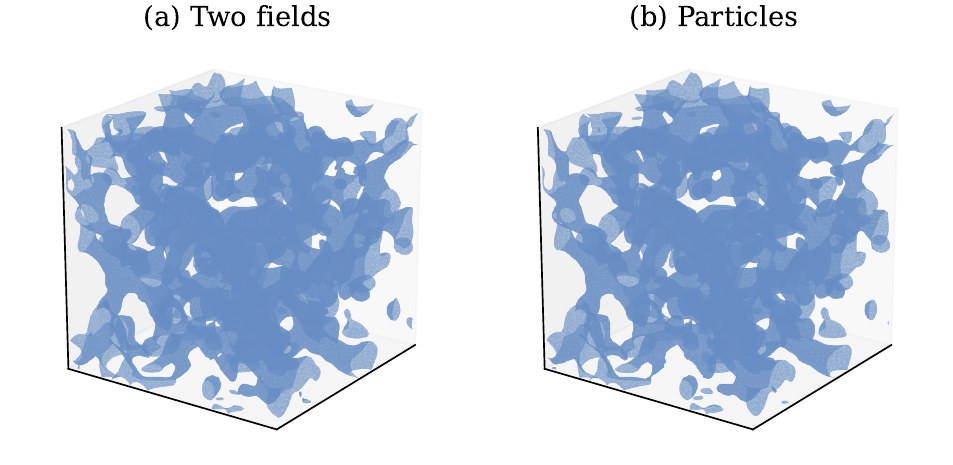}
\caption{The web produced by the two-field equations (left) and by the particles (right), with the original gravitational force.
Both runs use the same initial realisation and a $64^3$ mesh.
The surfaces show $\rho_s/\bar\rho=1.5$ at $a=1$, with the Gaussian width $\ell_s=1.5/64$ of Eq.~\eqref{eq:density_smoothing}. The isosurfaces are extracted with the marching-cubes implementation of Ref.~\cite{Lewiner2003}.
Each panel shows one unit periodic box, viewed from the same direction.}
\label{fig:v_web_matched}
\end{figure}
Fig.~\ref{fig:v_web_matched} compares the two densities at $a=1$.
Both descriptions produce a web of extended overdense structures enclosing depleted regions, with the main structures at corresponding locations.
The two-field and particle calculations give $\max(\rho_s/\bar\rho)=21.22$ and $23.96$, respectively.
Their `void fractions', defined as the fractions of the box with $\rho_s/\bar\rho<1$, are $0.678$ and $0.691$.
For a direct comparison of the spatial distributions, let superscripts $\mathrm f$ and $\mathrm p$ denote the field and particle calculations (in the following, the superscripts $\mathrm f$ and $\mathrm p$ are used with the same meaning).
Their correlation is
\begin{equation}
r_{\mathrm{fp}}=\frac{\sum_{\mathbf l}\delta^{\mathrm f}_{s,\mathbf l}\delta^{\mathrm p}_{s,\mathbf l}}{\left[\sum_{\mathbf l}(\delta^{\mathrm f}_{s,\mathbf l})^2\sum_{\mathbf l}(\delta^{\mathrm p}_{s,\mathbf l})^2\right]^{1/2}}=0.962\qquad(a=1).
\label{eq:web_correlation}
\end{equation}
The contrasts have zero spatial mean, so no mean subtraction is needed in this expression.
At $a=0.5$ and $0.7$ the same correlation is $0.989$ and $0.982$.
The two-field equations therefore reproduce the formation and spatial arrangement of the particle web.

The roles of $\mathcal A$ and $\mathcal I$ in Eq.~\eqref{eq:AI} can also be checked within the field equations by retaining only one at a time, while continuing to evolve $\chi$ with Eq.~\eqref{eq:chi_eq}.
Deleting either term disrupts the mass balance of Eq.~\eqref{eq:AI_mass}.
For this comparison we therefore impose the same total mass through
\begin{equation}
\partial_a\delta=R-(1+\delta)\int R\,\mathrm d^3r,
\qquad
R=\mathcal I\ \hbox{or}\ \mathcal A,
\qquad \int(1+\delta)\,\mathrm d^3r=1.
\label{eq:mass_fixed_ablation}
\end{equation}
Here $R$ is the retained rate. $R=\mathcal I$ deletes $\mathcal A$ and $R=\mathcal A$ deletes $\mathcal I$.
Integrating the right-hand side of Eq.~\eqref{eq:mass_fixed_ablation} gives $\int R-(\int R)\int(1+\delta)=0$.
The factor $\int R\,\mathrm d^3r$ is spatially uniform, so it changes $1+\delta$ everywhere at the same fractional rate.

Starting from the same $64^3$ initial fields, or their Fourier resampling to $16^3$ and $32^3$ with the mean retained, deleting $\mathcal A$ leaves the density close to uniform. The unsmoothed maxima of $\rho/\bar\rho$ at $a=1$ are $1.177$ on $16^3$, $1.272$ on $32^3$, and $1.373$ on $64^3$.
Deleting $\mathcal I$ instead concentrates the mass at one grid site, reaching $\max(\rho/\bar\rho)=M^3$ at $a=0.516$ on $16^3$, $0.440$ on $32^3$, and $0.290$ on $64^3$.
This is the largest density ratio the mesh can represent, $\rho/\bar\rho=1/h^{3}=M^{3}$, reached when the whole unit mass sits in a single cell.
Thus the two deletions remove the web in different ways. Without $\mathcal A$ the density stays near uniform, and without $\mathcal I$ it ends in a single condensation at the mesh scale.
These ablation tests exist only in the two-field representation because the particle calculation has one force law, and $\mathcal A$ and $\mathcal I$ cannot be varied separately.

\subsection{\label{sec:5E}Changing the competition from a web to spots}

The two-field equations also indicate how to change the competition.
$\mathcal A$ depends on the convergence $-\nabla^2\chi$, while $\mathcal I$ depends on the flow $\nabla\chi$ carrying matter across the spatial slopes of $\delta$.
Both are driven by the nonlocal source $\nabla^{-2}\delta$ in Eq.~\eqref{eq:chi_eq}.
Increasing the long-wavelength part of this source strengthens the flows through which condensations draw on a common supply.
The expected change is that the densest condensations continue to gain matter while the smaller ones between them lose their supply, turning a web into spots.

To strengthen the long-range inhibition we amplify the long-wavelength part of the source in the potential equation and leave its short-wavelength part unchanged.
The amplification is set by a dimensionless parameter $\eta\ge1$. 
Each mode is multiplied by $F_\eta(k)$ below, which tends to $\eta$ for $k\ell\ll1$ and to $1$ for $k\ell\gg1$. The choice $\eta=1$ leaves the source unchanged.
We therefore replace $\nabla^{-2}\delta$ in Eq.~\eqref{eq:chi_eq} by $\nabla^{-2}\mathcal F_\eta\delta$, where
\begin{equation}
\mathcal F_\eta\delta=\delta+(\eta-1)\mathcal S_\ell\delta,
\qquad
F_\eta(k)=1+(\eta-1)\mathrm{e}^{-\ell^2k^2/2},
\qquad 
\ell=\frac{1}{k_0},\quad k_0=8\pi.
\label{eq:range_kernel}
\end{equation}
The operator $\mathcal S_\ell$ is the Gaussian smoothing defined in Eq.~\eqref{eq:density_smoothing}, and $F_\eta(k)$ is the Fourier multiplier of $\mathcal F_\eta$.
We choose $k_0$ to be the characteristic wave number already appearing in the initial spectrum, Eq.~\eqref{eq:web_field}, so that $\ell=1/(8\pi)$ is $2.5$ cells on the $64^3$ mesh, and keep it fixed throughout the comparison.
The modified equations are
\begin{subequations}\label{eq:twofield_eta}
\begin{align}
\partial_a\delta&=-(1+\delta)\nabla^2\chi-\nabla\chi\cdot\nabla\delta,
\label{eq:twofield_eta_density}\\
\partial_a\chi&=-\frac{3}{2a}\chi-\frac{3}{2a^2}\nabla^{-2}\bigl[\delta+(\eta-1)\mathcal S_\ell\delta\bigr]-\frac12|\nabla\chi|^2+C(a).
\label{eq:twofield_eta_potential}
\end{align}
\end{subequations}
The zero-mean condition still fixes $C(a)=(1/2)\int|\nabla\chi|^2\,\mathrm d^3r$, because the filtered contrast also has zero mean.

Using the same approximation as in Eq.~\eqref{eq:growth},
\begin{equation}
\partial_a^2\delta=\frac{3}{2a}\nabla^2\chi+\frac{3}{2a^2}\mathcal F_\eta\delta=-\frac{3}{2a}\partial_a\delta+\frac{3}{2a^2}\mathcal F_\eta\delta.
\label{eq:eta_growth_derivation}
\end{equation}
For a nonzero Fourier mode this is
\begin{equation}
\partial_a^2\delta_{\mathbf k}+\frac{3}{2a}\partial_a\delta_{\mathbf k}-\frac{3F_\eta(k)}{2a^2}\delta_{\mathbf k}=0.
\label{eq:eta_growth_equation}
\end{equation}
The coefficients of Eq.~\eqref{eq:eta_growth_equation} are powers of $a$, with the exponent matching the order of the derivative, namely $1$, $a^{-1}$ and $a^{-2}$.
A power $\delta_{\mathbf k}\propto a^p$ therefore makes every term carry the same factor of $a$. With $\partial_a(a^p)=p\,a^{p-1}$ and $\partial_a^2(a^p)=p(p-1)a^{p-2}$, Eq.~\eqref{eq:eta_growth_equation} becomes
\begin{equation*}
a^{p-2}\Bigl[p(p-1)+\frac{3}{2}p-\frac{3}{2}F_\eta(k)\Bigr]=0.
\end{equation*}
Division by $a^{p-2}$ leaves $p(p-1)+3p/2-3F_\eta(k)/2=0$, whose roots are $p=\bigl(\pm\sqrt{1+24F_\eta(k)}-1\bigr)/4$. 
The growing solution is $\delta_{\mathbf k}\propto a^{p_\eta(k)}$, with
\begin{equation}
p_\eta(k)=\frac{\sqrt{1+24F_\eta(k)}-1}{4}.
\label{eq:kernel_growth}
\end{equation}
For the original force $p_1(k)=1$ at every wave number.
When $k\ell\ll1$, $F_\eta(k)\simeq\eta$. When $k\ell\gg1$, $F_\eta(k)\simeq1$.
Thus increasing $\eta$ makes long-wavelength flows grow faster relative to short-wavelength ones.
For the value used in the runs below, $\eta=4$, the long-wavelength exponent is $(\sqrt{97}-1)/4\simeq2.21$ while the short-wavelength one stays at $1$, which indicates that long-wavelength modes grow more than twice as fast as short-wavelength ones.
In the nonlinear equations these flows change both $\mathcal A$ and $\mathcal I$. They enhance convergence into dense regions and the removal of matter from their surroundings together.

The same $\eta$ can be put into the particle calculation.
Replace the gravitational potential in Eq.~\eqref{eq:particle_a_motion} by $\Phi^{(\eta)}$, with
\begin{subequations}\label{eq:matched_force}
\begin{align}
\mathbf x_i''+\frac{3}{2a}\mathbf x_i'&=-\frac{1}{a}\nabla\Phi^{(\eta)}(\mathbf x_i),
\label{eq:matched_particle_motion}\\
\nabla^2\Phi^{(\eta)}&=\frac{3}{2a}\mathcal F_\eta\delta,
\qquad
\Phi_{\mathbf k}^{(\eta)}=-\frac{3}{2a}\frac{F_\eta(k)}{k^2}\delta_{\mathbf k}
\quad(\mathbf k\ne\mathbf0).
\label{eq:matched_particle_potential}
\end{align}
\end{subequations}
The zero mode of $\Phi^{(\eta)}$ is set to zero.
The expression for $\Phi^{(\eta)}_{\mathbf k}$ follows by replacing $\nabla^2$ with $-k^2$ and using the definition of $F_\eta$.
Conversely, inserting $\Phi^{(\eta)}=3/2a\nabla^{-2}\mathcal F_\eta\delta$ into the integrated potential equation before eliminating $\Phi$, $\partial_a\chi=-3\chi/(2a)-\Phi/a-|\nabla\chi|^2/2+C(a)$ gives the modified source in Eq.~\eqref{eq:twofield_eta_potential}.
The change can therefore be implemented by multiplying each nonzero density mode by the same $F_\eta(k)$ before solving for the force in either calculation.
The particle calculation divides by $k^2$ as in Eq.~\eqref{eq:poisson_k}, while the field calculation divides by $\kappa_{\rm grid}^2(\mathbf k)$ as in Eq.~\eqref{eq:field_initial_potential}. 
Each retains the spatial differences of its original solver.
The initial positions, masses, velocities, density phases and $\chi_0$ of Eq.~\eqref{eq:field_initial_potential} are held fixed as $\eta$ changes.
In particular, the initial velocities are not reset to the new growing exponent of Eq.~\eqref{eq:kernel_growth}.
Both descriptions conserve total mass for every $\eta$.

Changing $\eta$ changes the gravitational force law of the toy universe itself.
For each nonzero Fourier mode, Eq.~\eqref{eq:matched_particle_potential} amounts to replacing $G$ by the effective coupling $G_{\rm eff}(k)\equiv G F_\eta(k)$, which approaches $G$ at short wavelengths and $\eta G$ at long wavelengths.
Related changes of gravity have been studied as tests of departures from the Newtonian force law, motivated in part by possible gravitational explanations of cosmic acceleration, with their parameters constrained by galaxy clustering~\cite{Sealfon:2004gz,Shirata:2005yr}.
For example, the Yukawa modification considered in Ref.~\cite{Shirata:2005yr} has Fourier multiplier $F_{\rm Y}(k,a)=1+\alpha/[1+(k\lambda/a)^2]$, where $\alpha$ is the dimensionless change in the long-wavelength coupling and $\lambda$ is a fixed physical length.
For $\alpha=\eta-1$ it has the same short- and long-wavelength limits as our prescription, but a different interpolation and a fixed physical scale, whereas our $\ell$ is fixed in comoving coordinates.
The Gaussian form in Eq.~\eqref{eq:range_kernel} is our choice for the present mechanism test.
The lengths and density profiles of cosmic filaments have also been investigated as probes of modified gravity~\cite{Ho:2018byw}.
In this spirit, the web--spot response below provides a morphology diagnostic of this force family at fixed initial conditions and epoch.
A corresponding comparison with the observed web could constrain such enhancements in a cosmological application with specified physical scales, initial density spectrum and expansion history.

\begin{figure}[htbp]
\centering
\includegraphics[width=0.98\linewidth]{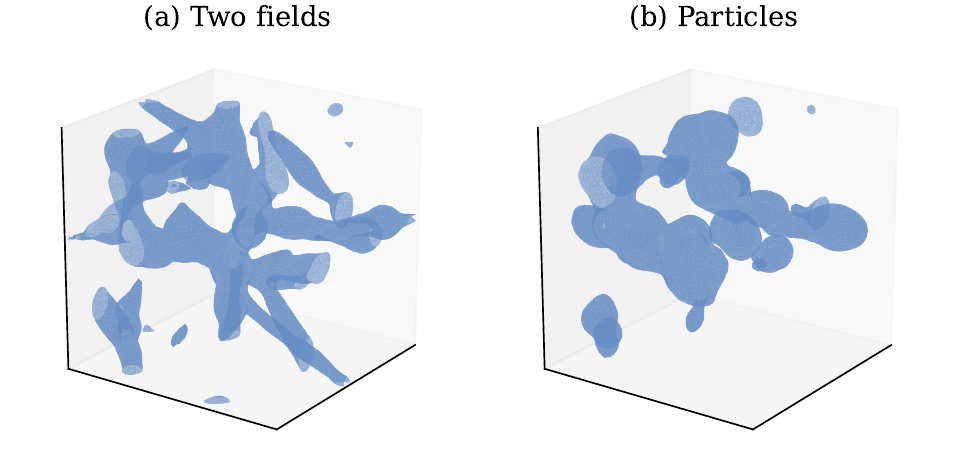}
\caption{The spot-dominated configuration at $\eta=4$ in the two-field model (left) and in the particle model (right).
The force multiplier $F_\eta(k)$, its length $\ell=1/(8\pi)$, and the initial realisation are identical in the two calculations.
As in Fig.~\ref{fig:v_web_matched}, the surfaces show $\rho_s/\bar\rho=1.5$ at $a=1$ after smoothing over $1.5$ cells on a $64^3$ mesh, with the same box and viewing direction.
The two-field model retains more extended bridges, whereas the particle concentrations are rounder.}
\label{fig:v_spot_matched}
\end{figure}
At $\eta=4$, the field calculation and the particle calculation both collect matter into fewer dominant condensations and leave larger depleted regions, as shown in Fig.~\ref{fig:v_spot_matched}.
We call this a spot-dominated configuration. Most of the mass is concentrated in a few condensations, although sheets and bridges remain, especially in the field calculation.

\begin{table}[htbp]
\centering
\caption{Web and spot-dominated configurations from the same initial realisation at $a=1$.
All quantities use the common smoothed density $\rho_s$.}
\label{tab:web}
\setlength{\tabcolsep}{7pt}
\begin{tabular}{lrrrr}
\hline
Description & $\eta$ & $\max(\rho_s/\bar\rho)$ & Void fraction & Dense nodes\\
\hline
Two fields & 1 & 21.22 & 0.678 & 58\\
Particles  & 1 & 23.96 & 0.691 & 62\\
Two fields & 4 & 534.21 & 0.904 & 8\\
Particles  & 4 & 204.37 & 0.900 & 11\\
\hline
\end{tabular}
\end{table}
Table~\ref{tab:web} compares this result with the original web at the same $a=1$.
The void fraction rises from $0.678$ to $0.904$ in the field calculation and from $0.691$ to $0.900$ in the particle calculation.
We define a `dense node' as a group of at least eight grid sites with $\rho_s/\bar\rho\ge5$, connected through faces, edges or corners, including across the periodic boundary.
The number of dense nodes falls from $58$ to $8$ in the field calculation and from $62$ to $11$ in the particle calculation.
The spatial correlation of Eq.~\eqref{eq:web_correlation} is $0.842$ at $a=0.5$, $0.671$ at $a=0.7$ and $0.326$ at $a=1$ for these $\eta=4$ runs.
Thus the two descriptions agree on the change toward fewer condensations and a larger depleted volume, while the late-time density of individual condensations agrees less closely than in the web case. $\max(\rho_s/\bar\rho)$ is $534.21$ and $204.37$.

\subsection{\label{sec:5F}Agreement and redistribution in the modified runs}

\begin{figure}[htbp]
\centering
\includegraphics[width=0.98\linewidth]{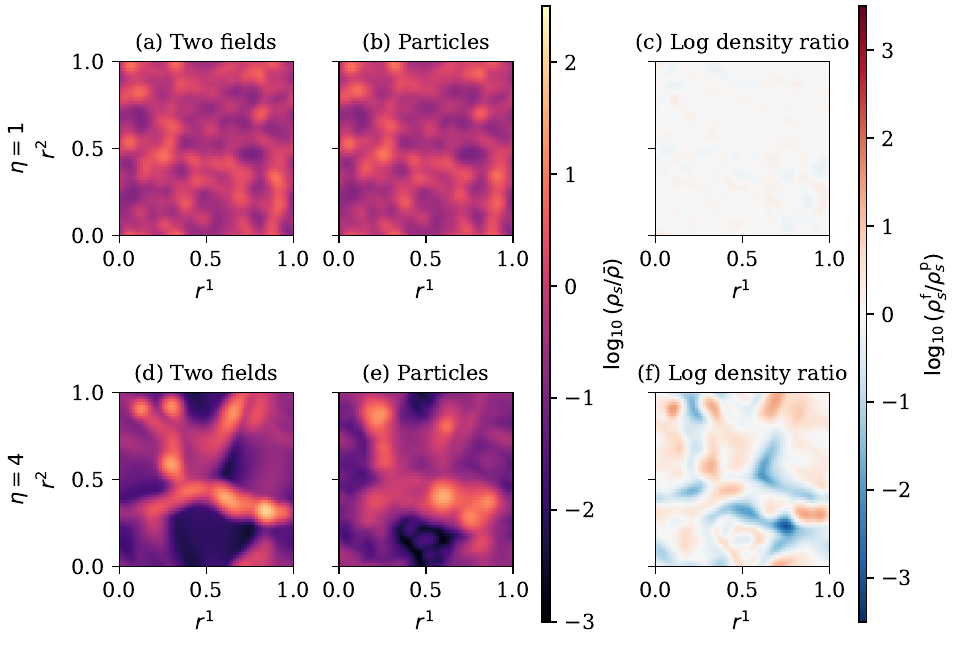}
\caption{Density slices at $a=1$, taken at the same coordinate $r^3=1/2$.
The upper row has $\eta=1$ and the lower row $\eta=4$.
Columns one and two show $\log_{10}(\rho_s/\bar\rho)$ in the field and particle calculations, with one colour scale for all four panels.
The last column shows their logarithmic density ratio $\log_{10}(\rho_s^{\mathrm f}/\rho_s^{\mathrm p})$. Red means the field calculation is denser, blue means the particle calculation is denser, and white means the two agree.
The same positions are compared without a relative shift or a change of density normalisation.}
\label{fig:v_density_slices}
\end{figure}
The agreement can be examined beyond the single isodensity level used in Figs.~\ref{fig:v_web_matched} and \ref{fig:v_spot_matched}.
We compare the field and particle densities at the same positions, in Fourier amplitude and phase, and at the three times $a=0.5$, $0.7$ and $1$.
All comparisons use the same $64^3$ realisation and the Gaussian smoothing of Eq.~\eqref{eq:density_smoothing}.
Fig.~\ref{fig:v_density_slices} takes the fixed central plane $r^3=1/2$ in each box.
The first two columns show the field and particle densities on a common colour scale. 
The third shows $\log_{10}(\rho_s^{\mathrm f}/\rho_s^{\mathrm p})$ at each position, which vanishes where the densities agree.
On the $\eta=1$ slice the two densities differ by at most a factor $10^{0.277}=1.9$ at any position.
For $\eta=4$ the largest ratio on the slice is $10^{3.054}=1.1\times10^3$. The condensations also sit at different positions and have different shapes in the two calculations.

To distinguish amplitude agreement from agreement of the spatial arrangement, we next compare the Fourier modes of the same smoothed contrasts.
Let $\mathcal B_j$ be the set of wavevectors in the $j$-th spherical Fourier shell of this three-dimensional unit box. The shell contains $N_j$ modes. The shells of Sec.~\ref{sec:pure} are annuli in the two-dimensional lattice Fourier plane.
The shells have width $2\pi$, start at $k=\pi$, and end at $k=\pi M$, with the last shell shortened at that limit.
The latter is the largest wave number along a grid axis, corresponding to a wavelength of two grid spacings.
Writing $\delta_{s,\mathbf k}$ for the transform of $\delta_{s,\mathbf l}$ under Eq.~\eqref{eq:poisson_k}, the mean squared Fourier amplitude in a shell and the mean product of the two amplitudes are
\begin{subequations}\label{eq:density_spectral_comparison}
\begin{align}
P_j^{\mathrm{f,p}}&=\frac{1}{N_jM^6}\sum_{\mathbf k\in\mathcal B_j}|\delta^\mathrm{f,p}_{s,\mathbf k}|^2,\\
c_j&=\frac{1}{\sqrt{P_j^{\mathrm f}P_j^{\mathrm p}}}\frac{1}{N_jM^6}\sum_{\mathbf k\in\mathcal B_j}\operatorname{Re}\!\left[\delta^{\mathrm f}_{s,\mathbf k}(\delta^{\mathrm p}_{s,\mathbf k})^*\right].
\end{align}
\end{subequations}
Here the star denotes complex conjugation and $\operatorname{Re}$ the real part.
The factor $M^{-6}$ puts the shell means in the normalisation of that transform, for which $M^{-6}\sum_{\mathbf k}|\delta_{s,\mathbf k}|^2$ is the mean of $\delta_s^2$ over the $M^3$ sites.
The ratio $P_j^{\mathrm f}/P_j^{\mathrm p}$ measures the relative squared amplitudes, and $c_j$ their normalised Fourier correlation, which depends on both relative phases and relative amplitudes within the shell.
If the field and particle densities differ in one shell only by a common positive amplitude factor, $c_j=1$ even when the power ratio differs from one. A displacement of the condensations changes the relative phases and can reduce $c_j$ without changing either power.

\begin{figure}[htbp]
\centering
\includegraphics[width=0.96\linewidth]{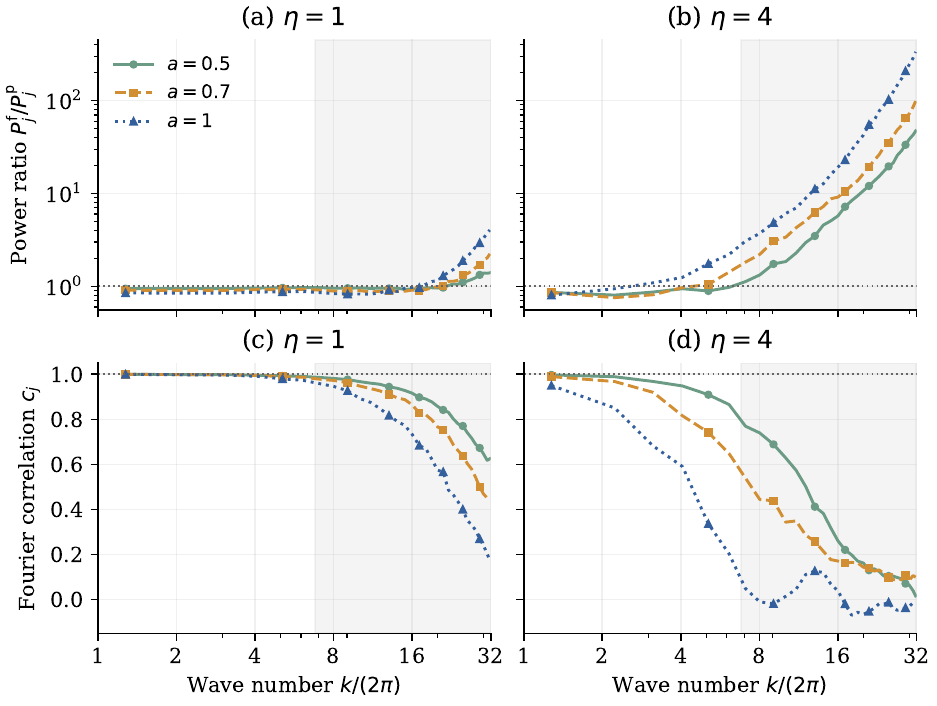}
\caption{Comparison of Fourier amplitudes and spatial arrangement for the same density fields.
The upper row gives the power ratio of the field and particle calculations. The lower row gives the Fourier correlation $c_j$ of Eq.~\eqref{eq:density_spectral_comparison}.
Columns correspond to $\eta=1$ and $4$. The three times are distinguished by line style and marker as well as by colour.
The horizontal coordinate is the mean $k/(2\pi)$ in each shell. The dotted lines mark unity.
The shaded region has $k\ell_s\ge1$, where the smoothing of Eq.~\eqref{eq:density_smoothing} suppresses the absolute amplitudes. The smoothing factor cancels in a single-mode power ratio. In a shell sum it weights modes differently, so neither the shell power ratio nor $c_j$ is generally invariant under smoothing.}
\label{fig:v_scale_comparison}
\end{figure}
Fig.~\ref{fig:v_scale_comparison} shows both quantities at $a=0.5$, $0.7$ and $1$.
Evaluating the sums of Eq.~\eqref{eq:density_spectral_comparison} over all modes of the band $0<k/(2\pi)\le4$ at $a=1$, with the common normalisation cancelling in both quantities, gives a power ratio of $0.854$ and a Fourier correlation of $0.996$ for the web, and $1.031$ and $0.759$ for $\eta=4$. At $\eta=4$ the total power in this band therefore differs by $3.1\%$, while the spatial arrangement does not.
At shorter wavelengths the field calculation carries more power. In $8<k/(2\pi)\le16$ at $a=1$ the ratio is $5.96$ and $c_j$ is $0.024$ for $\eta=4$, against $0.836$ and $0.898$ for the web. The loss of agreement is therefore at short wavelengths and at late times.

To measure how density is redistributed, we use two quantities alongside the void fraction. The root mean square (RMS) smoothed contrast is $\sigma_s$, and $f_{\rm dense}$ is the fraction of the total mass in sites with $\rho_s/\bar\rho\ge5$,
\begin{equation}
\sigma_s=\left[\frac{1}{M^3}\sum_{\mathbf l}\delta_{s,\mathbf l}^2\right]^{1/2},
\qquad
f_{\rm dense}=\frac{\displaystyle\sum_{\mathbf l:\,1+\delta_{s,\mathbf l}\ge5}(1+\delta_{\mathbf l})}{\displaystyle\sum_{\mathbf l}(1+\delta_{\mathbf l})}.
\label{eq:density_history_diagnostics}
\end{equation}
The mass fraction uses the smoothed density to select sites and their unsmoothed density to count mass.

\begin{figure}[htbp]
\centering
\includegraphics[width=0.98\linewidth]{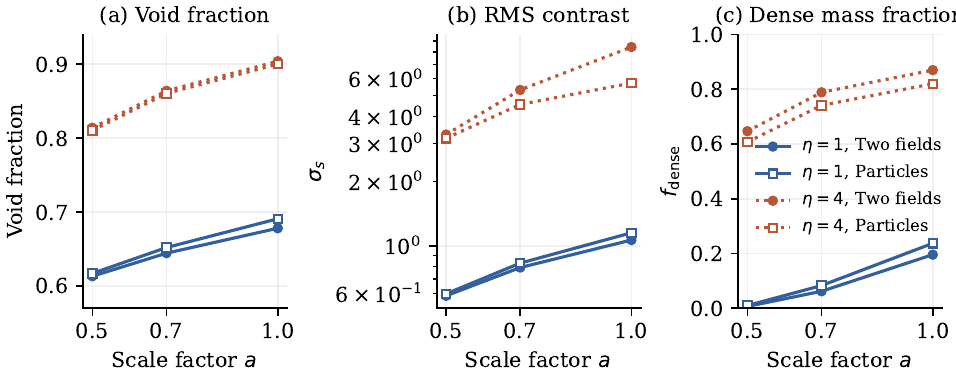}
\caption{Evolution of the density statistics in the field and particle calculations.
Panels show (a) the void fraction, (b) $\sigma_s$ and (c) $f_{\rm dense}$.
The last two quantities are defined in Eq.~\eqref{eq:density_history_diagnostics}.
Solid lines denote $\eta=1$ and dotted lines $\eta=4$. Filled circles denote the field calculation and open squares the particle calculation.
Markers are the three times $a=0.5$, $0.7$ and $1$ from the same realisation. 
Lines connect them to guide the eye.}
\label{fig:v_comparison_history}
\end{figure}
Apart from that, Fig.~\ref{fig:v_comparison_history} compares how density is redistributed as $a$ increases.
Both calculations develop a larger depleted volume and concentrate more mass in dense regions when $\eta$ changes from $1$ to $4$.
At $a=1$, $f_{\rm dense}$ is $0.196$ and $0.237$ in the field and particle web, and $0.870$ and $0.820$ in the spot case.
The spot void fractions differ by at most $0.004$ at the three times, while $\sigma_s$ in the field calculation grows above the particle value, reaching $8.435$ against $5.729$ at $a=1$.
These comparisons separate the shared redistribution of matter from the differences in the density of the final condensations.

\begin{table}[htbp]
\centering
\caption{Regional activation and inhibition in the field calculation at $a=1$.
Each rate is first smoothed over $1.5$ cells and then averaged over the sites in the stated density interval.
The net rate is the mean of $(\partial_a\delta)_h$. 
Its last digit can differ from the sum of the two rounded entries.}
\label{tab:AI_budget}
\setlength{\tabcolsep}{7pt}
\begin{tabular}{ccrrr}
\hline
$\eta$ & Density interval & Activation & Inhibition & Net rate\\
 & $\rho_s/\bar\rho$ & $\mathcal A_h$ & $\mathcal I_h$ & $(\partial_a\delta)_h$\\
\hline
1 & $[1.5,5)$ & 22.554 & $-21.693$ & $+0.861$\\
4 & $[1.5,5)$ & 197.661 & $-201.129$ & $-3.468$\\
1 & $[5,\infty)$ & 132.278 & $-126.451$ & $+5.826$\\
4 & $[5,\infty)$ & 5365.799 & $-5353.796$ & $+12.002$\\
\hline
\end{tabular}
\end{table}
To see how the change from the web to the spot case follows the activation--inhibition competition, we measure $\mathcal A$ and $\mathcal I$ on the fields already evolved.
Let $(\partial_a\delta)_h$ denote the rate that the flux of Eq.~\eqref{eq:flux_update} applies, that is, the right-hand side of that equation.
Using the Laplacian of Eq.~\eqref{eq:field_laplacian}, the corresponding decomposition is
\begin{equation}
\mathcal A_h=-(1+\delta)\Delta_h\chi,
\qquad \mathcal I_h=(\partial_a\delta)_h-\mathcal A_h,
\qquad (\partial_a\delta)_h=\mathcal A_h+\mathcal I_h.
\label{eq:discrete_competition}
\end{equation}
The remainder $\mathcal I_h$ in Eq.~\eqref{eq:discrete_competition} includes the density interpolation used in the flux, so the sum is the rate actually applied in the evolution.
We smooth $(\partial_a\delta)_h$, $\mathcal A_h$ and $\mathcal I_h$ with the same $\mathcal S_{\ell_s}$ used for the density, then average each over the sites in a specified interval of $\rho_s/\bar\rho$.
Table~\ref{tab:AI_budget} gives these arithmetic means at $a=1$, in units of contrast change per unit scale factor.

For the original web, $\mathcal A_h$ exceeds $|\mathcal I_h|$ in both density intervals.
At $\eta=4$ both rates are larger in magnitude, but their balance changes. The sites with $1.5\le\rho_s/\bar\rho<5$ have a negative net rate, $-3.468$, while those with $\rho_s/\bar\rho\ge5$ keep growing at $+12.002$.
Growth is therefore concentrated in the sites that are already dense.
We assign sites to intervals of $\rho_s/\bar\rho$ and sum their unsmoothed mass. The fraction in $\rho_s/\bar\rho\ge5$ rises from $0.196$ to $0.870$.
The fraction in $1\le\rho_s/\bar\rho<5$ falls from $0.575$ to $0.092$.
The same flow therefore amplifies the dominant condensations and depletes the material that supports the structures between them.
The change of relative growth rates, the regional density budget and the matched particle force together give the route from the web to the spot-dominated configuration.

\section{\label{sec:6}Summary}

The repeated appearance of spots, bands, and networks in unrelated systems motivated a question about the Turing mechanism. 
Are the conditions for Turing-like pattern formation easy to realise? 
Our results support a constructive answer. 
Local amplification and spatially extended suppression can arise from simple counting rules or from the redistribution of a conserved supply, and both settings produce changes between familiar pattern geometries.

In the cellular automaton, adding a distant neighbour count to the Game of Life produces a map of spots, stripes, coexistence, and disordered states. 
The scan uses $2000$ realisations at each of the $18\times18$ threshold pairs. 
The more direct test is the rule of Sec.~\ref{sec:pure}, in which a nearby count favours activity and a distant count suppresses it, with no separate birth or survival conditions.
It still produces islands, stripes, and holes. 
Thus these morphologies do not require the particular arithmetic of the Game of Life.

At $A=1$ in this rule, raising the inhibition threshold $I$ from $4$ to $16$ increases the mean number of consecutive live cells along a row or column from $1.4$ to $6.6$ lattice spacings. Restricting the distant count to one axis selects bands along the other, at densities differing by less than $0.2\%$.

The self-gravitating toy universe supplies a different realisation. 
Convergence increases density in contracting regions, while the same mass flow depletes their surroundings.
The two-field description reproduces the smoothed particle density with spatial correlation $0.962$ at scale factor $a=1$ under the original force law. 
Strengthening its long-wavelength component moves both calculations from a web toward fewer dense condensations. Within the field description, deleting either convergence or transport while preserving total mass removes the web, producing a nearly uniform density or a single condensation, respectively. 
Convergence therefore plays the part of the activation, since a dense region attracts matter and grows denser, and redistribution plays the part of the inhibition, since the same mass flow depletes the surroundings of that region. 
A reaction--diffusion system needs a second chemical species to play the inhibitor, whereas this toy universe has matter only, and its inhibition is the transport of that same matter, so no additional species is introduced.

Read together, the two tests say that little is required.
The mechanism survived the removal of continuous amplitudes, of a diffusion operator, and of a uniform steady state that a finite wavelength could destabilise. It survived being rewritten as mass conservation in a self-gravitating system, where the opposition is not a second species that diffuses faster but the fact that a growing concentration takes its matter from its surroundings.
What the tests do not show is as specific as what they do.
Neither shows that the textures listed at the outset, the flux tubes of a confining gauge theory or the filaments of the cosmic web, are produced by these ingredients. That a mechanism is easy to realise does not say what any particular system realises.
The automaton spectra have maxima near the box scale or the lattice cutoff, as quantified in Sec.~\ref{sec:pure}, and the corresponding real-space feature sizes vary with the thresholds.
The morphology map is drawn at one lattice size and several of its thresholded observables depend on system size, so its boundaries do not transfer to another $L$ without rescaling.
The toy universe was tested at one mesh, one box and one initial realisation, and its agreement with the particles fails at short wavelengths and late times.
Each of these four is also the next thing to test.

\begin{acknowledgments}
This work was supported in part by the National Natural Science Foundation of China under Grants No. 12575106 and 12147214, and by the Basic Scientific Research Project of the Education Department of Liaoning Province under Grant No. LJ212510165024.
\end{acknowledgments}
\begin{samepage}
\medskip
\noindent\textbf{Data availability.} The data and the programs used in this paper are available at \url{https://www.modelscope.cn/datasets/nbalexis/turing_pattern_game_of_life_phase_data_and_gravity_web_data}.

\medskip
\noindent\textbf{Use of generative AI.} The derivation and the numerical calculations of Sec.~\ref{sec:5} were carried out with the assistance of the language models \mbox{deepseek-v4.1-flash} and \mbox{gpt6-aster}. The text of the paper was polished with \mbox{deepseek-v4.1-flash} and \mbox{qwen3.8-flash}. The authors have verified all results and take full responsibility for the content of this paper.
\end{samepage}

\bibliography{turing}

\end{document}